 \documentclass[final,3p,times]{elsarticle}

\usepackage{amssymb, bm}
\usepackage{amsthm, amsmath}
\usepackage{multirow}
\usepackage{mathtools,xparse}
\usepackage{float}
\restylefloat{table}
\usepackage{chemformula}
\usepackage{graphicx,caption,subcaption}
\usepackage{hyperref}
\usepackage[T1]{fontenc} 
\usepackage{xcolor} 

\graphicspath{{figures/case1/},{figures/case2/},{figures/}}
\biboptions{numbers,sort&compress}

\newcommand{\abs}[1]{\lvert#1\rvert}
\newcommand{\norm}[1]{\lVert#1\rVert}

\newcommand{\sprod}[2][\Omega]{\Big(#2\Big)_{#1}}
\newcommand{\dprod}[2][\Gamma]{\Big\langle#2\Big\rangle_{#1}}

\newcommand{\Div}{{\nabla}\cdot}

\newcommand{\Grad}{\nabla}
\newcommand{\pd}[2]{\frac{\partial{#1}}{\partial{#2}}}

\newcommand{\RR}{\mathbb{R}}
\newcommand{\testwh}{\ensuremath{\bm{\mathcal{W}}^k_h}}
\newcommand{\testqh}{\ensuremath{\bm{\mathcal{Q}}^k_h}}
\newcommand{\elempart}{\mathcal{T}_h}
\newcommand{\boundpart}{\partial \elempart}

\newcommand{\eltwo}{\ensuremath{\mathcal{L}_2}}

\newcommand{\Polyk}{\ensuremath{\mathcal{P}^{k}}}
\newcommand{\ndofs}{\ensuremath{\texttt{n}_{\texttt{dof}}}}
\newcommand{\ntime}{\ensuremath{\texttt{n}_{\texttt{time}}}}

\newcommand{\tscheme}{\ensuremath{\texttt{t}_{\texttt{scheme}}}}
\newcommand{\bu}{\bm{u}}
\newcommand{\bv}{\bm{v}}
\newcommand{\bn}{\bm{n}}

\newcommand{\br}{\bm{r}}
\newcommand{\bg}{\bm{g}}
\newcommand{\ba}{\bm{a}}
\newcommand{\bomega}{\bm{\omega}}

\newcommand{\cv}{c_v}
\newcommand{\cp}{c_p}
\newcommand{\kB}{k_B}

\newcommand{\logr}{\varrho}
\newcommand{\sr}{\sqrt{\rho}}
\newcommand{\srv}{\sr \bv}
\newcommand{\srT}{\sr T}

\newcommand{\bq}{\bm{q}}
\newcommand{\bF}{\bm{F}}
\newcommand{\bs}{\bm{s}}

\newcommand{\bhF}{\widehat{\bF}}
\newcommand{\bhu}{\widehat{\bu}}

\newcommand{\buh}{\bu_h}
\newcommand{\bqh}{\bq_h}
\newcommand{\bhuh}{\bhu_h}
\newcommand{\bhFh}{\bhF_h}

\newcommand{\qeuv}{q_{_\text{EUV}}}
\newcommand{\modtwopi}[1]{\mod_{2 \pi}}
\newcommand{\btau}{\boldsymbol{\tau}}
\newcommand{\bsigma}{\boldsymbol{\sigma}}
\newcommand{\beps}{\boldsymbol{\varepsilon}}

\newcommand{\lamax} {\lambda_{\texttt{max}}}
\newcommand{\Id}{\mathbf{I}}
\newcommand{\Idm}[1]{\Id_{#1}}

\newcommand{\Fr}{\small{\textsl{Fr}}}
\newcommand{\Gr}{\small{\textsl{Gr}}}
\newcommand{\N}{\small{\mathcal{N}}}
\newcommand{\K}{\small{\mathcal{K}}}

\journal{arXiv}

\begin{document}

\begin{frontmatter}



\author{Jordi Vila-P\'erez\corref{cor1}}
\cortext[cor1]{Corresponding author}
\ead{jvilap@mit.edu}
\author{Ngoc Cuong Nguyen}
\ead{cuongng@mit.edu}
\author{Jaume Peraire}
\ead{peraire@mit.edu}

\address{Department of Aeronautics and Astronautics, Massachusetts Institute of Technology, Cambridge, 02139, Massachusetts, USA}


\title{A high-order discontinuous Galerkin approach for physics-based thermospheric modeling}

\begin{abstract}

The accurate prediction of aerodynamic drag on satellites orbiting in the upper atmosphere is critical to the operational success of modern space technologies, such as satellite-based communication or navigation systems, which have become increasingly popular in the last few years due to the deployment of constellations of satellites in low-Earth orbit.
As a result, physics-based models of the ionosphere and thermosphere have emerged as a necessary tool for the prediction of atmospheric outputs under highly variable space weather conditions.

This paper proposes a high-fidelity approach for physics-based space weather modeling based on the solution of the Navier-Stokes equations using a high-order discontinuous Galerkin method, combined with a matrix-free strategy suitable for high-performance computing on GPU architectures.
The approach consists of a thermospheric model that describes a chemically frozen neutral atmosphere in non-hydrostatic equilibrium driven by the external excitation of the Sun.
A novel set of variables is considered to treat the low densities present in the upper atmosphere and to accommodate the wide range of scales present in the problem. 
At the same time, and unlike most existing approaches, radial and angular directions are treated in a non-segregated approach.

The study presents a set of numerical examples that demonstrate the accuracy of the approximation and validate the current approach against observational data along a satellite orbit, including estimates of established empirical and physics-based models of the ionosphere-thermosphere system.
Finally, a 1D radial derivation of the physics-based model is presented and utilized for conducting a parametric study of the main thermal quantities under various solar conditions.

\end{abstract}

%

\begin{keyword}
	Discontinuous Galerkin \sep space weather \sep physics-based model \sep thermosphere \sep atmospheric drag \sep high-order.
\end{keyword}

\end{frontmatter}


\section{Introduction}
The increasing interest in mega-constellations of satellites for space-based communication, connectivity or navigation, has {unveiled immense} 
economic opportunities {but also underscored the necessity for a more comprehensive understanding of the thermosphere \cite{Portillo2019}.} 
The rapid escalation in the number of on-orbit active satellites across a wide range of altitudes has {accentuated the need to } predict their trajectories with the goal of minimizing possible deorbits and mitigating the risk of collisions \cite{Radtke2017}.
For this reason, the accurate modeling of atmospheric drag has become one of the major challenges for the space industry nowadays since it heavily impacts trajectory predictions at low-Earth orbits (LEO) \cite{Vallado2014,Fang2022}.
However, predicting the dynamics of the thermosphere is a complex task due to the intricate interactions between various physical processes that interact on different spatial and temporal scales and which are amplified by the unpredictable activity of the Sun.
For this reason, developing new methods for estimating the state of the upper atmosphere is considered a crucial step for enabling the success of many space-based applications.

Traditionally, empirical models of the Earth's upper atmosphere have been used to interpret the thermospheric response to different space weather events and to provide continuous coverage of sparse observational data \cite{He2018}.
Several empirical models have been developed during the past decades, such as HASDM (High Accuracy Satellite Drag Model) \cite{Storz2005,Licata2021}, the current standard used by the US Air Force that leverages the Jacchia-Bowman series of models \cite{Jacchia1970,Bowman2008} with data assimilative tools; or the NRLMSISE (referred to as MSIS, for Mass Spectrometer and Incoherent Radar) \cite{Hedin1983,Hedin1987,Hedin1991,Picone2002,Emmert2021}, an empirical model initially developed at NASA and later adopted by the US Naval Research Laboratory.
However, these methods offer a limited forecasting capability due to the scarce availability of data and observations, especially in remote and high-latitude regions, which limits their effectiveness for predicting space weather events and, especially, for conjunction assessment.

As a result, several physics-based models have been proposed to improve modeling and prediction power.
This is the case of the Global Ionosphere-Thermosphere model (GITM) \cite{Ridley2006}, the Thermosphere-Ionosphere-Electrodynamics General Circulation model (TIE-GCM) \cite{Roble1982,Roble1988,Richmond1992,Qian2014}, the coupled Whole Atmosphere Model-Ionosphere Plasmasphere Electrodynamics (WAM-IPE) \cite{Akmaev2008,FullerRowell2008,Jackson2019}, or the Whole Atmosphere Community Climate Model with thermosphere and ionosphere extension (WACCM-X) \cite{Liu2010,Liu2018,Liu2018a}, among others.

Physics-based models of the Earth's upper atmosphere are typically based on finite difference or finite volume discretizations of the conservation equations of mass, momentum, and energy for several neutral and charged components. However, the large number of grid points needed to resolve the space and time scales present in the solution of these equations, combined with the complexity of the multi-parameter inputs, can make these models computationally expensive.
To this end, several simplifications are commonly employed to make these models tractable, such as hydrostatic equilibrium assumptions, decoupled angular and vertical scales, chemical equilibrium of certain components, or simplified eddy diffusion terms for under-resolved scales. Unfortunately, these simplifications come at the cost of increased uncertainty. Moreover, the requirement of these methods for structured grids and spherical coordinate systems can present difficulties. These include dealing with singularities at the poles and maintaining accuracy when the grid stretching is large.

This paper introduces a high-order discontinuous Galerkin (DG) method for space weather applications to address these challenges. The goal is to capitalize on the prediction capabilities of physics-based models and provide high-fidelity approximations with highly resolved physics.
Indeed, DG methods have become an attractive numerical alternative for solving complex systems of equations \cite{Cockburn:00,Hesthaven2010} based on the reduced computational cost to reach desired levels of accuracy and their ability to deal with unstructured and adapted meshes.
Among the benefits they offer as high-order alternatives, DG methods provide local conservation, a stable discretization of the convection operator, well-defined stabilization, and the ability to preserve high-order accuracy on complex geometries and unstructured meshes.
To this end, the family of DG methods has been widely studied over the past years, and different variations, such as the LDG method \cite{Cockburn1998}, the compact DG (CDG) method \cite{PerairePersson08} or the hybrizable DG (HDG) method \cite{Fernandez2017a,Nguyen2012,VilaPerez2020}, have emerged.

A standout characteristic of DG schemes is their ability to utilize parallel computing architectures effectively \cite{Roca-RNP:2013}, and various numerical strategies that implement different nonlinear solvers have been suggested for these methods. These aim to enable the  fast and robust solution of large systems of equations \cite{Fernandez2017a}.
In particular, Jacobian-free approaches have gained considerable attention and have been adopted in multiple production codes in recent years, based on the simplicity of implementation and reduced computational cost and memory demands compared to classic implicit methods \cite{Kronbichler2019,ExaDG2020,Arndt2021}.

This is the case of Exasim, an open-source code for generating high-order DG codes for the numerical solution of wide-ranging classes of partial differential equations (PDEs) \cite{Terrana2020,Nguyen2022,VilaPerez2022}.
The solver uses an implicit matrix-free strategy, combined with scalable reduced basis preconditioners and Newton-GMRES solvers. This makes it suitable for high-performance computing on graphic processor units (GPU) architectures.
In addition, the software contains a code generator module that produces C++ and CUDA code from mathematical functions specified in a high-level interface. This simplifies the user experience and yields a high-performance code capable of operating across different platforms using multiple processors.
These features make Exasim an ideal tool for computational research, allowing it to tackle large-scale problems and easily discretizing complex physical systems.

This paper presents a high-order DG approach for space weather modeling, which describes the behavior of neutral components in the upper atmosphere under the influence of the Sun's external excitation.
The numerical approach is suitable for GPU architectures. It allows the efficient estimation of neutral densities in the upper atmosphere, which constitutes the main driver for the computation of atmospheric drag and orbit determination.
Unlike most physics-based methods, the proposed model uses a simplified description of the thermosphere, currently excluding ionization effects and charged particles. While this reduces the size of the system of equations solved, the incorporation of ionization effects will be considered in future iterations of this work. This work aims to bring modern numerical discretization strategies to the description of the complex thermosphere physical system, thus improving the prediction capabilities of existing methods.
It is worth noting that due to the inherent uncertainty in upper atmosphere models, contemporary space weather research trends underscore the importance of integrating sophisticated methods for uncertainty quantification, data assimilation, and machine learning \cite{Matsuo2012,Morozov2013,Camporeale2019,Gondelach2022,Briden2022}. These techniques can significantly boost the precision and trustworthiness of predictions. Exploration of these aspects is planned for future research.

The remainder of this work is organized as follows. Section~\ref{sc:model} describes the physics-based space weather model used to characterize the thermosphere. The proposed methodology is used to model an atmosphere in non-hydrostatic equilibrium, driven by the external excitation of the Sun.
Then, the discontinuous Galerkin approach and the description of the matrix-free numerical solution are detailed in Section~\ref{sc:method}.
A set of numerical studies showing the approximation capabilities of the current approach and a parametric study under different solar conditions are presented in Section~\ref{sc:studies} to validate and demonstrate the applicability of the current approach.
Finally, Section~\ref{sc:conclusion} summarizes the main conclusions of this work.

\section{Physics-Based Space Weather Modeling of the Upper Atmosphere} \label{sc:model}
This section details the derivation of the physics-based space weather model devised for the characterization of neutral quantities in the upper atmosphere, specifically between 100km and 600km of altitude.
The model describes the conservation of mass, momentum and energy, subject to the effect of external solar heating. Furthermore, a non-conservative variant of the equations is devised, introducing a new set of variables tailored to accommodate the large density variations inherent in the problem. Finally, the non-dimensional form of the model is presented.

\subsection{Governing equations} \label{sc:governingEqs}
The proposed physics-based model stems from the compressible Navier-Stokes equations in the Earth's rotating frame of reference and describes the behavior of a chemically frozen neutral gas mixture in non-hydrostatic equilibrium.
The system of conservation laws, expressed in a non-conservative form, reads
\begin{subequations} \label{eq:NS1}
	\begin{align} 
		\pd{\rho}{t} + \Div \left( \rho \bv \right) &= 0, \\
		\rho \left(  \pd{\bv}{t} +  \bv \cdot \Grad \bv \right)  +  \Grad p - \Div \btau &= \rho \ba, \\
		\rho \left( \pd{\cv T}{t} +  \bv \cdot \Grad \left( \cv T \right) \right) + p \, \Div \bv  &= \qeuv +  \Div \left(\kappa \Grad T\right) + \btau : \Grad \bv.
	\end{align}
\end{subequations}
Here, $\rho$, $\bv$ and $T$ denote density, velocity, and temperature, respectively, and $p$ stands for the thermodynamic pressure.
In addition, an equation of state relating density, temperature, pressure, and fluid properties completes the system.
In this instance, an ideal gas law for the gas mixture is considered, thus
\begin{equation} \label{eq:eos}
	p = \rho \kB T / m,
\end{equation}
where $\kB$ is the Boltzmann constant, and $m$ is the molecular mass of the gas.
In addition, the relationships $\gamma = \cp/\cv$ and $\cp - \cv = \kB/m$ hold, where $\gamma$ is the ratio of specific heats at constant pressure, $c_p$, and constant volume, $c_v$. For the gas mixture considered, we take $\gamma=5/3$.

On the other hand, the acceleration vector $\ba = \bg - \bomega \times \left(\bomega \times \br \right) - 2 \bomega \times \bv$ contains the gravitational, centripetal, and Coriolis contributions, respectively, where $\bomega$ stands for the angular velocity of the Earth and $\br$ is the position vector.
In addition, the gravitational acceleration is given by $\bg = - G M \br / \norm{\br}^3$, being $G$ the gravitational constant and $M$ the mass of the Earth.

Finally, $\kappa$ stands for the thermal conductivity, whereas $\btau$ denotes the viscous stress tensor, which, under Stokes' hypothesis (i.e., zero bulk viscosity), can be expressed in terms of the strain rate tensor, $\beps$, namely
\begin{equation} \label{eq:stress}
	\btau = \mu \beps = \mu \left[ \left(\Grad \bv + \Grad^T \bv \right) - \frac{2}{3} (\Div \bv) \Idm{} \right],
\end{equation}
being $\mu$, the dynamic viscosity of the fluid and $\Idm{}$, the identity matrix.

The energy equation incorporates the external heating source term, $\qeuv$, that models the energy absorption due to photoionization caused by solar extreme ultraviolet (EUV) radiation and is  described in section~\ref{sc:EUV}.

Note that the system~\eqref{eq:NS1} is expressed in Cartesian coordinates, which defines a computational domain consisting of a spherical shell comprised between the lower and top boundaries at 100km and 600km of altitude, respectively.

\subsubsection{Neutral gas with chemically frozen composition} \label{ssc:chemistry}
The proposed model describes a neutral gas mixture with a chemically frozen composition.
That is, the mass fractions of the neutral species, $\chi_s = \rho_s/\rho$, are assumed to be known \emph{a priori} for all the species, $s = 1, \dots n_{\texttt{species}}$, as a function of altitude, and the total density $\rho = \sum_s \rho_s$ is obtained as the sum of the partial mass densities.

\begin{figure}[htbp]
	\centering
	\begin{subfigure}[t]{0.45\textwidth} \centering
		\includegraphics[width=\textwidth]{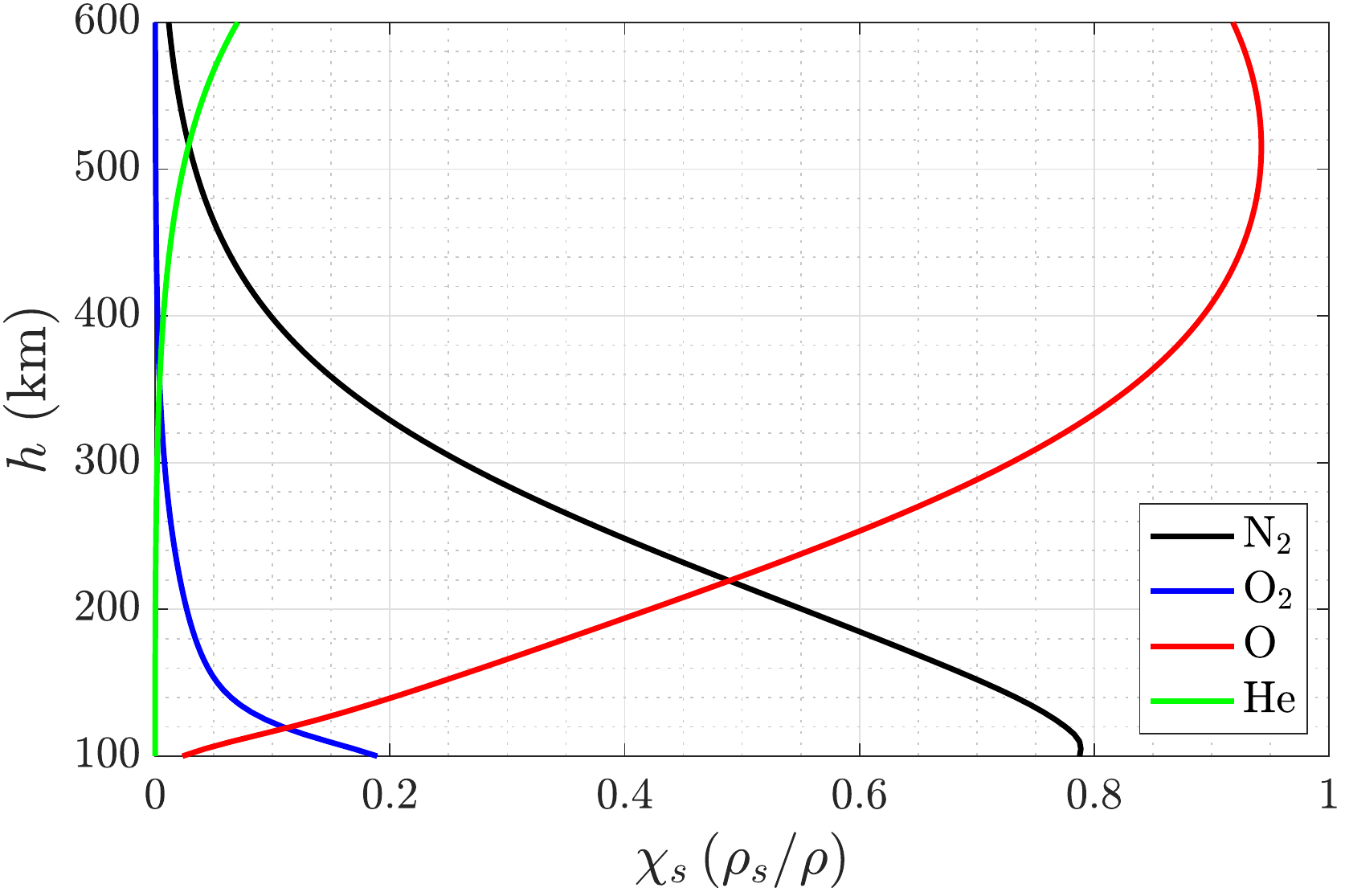}
		\caption{Partial densities of the main components.}
		\label{fig:compositionVsH}
	\end{subfigure}  \qquad
	\begin{subfigure}[t]{0.45\textwidth} \centering
		\includegraphics[width=\textwidth]{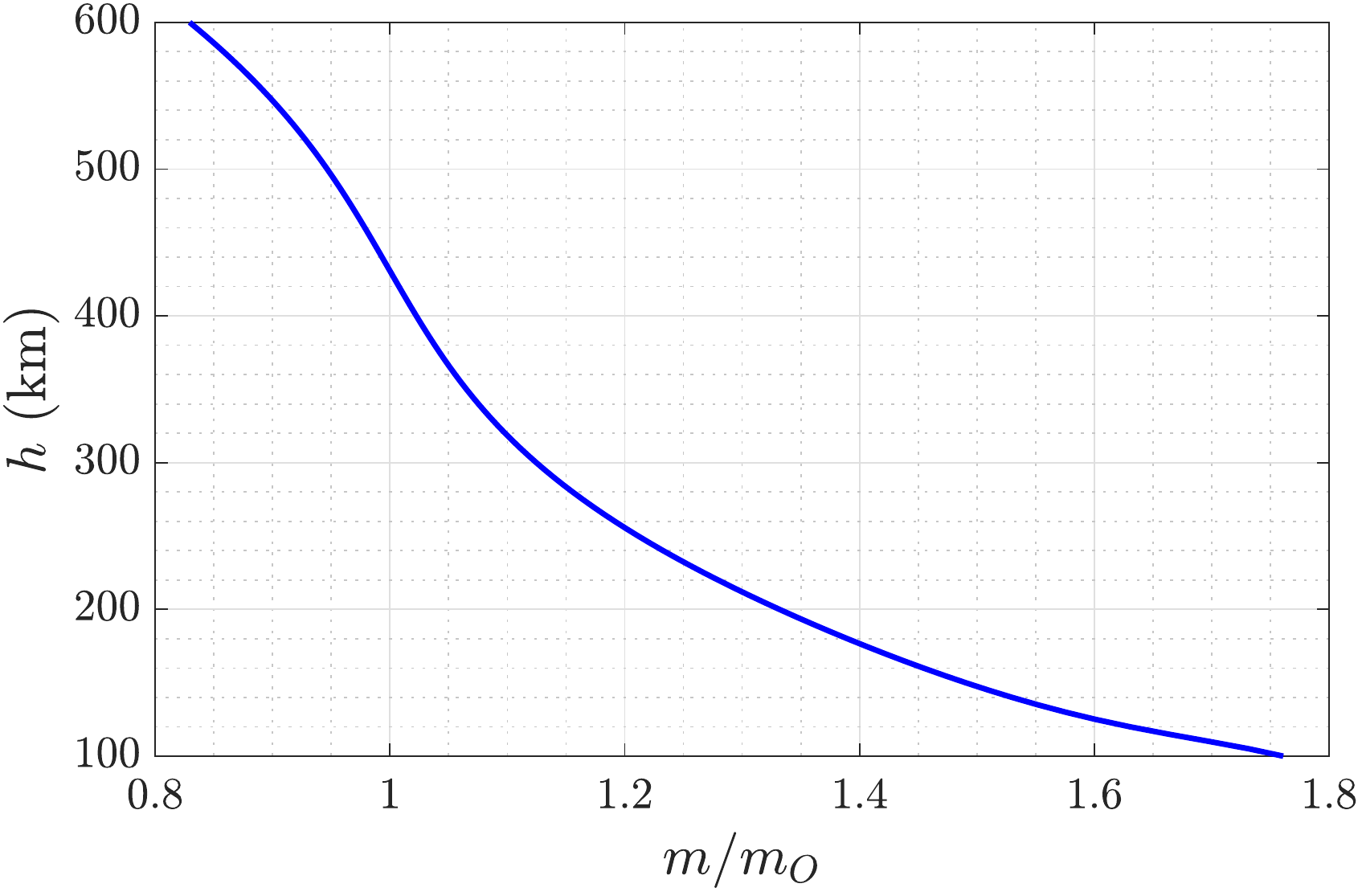}
		\caption{Molecular mass of the gas mixture.}
		\label{fig:massVsH}
	\end{subfigure} 
	\caption{Typical evolution of the mass fractions (left) and molecular mass of the mixture (right) with altitude.}
	\label{fig:DensityMass}
\end{figure}

The four species considered in this work are \ch{O}, \ch{N2}, \ch{O2} and \ch{He}.
To this end,  the mass fraction of these species is inferred from MSIS in a preprocessing stage of the computation. The data is retrieved as a function of the radial position, performing an average over latitude and longitude, as depicted in Figure~\ref{fig:compositionVsH}.
Finally, the radially distributed data is used to fit an exponential expression of the form
\begin{equation}
	\chi_s = \sum_{j = 1}^{j_{\max}} a_j^s \exp \left( b_j^s \, \norm{\br} \right), \quad \text{ with } j_{\max} = 2,
\end{equation}
via a non-linear least squares method.
This particular approximation is performed for all the species except for one (atomic oxygen, \ch{O}), which is computed to preserve the total sum equal to one, namely $\sum_s \chi_s = 1$.

Note that the gas mixture's variable composition with altitude significantly affects the fluid properties, namely $\cp$, $\cv$, and the equation of state~\eqref{eq:eos}.
In particular, the molecular mass of the mixture is computed as $m = \rho / \sum_{s} \left( \rho_s / m_s \right) = \left[ \sum_s \left( \chi_s / m_s \right) \right]^{-1}$, with $m_s$ being the molecular mass of each of the species. The evolution of $m$ with altitude for a typical case is shown in Figure~\ref{fig:massVsH}.

\subsubsection{Physical modeling of viscous and conduction terms}	 \label{ssc:viscosity}
The viscosity and thermal conductivity coefficients in equations~\eqref{eq:NS1} and ~\eqref{eq:stress} can be expressed as the sum of molecular and eddy contributions, namely
\begin{equation} 
	\mu = \mu_m + \rho \nu_{\text{eddy}}, \qquad \kappa = \kappa_m + \rho \cp \alpha_{\text{eddy}}.
\end{equation}
Here, the molecular contributions of the dynamic viscosity and thermal conductivity are functions of the temperature and owe their expressions to basic analyses on collision theory and to a combination of theoretical studies and measured data ~\cite{Schunk2009,Ridley2006,Pawlowski2009,Ponder2023}.
These two quantities typically carry a substantial degree of uncertainty, both in their formulation and their magnitude. In this study, their nominal  expression is 
\begin{equation} 
	\mu_m = 6.5 \cdot 10^{-4} \sqrt{\frac{T m}{k}}, \qquad \kappa_m = \sum_s \left( \frac{n_s}{\sum_k n_k} \alpha_s \right)  T ^{0.69}, \quad \text{ (for SI units)}
\end{equation}
with $\alpha_{\ch{N2}} = \alpha_{\ch{O2}} = 3.6 \cdot 10^{-4}$, $\alpha_{\ch{O}} = \alpha_{\ch{He}} = 5.6 \cdot 10^{-4}$, and $n_s = \rho_s/m_s$ being the number density of species $s$.
Note that the nominal value of viscosity is taken 2 to 5 times higher than the usual nominal values for molecular viscosity. This choice is made in order to account for missing dissipative effects such as ion drag or other collision-related phenomena.

On the other hand, the eddy viscosity and the eddy conductivity account for the mixing and chemistry effects taking place at lower altitudes and are expressed in terms of the kinematic viscosity, $\nu_{\text{eddy}}$, and thermal diffusivity, $\alpha_{\text{eddy}}$, respectively.
These two terms, $\nu_{\text{eddy}}$ and $\alpha_{\text{eddy}}$, are assumed constant and represent an offset on the total kinematic viscosity and thermal diffusivity of the system.
Their effect is mostly relevant at lower altitudes, where mixing and chemistry effects are stronger, and their value is taken as $\nu_{\text{eddy}}/\nu_0 = 200$ and $\alpha_{\text{eddy}}/\alpha_0 = 10$, where $\nu_0$ and $\alpha_0$ are the reference molecular kinematic viscosity and thermal diffusivity at the lowest altitude in the computational domain.

\subsection{Physical modeling of the solar EUV heating} \label{sc:EUV}
The effect of solar EUV radiation on the system is taken into account by means of the source term $\qeuv$, in the energy equation~\eqref{eq:NS1}.
This source term is based on the $F_{10.7}$ index and employs the EUVAC model~\cite{Richards1994}, following the approach described in GITM~\cite{Ridley2006}, and describes the amount of energy being transferred to the system due to exothermic recombination from photoionization.
In particular, its expression is given by
\begin{equation} \label{eq:EUVcontinuous}
	\qeuv = \epsilon \sum_{s}^{n_{\texttt{species}}} n_s \int_{\lambda_{\min}}^{\lambda_{\max}} \phi_s(\lambda) I_s(\lambda) E(\lambda) d\lambda ,
\end{equation}
where $\epsilon$ is a heating efficiency, $\phi_s$ is the photoabsorption cross section, $I_s$ is the flux of photons and $E(\lambda) = hc/\lambda$ is the photon energy associated to the wavelength $\lambda$, being $h$ the Planck's constant and $c$ the speed of light.
In addition, $\lambda_{\min} \sim 5${\AA} and $\lambda_{\max} \sim 1000${\AA} denote the lower and upper ionization thresholds used for the computation of $\qeuv$.

For practical purposes, the computation of the EUV source term~\eqref{eq:EUVcontinuous} is carried out in a discrete manner, namely
\begin{equation}
	\qeuv = \epsilon \sum_{s}^{n_{\texttt{species}}} n_s \sum_{j=1}^{n_{\lambda}} \phi_s^j I_s^j \frac{h c}{\lambda_j},
\end{equation}
taking advantage of reported experimental measurements of the absorption cross section for different species, $\phi_s^j$, at discrete wavelength bins~\cite{Torr1979}, compiled in Table~\ref{tb:EUVcrosssections} in~\ref{app:EUVcrosssections}. In addition, the photon flux $I_s^j$ is described by
\begin{equation}
	I_s^j =  I_0^j e^{-{\tau_s^j}}
\end{equation}
where $I_0^j$ is the normalized EUV photon flux and $\tau_s^j$ stands for the optical depth of the EUV irradiance flux. Indeed, the normalized solar EUV flux $I_0^j$ is computed according to the EUVAC model \citep{Richards1994} employing the daily average of the $F_{10.7}$ solar flux index and its 81-day mean, $F_{10.7}^{81}$, that is $\bar{F}_{10.7} = (F_{10.7} + F_{10.7}^{81})/2$. As a result,
\begin{equation}
	I_0^j = \frac{1}{d^2} F_{74113}^j \max \left \lbrace 0.8, 1 + A_j (\bar{F}_{10.7} - 80) \right \rbrace,
\end{equation}
where $F_{74113}^j$ and $A_j$ are obtained directly from \citep{Richards1994} (Table~\ref{tb:EUVcrosssections}) and symbolize the normalized EUV flux spectrum and a scaling factor, respectively, whereas $d$ is the normalized distance to the Sun.

On the other hand, the optical depth, $\tau_s^j$, is computed as
\begin{equation}
	\tau_s^j =  \int_{\infty}^{\br} n_s \phi_s^j d \ell,
\end{equation}
and accounts for the columnar content of each constituent extending from a certain point towards the Sun.
This quantity is computed by means of Chapman integrals~\cite{Chapman1931} and allows to determine the intensity of the solar irradiance flux at a given spatial position.
In practice, its calculation is performed using the approximation formulated by Smith and Smith~\cite{Smith1972}.
Moreover, in this particular study, it is computed assuming an exponential density law, corresponding to an isothermal atmosphere with constant mixture properties.
In this manner, the optical depth can be simplified and expressed as
\begin{align}
	\tau_s^j = \begin{cases} 
		N_s^j \exp \left( y^2 \right) \text{erfc}(y) \sqrt{\pi\tilde{r}/2}, &\text{if } \cos \chi > 0 \\
		N_s^j \left[ 2 \exp \left( \tilde{r}(1-|\sin \chi|) \right) - \exp \left( y^2 \right) \text{erfc}(y)\right] \sqrt{\pi\tilde{r}/2}, &\text{if } \cos \chi \leq 0, \text{ and } \abs{\sin \chi} > R_0/\norm{\br}\\
		\tau_\infty, &\text{otherwise,}
	\end{cases}
\end{align}
where $N_s^j = n_s \phi_s^j H$ denotes the molecular content, with $H = \kB T/(mg)$ being the scale height, $\tilde{r} = \|\bm{r}\|/H$, $y = \sqrt{\tilde{r}/2} |\cos \chi |$, $\text{erfc} = 1 - \text{erf}$ is the complementary error function and $\tau_\infty$ is an arbitrarily big number such that $e^{-{\tau_\infty}} \to 0$.
In addition, $R_0$ is the radius of the lower boundary, that is $R_0 = R_{\text{Earth}}+ h_0$, the sum of the Earth's radius and the height of the lower boundary of the computational domain, $h_0=100$km.

We note that the optical depth has a different expression depending on the solar zenith angle $\chi$, which can be obtained based on the north declination of the Sun $\delta$, the latitude $\theta$, and the local time $\tilde{\phi}$, as
\begin{equation}
	\chi = \arccos \left( \sin \delta \sin \theta + \cos \delta \cos \theta \cos (\tilde{\phi}-\pi)\right).
\end{equation}
In particular, the local time $\tilde{\phi} = \mod_{2\pi} \left(\phi_0 + \phi_t\right)$ is computed in terms of the longitude, $\phi_0$, corrected with the effect of the Earth's rotation, $\phi_t = \omega t$.
In addition, the north declination of the Sun is corrected according to season of the year, which can be computed using Kepler's equations, or simply approximated by means of
\begin{equation}
	\delta = \arcsin \left( \sin(-\delta_0) \cos \left(\frac{2\pi}{365.24} (N_d+9) +  2 e \sin \left(\frac{2\pi}{365.24}(N_d-3)\right) \right)\right),
\end{equation}
where $\delta_0= 23.44^{\circ}$ is the maximum declination of the Sun at summer solstice, $e \simeq 0.0167$ is the eccentricity of the Earth's orbit and $N_d$ is the day of the year, being $N_d=1$ on January 1st.

\subsection{Initial and boundary conditions}
The simulations are initiated using a postproccesed MSIS solution, which provides a realistic estimate of the density and temperature fields. The initial condition is assumed to be in locally hydrostatic equilibrium so as to accommodate an initial rest state ($\bv = \bm{0}$). In consequence, the density and temperature states are obtained assuming hydrostatic conditions employing pressure data computed from MSIS.

The system of partial differential equations is finally closed with the prescription of boundary conditions.
On the one hand, the lower boundary ($h = 100$km) is characterized by means of a no-slip, isothermal boundary condition. To this end, a constant temperature obtained from averaging in longitude and latitude the MSIS output is considered.
On the other hand, an adiabatic condition is imposed at the upper boundary ($h = 600$km of altitude).
Such condition is combined with the imposition of zero normal stresses in inflow regions, whereas no additional constrains are imposed for outflows.

\subsection{A non-conservative formulation for exponential density and reduced stiffness} \label{sc:formulation}
In order to handle the exponential variations of density with altitude, the density field is normalized using a reference value  $\rho_0$, and the natural logarithm of the normalized density, that is $\logr = \log(\rho/\rho_0)$, is employed as a primal dependent variable of the system. In addition, velocity and temperature are also scaled with $\sqrt{\logr}$ to avoid dividing by small values of density in the dissipation and heat conduction terms.
As a result, the system of conservation equations~\eqref{eq:NS1} is formulated in terms of the set of primal dependent variables, $\bu = \left(\logr, \srv, \srT \right)$ and can be expressed in flux form, namely $\partial \bu/ \partial t + \Div \bF (\bu, \Grad \bu) = \bm{s} (\bu, \Grad \bu)$, as
\begin{subequations} \label{eq:compactEqs}
		\begin{align} 
			&\pd{\logr}{t} + \Div \left( \logr \bv \right) = (\logr - 1) \Div \bv, \\
			&\pd{\srv}{t} +  \Div \left( \srv \otimes \bv + (\gamma - 1) \cv \srT \Idm{d} - \frac{\mu}{\sr} \beps \right) = \sr \ba + \frac{1}{2}\left( \Div \bv \right) \srv - \frac{\gamma - 1}{2} \cv \srT \Grad \logr + \frac{1}{2}\frac{\mu}{\sr} \beps \cdot \Grad \logr, \\
			&\begin{multlined}[b] \pd{\srT}{t} +  \Div \left( \srT \bv- \frac{\kappa}{\sr \cv} \Grad T \right) = \frac{\qeuv}{\sr \cv} + \left( \frac{3}{2} - \gamma \right) \srT \left( \Div \bv \right) + \frac{1}{2} \frac{\kappa}{\sr \cv} \Grad T \cdot \Grad \logr + \frac{\mu}{\sr \cv} \beps : \Grad \bv \\ - \frac{1}{\cv}\left( \srT \bv - \frac{\kappa}{\sr \cv} \Grad T \right) \cdot \Grad \cv \end{multlined}.
		\end{align}
\end{subequations}

\subsection{Non-dimensional description}
The system of equations~\eqref{eq:compactEqs} is written in nondimensional form by introducing reference quantities at the lower boundary, indicated here with the subscript zero. Thus, the dimensionless variables are given by
\begin{equation}
	\begin{aligned}
		x^* &= \frac{x}{H_0}, \quad &v^* &= \frac{v}{a_0}, \quad &t^* &= \frac{t}{H_0/a_0}, \quad &\rho^* &= \frac{\rho}{\rho_0}, \quad &T^* &= \frac{T}{T_0},\quad &p^* &= \frac{p}{\rho_0 a_0^2}, \quad &m^* &= \frac{m}{m_O},\\
		\mu^* &= \frac{\mu}{\mu_0}, \quad &\kappa^* &= \frac{\kappa}{\kappa_0}, \quad &g^* &= \frac{g}{g_0}, \quad & \omega^* &= \frac{\omega}{\omega_0}, \quad &\phi^* &= \frac{\phi}{H_0^2}, \quad &I^* &= \frac{I}{a_0/H_0^3}, \quad &\lambda^* &= \frac{\lambda}{\lambda_0},
	\end{aligned}
\end{equation}
where $H_0 = k_B T_0 / (m_O g_0)$ is the reference scale height, $a_0 = \sqrt{\gamma k_B T_0/m_O}$ is the reference speed of sound, $\lambda_0 = 1$nm is a reference wavelength, and $m_O$, is the molecular mass of atomic oxygen, which is used as a reference value.
The nondimensional form of the state equation becomes, $p^* = \rho^* T^*/(\gamma m^*) =  \rho^* \cv^* T^*/\gamma$, given that $\cv^* = 1/m^*$.

Dropping the asterisk superscritps, for simplicity, the dimensionless system of equations~\eqref{eq:compactEqs} can be rewritten in flux form as
\begin{subequations} \label{eq:fluxSWNondimensional}
	\begin{align} 
		&\pd{\logr}{t} + \Div \left( \logr \bv \right) = (\logr - 1) \Div \bv, \\
		&\pd{\srv}{t} +  \Div \left( \srv \otimes \bv + \frac{1}{\gamma m} \srT \Idm{d} - \frac{1}{\sqrt{\gamma \Gr}} \frac{\mu}{\sr} \beps\right) = \sr \ba + 	\frac{1}{2}\left( \Div \bv \right) \srv - \frac{1}{2 \gamma m}  \srT \Grad \logr + \frac{1}{2} \frac{1}{\sqrt{\gamma \Gr}} \frac{\mu}{\sr} \beps \cdot \Grad \logr, \\
		& \begin{multlined}[b] \pd{\srT}{t} +  \Div \left( \srT \bv- \frac{1}{\Pr} \sqrt{\frac{\gamma}{\Gr}} \frac{\kappa}{\sr \cv} \Grad T \right) = \frac{\gamma(\gamma-1)}{\K}\frac{\qeuv}{\sr \cv} + \left( \frac{3}{2} - \gamma \right) \srT \left( \Div \bv \right) + \frac{1}{2 \Pr} \sqrt{\frac{\gamma}{\Gr}} \frac{\kappa}{\sr \cv} \Grad T \cdot \Grad \logr \\ + (\gamma-1) \sqrt{\frac{\gamma}{\Gr}} \frac{\mu}{\sr \cv} \beps : \Grad \bv - \frac{1}{\cv}\left( \srT \bv - \frac{1}{\Pr} \sqrt{\frac{\gamma}{\Gr}} \frac{\kappa}{\sr \cv} \Grad T \right) \cdot \Grad \cv \end{multlined}.
	\end{align}
\end{subequations}

\noindent
Additionally, the dimensionless acceleration vector is given by 
\begin{equation}
	\bm{a} = \frac{1}{\gamma}\bm{g}  - \frac{Fr^2}{\gamma}  \bm{\omega} \times (\bm{\omega} \times \bm{r})  - \frac{2 Fr}{\sqrt{\gamma}} \bm{\omega} \times \bm{v}.
\end{equation}

The nondimensional numbers Grasshoff, $Gr$, Prandtl, $Pr$,  and Froude, $Fr$, as well as the ratio of kinetic to photoionization energy, $\K$, required for the EUV heating source terms are obtained as 
\begin{align} \label{eq:nondimensional}
	\Gr &= \frac{g_0 H_0^3}{(\mu_0/\rho_0)^2}, &\Pr &= \frac{\mu_0 {\cp}_O}{\kappa_0}, &\Fr &= \sqrt{\frac{\omega_0 ^2 H_0}{g_0}}, &\K &= \frac{\gamma \kB T_0}{h c / \lambda_0}.
\end{align}

Finally, the dimensionless term $\N = \rho_0 H_0^3 / m_O$, which measures the amount of particles in a reference volume, nondimensionalizes $N_s^j = n_s \phi_s^j H$ in the expression of the optical depth in the EUV term.

\section{Numerical Methodology} \label{sc:method}

The numerical methodology employed to solve the system ~\eqref{eq:fluxSWNondimensional} is described below. The method has  been implemented in the  open-source software Exasim \cite{VilaPerez2022}.

\subsection{Discontinuous Galerkin discretization}
Let $\Omega \subseteq \mathbb{R}^{d}$, with $d=3$, represent the physical domain with $\partial \Omega$ its Lipschitz boundary and $t_f>0$ denote a final time of interest.
Consider $\elempart$, a partition of $\Omega$ into disjoint, $k$-th degree curved elements $K$. Moreover, let $\boundpart := \{ \partial K : K \in \elempart \}$ be the collection of element boundaries in $\elempart$ and $\bn$ the outward unit normal vector at the element boundary $\partial K$.
Finally, let $\eltwo(D)$ denote the space of square-integrable functions and $\Polyk(D)$, the set of polynomials of degree at most $k$ on the domain $D\subset \RR^d$.

Then, the following discontinuous finite element spaces are introduced
\begin{subequations}
\begin{align}
	\testqh & = \left\lbrace \bm{p} \in \left[\eltwo (\elempart) \right]^{m\times d} : \ \bm{p}|_K \in \left[ \Polyk(K) \right]^{m\times d}, \ \forall K \in \elempart  \right\rbrace, \\
	\testwh & = \left\lbrace \bm{w} \in \left[\eltwo (\elempart) \right]^m : \ \bm{w}|_K \in \left[ \Polyk(K) \right]^m, \ \forall K \in \elempart  \right\rbrace,
\end{align}
\end{subequations}
where $m=d+2$ stands for the number of equations in the system of conservation laws and $d$ is the spatial dimension.
Finally, the inner products associated with these finite element spaces are defined as 
\begin{subequations}
\begin{align}
	\sprod[\elempart]{\bv,\bm{w}} &= \sum_{K \in \elempart} \sprod[K]{\bv,\bm{w}} = \sum_{K \in \elempart} \int_{K} \bv \cdot \bm{w}, \\
	\sprod[\elempart]{\bm{p},\bm{r}} &= \sum_{K \in \elempart} \sprod[K]{\bm{p},\bm{r}} = \sum_{K \in \elempart} \int_{K} \bm{p} : \bm{r}, \\
	\dprod[\boundpart]{\bv,\bm{w}} &= \sum_{K \in \elempart} \dprod[\partial K]{\bv,\bm{w}} = \sum_{K \in \elempart} \int_{\partial K} \bv \cdot \bm{w},
\end{align}
\end{subequations}
for $\bv, \bm{w} \in \testwh$ and $\bm{p}, \bm{r} \in \testqh$, with $\cdot$ and $:$ denoting the scalar product and the Frobenius inner product, respectively.

As a result, the DG discretization of the proposed space weather model~\eqref{eq:fluxSWNondimensional} consists of finding $\left( \bqh(t),\buh(t) \right) \in \testqh \times \testwh$ such that
\begin{subequations}
	\label{eq:DGformulation}
	\begin{alignat}{2}
		\sprod[\elempart]{\bqh,\bm{p}} + \sprod[\elempart]{\buh, \Div \bm{p}} -  \dprod[\boundpart]{\bhuh,\bm{p}\cdot\bn}  & =  0, \\
		\sprod[\elempart]{\pd{\buh}{t},\bm{w}} - \sprod[\elempart]{\bF(\buh,\bqh), \Grad \bm{w}} +  \dprod[\boundpart]{\bhFh (\bhuh,\buh,\bqh)\cdot \bn,\bm{w}}  & =  \sprod[\elempart]{\bs(\buh,\bqh),\bm{w}},
	\end{alignat}
	for all $(\bm{p},\bm{w}) \in \testqh \times \testwh$ and all $t \in (0,t_f)$, satisfying the initial condition
	\begin{alignat}{2}
		\sprod[\elempart]{\buh|_{t=0} - \bu_0,\bm{w}_0}  & =  0,
	\end{alignat}
\end{subequations}
for all $\bm{w}_0 \in \testwh$.

In the above equations, the numerical trace $\bhuh$ and the numerical flux $\bhFh$ approximate the state $\bu$ and the physical flux $\bF$ on the element faces, respectively. Different types of DG methods can be implemented depending on their definition.
In particular, the LDG method~\cite{Cockburn1998} is implemented in Exasim, yielding the following expression of the numerical trace and flux
\begin{equation}
	\label{eq:LDG2}
	\begin{split}
		\bhuh  &= \frac{1}{2} (\buh + \buh^{-}), \\
		\bhFh \cdot \bn & = \frac{1}{2} \left( \bF(\buh , \bqh)  + \bF(\buh^- , \bqh^-)  \right) \cdot \bn + \bsigma \cdot ( \buh - \bhuh) ,
	\end{split}
\end{equation}
where $(\bm u_h , \bm q_h)$ and $(\bm u_h^- , \bm q_h^-)$ are the numerical solutions on the element $K$ and its neighboring element $K^-$, respectively.
In addition, the stabilization tensor $\bsigma = \bsigma_c + \bsigma_d$ includes the convective and diffusive contributions.
On the one hand, the convection-stabilizing term is given by $\bsigma_c = \lamax^0 \Id$, with $\lamax^0$ being a maximum wave-speed velocity computed at reference conditions, which defines a sort of global Lax-Friedrichs numerical flux.
On the other hand, the diffusive stabilization term $\bsigma_d$ is defined by means of the block-diagonal matrix
\begin{equation}
	\bsigma_d = \begin{pmatrix} 0 & &\\ & \frac{1}{\sqrt{\gamma \Gr}} \frac{\mu}{\sr} \Id_d & \\ & & \frac{1}{\Pr} \sqrt{\frac{\gamma}{\Gr}} \frac{\kappa}{\sr \cv} \end{pmatrix},
\end{equation}
where $\Id_d$ is the $d \times d$ identity matrix.

Finally, the semi-discrete system of PDEs~\eqref{eq:DGformulation} is further discretized in time using diagonally implicit Runge-Kutta (DIRK) schemes \cite{Alexander77}, which provide high-order accuracy, low dissipation and dispersion, and a strong stability for stiff problems.

\subsection{Jacobian-free Newton-Krylov solver}
The nonlinear system of equations resulting from the temporal discretization of Eq. \eqref{eq:DGformulation} at the $n^{\textnormal{th}}$ timestep, $\bm R(\bm u_n) = 0$, is solved using Newton's method.
This leads to the linear system of equations
\begin{equation}
	\label{eq:NewtonSolve}
	\bm J(\bm u_n^m) \delta \bm u_n^m = -\bm R(\bm u_n^m),
\end{equation}
with $\bm J(\bm u_n^m) = \partial  \bm R(\bm u_n^m) / \partial \bm u$ being the Jacobian matrix and $m$ denoting the $m^{\textnormal{th}}$ Newton's iteration, which is solved until convergence.

In addition, for every iteration of Newton's method, the linear system~\eqref{eq:NewtonSolve} is solved employing a GMRES method.
To this end, a matrix-free scalable preconditioner is applied to accelerate the convergence rate of GMRES.
The preconditioner constructs a low-rank approximation to the Jacobian matrix $\bm J(\bm u_n^m)$ employing a reduced basis (RB) space, $\bm W_n = \text{span} \left\lbrace \delta \bm u_n^{m-j}, 1 \leq j \leq n_{rb} \right\rbrace$, to approximate $\delta \bm u_n^m$.
In addition, the RB is used at the same time to construct an initial guess to accelerate the GMRES convergence.

The numerical algorithm ultimately requires Jacobian-vector products $\bm J(\bm u_n) \bm W_n$, which can be expensive if the Jacobian matrix needs to be constructed.
In order to avoid this step, the solver implements a matrix-free strategy which computes directly the product of the Jacobian matrix with a vector employing a finite difference approximation.
In summary, this matrix-free Newton-GMRES strategy, combined with the RB preconditioner, only requires residual vector evaluations, as detailed in \cite{Terrana2020,Nguyen2022,VilaPerez2022}.

\subsection{Cube-sphere discretization of the computational domain}
One of the main advantages of DG methods relies on their seamless implementation on general unstructured and adapted meshes without compromising the accuracy of the approximation.
In this manner, the presented formulation ~\eqref{eq:compactEqs}, expressed in Cartesian coordinates, allows to naturally employ both unstructured and cube-sphere meshes that avoid pole singularities.
Based on all that, we exploit the flexibility of DG methods and consider cube-sphere meshes of the spherical shell comprised between $100$km and $600$km of altitude.
The grids feature a radial mesh size distribution with geometric growth rate, which is defined by the height of the first layer of elements and the total number of elements in the radial direction.

\begin{figure}[htbp]
	\centering
	\begin{subfigure}[t]{0.32\textwidth} \centering
		\includegraphics[width=0.85\textwidth]{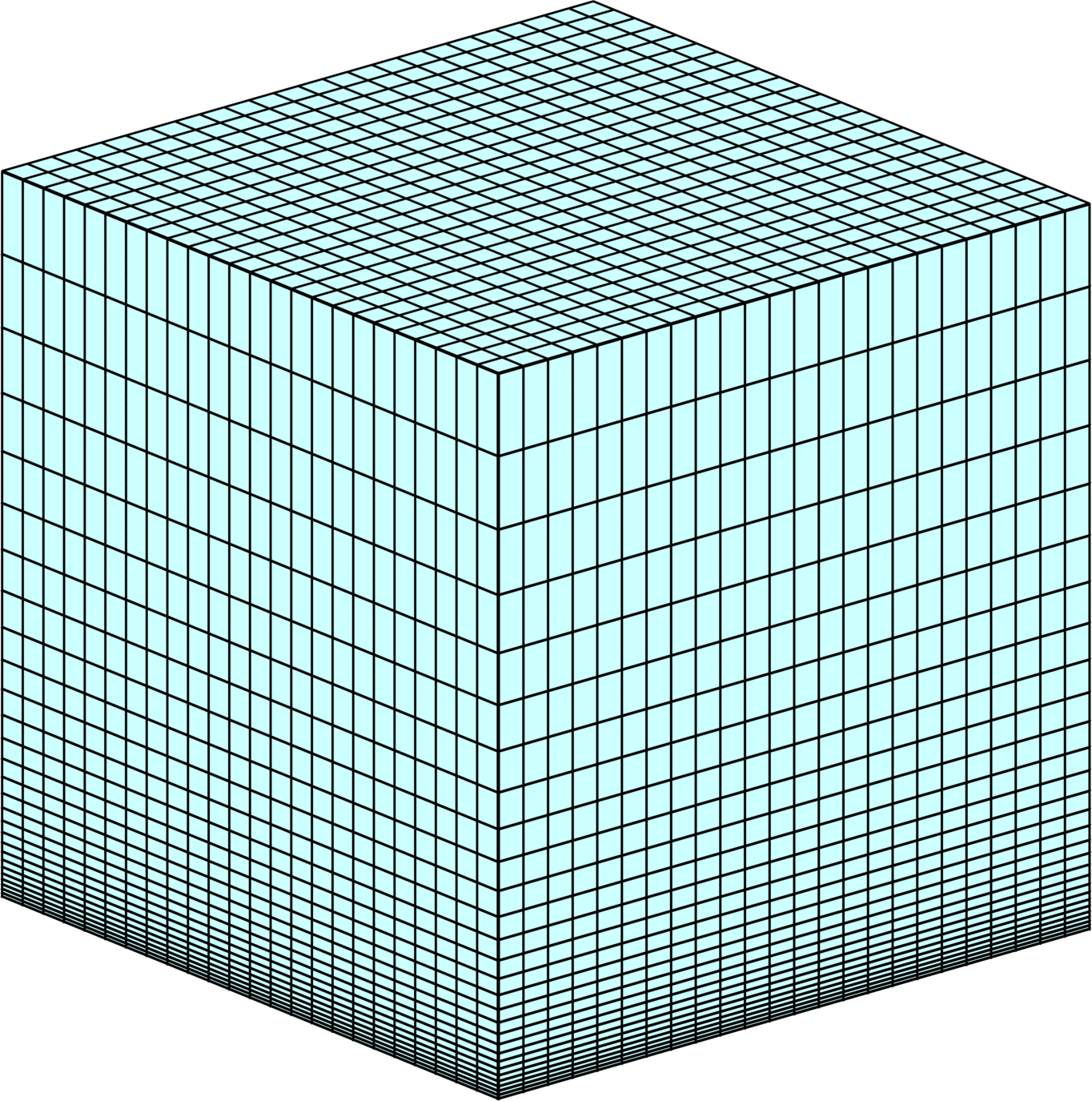}
		\caption{Grid of an individual cube}
		\label{fig:cubeSingle}
	\end{subfigure} 
	\begin{subfigure}[t]{0.32\textwidth} \centering
		\includegraphics[width=0.85\textwidth]{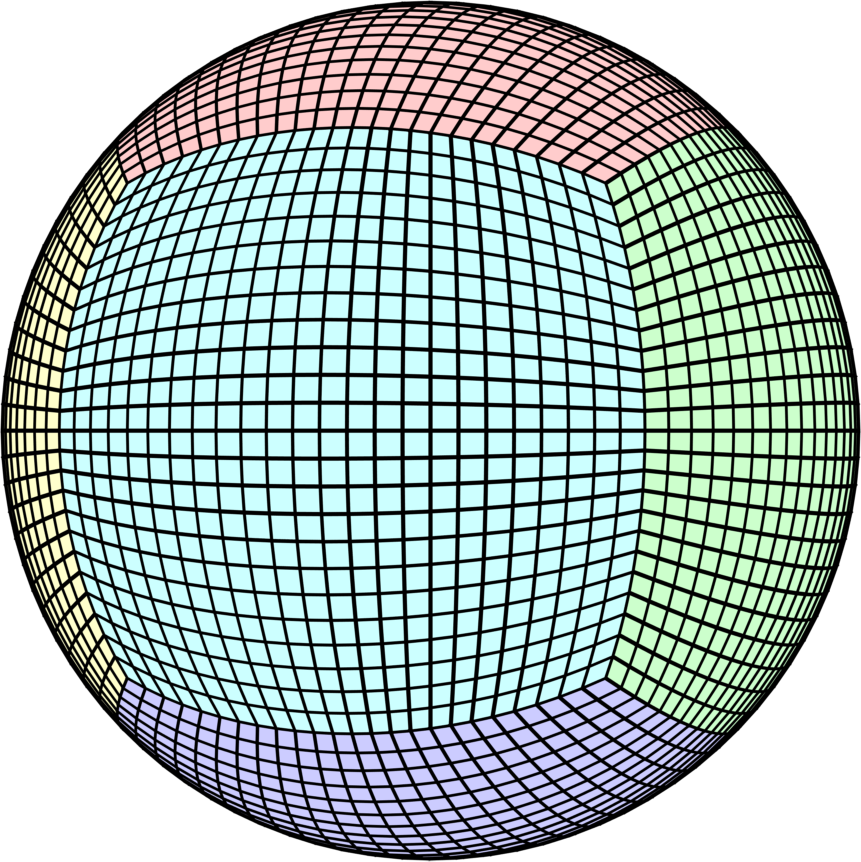}
		\caption{Default mesh with $2.5^{\circ}$ of angular resolution}
		\label{fig:cubeSphereH2}
	\end{subfigure} 
	\begin{subfigure}[t]{0.32\textwidth} \centering
		\includegraphics[width=0.85\textwidth]{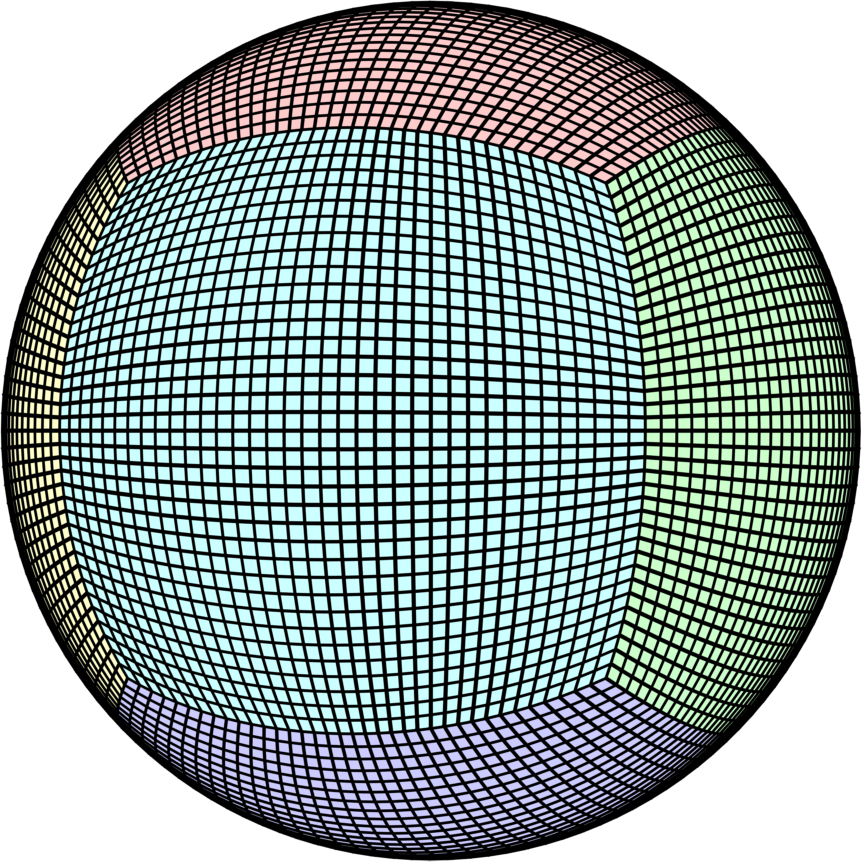}
		\caption{Refined mesh  with $1.25^{\circ}$ of angular resolution}
		\label{fig:cubeSphereH3}
	\end{subfigure} 
	\caption{Cube-sphere grids used for the computational studies and detail of the radial mesh refinement on one individual cube grid.}
	\label{fig:meshes}
\end{figure}

In particular, two different meshes are used in this study, both of them consist of 28 elements in the radial direction, with a first element of 2.5km of height.
A mesh with $36\times36$ elements the local horizontal directions in each individual cube, i.e. with 217,728 total elements and $2.5^{\circ}$ of angular resolution, is used as a default grid and is employed for most of the numerical examples.
In addition, a refined grid with $1.25^{\circ}$ of angular resolution (72 elements on each direction in each cube, for a total of 870,912 elements) is also used in the first computational study in order to verify the convergence properties of the approximation.
The two meshes, together with an sketch of the grid structure of one individual cube, are shown in Figure~\ref{fig:meshes}.

\section{Computational studies} \label{sc:studies}
This section presents a set of computational examples that illustrate the capabilities of the proposed numerical approach.
Several cases corresponding to different levels of solar activity during equinox and solstice conditions are considered.

\subsection{Numerical analysis of the 2014 spring equinox} \label{ssc:case1}
\begin{figure}[htbp]
	\centering
	\begin{subfigure}[t]{0.2933\textwidth} \centering
		\includegraphics[width=0.9\textwidth]{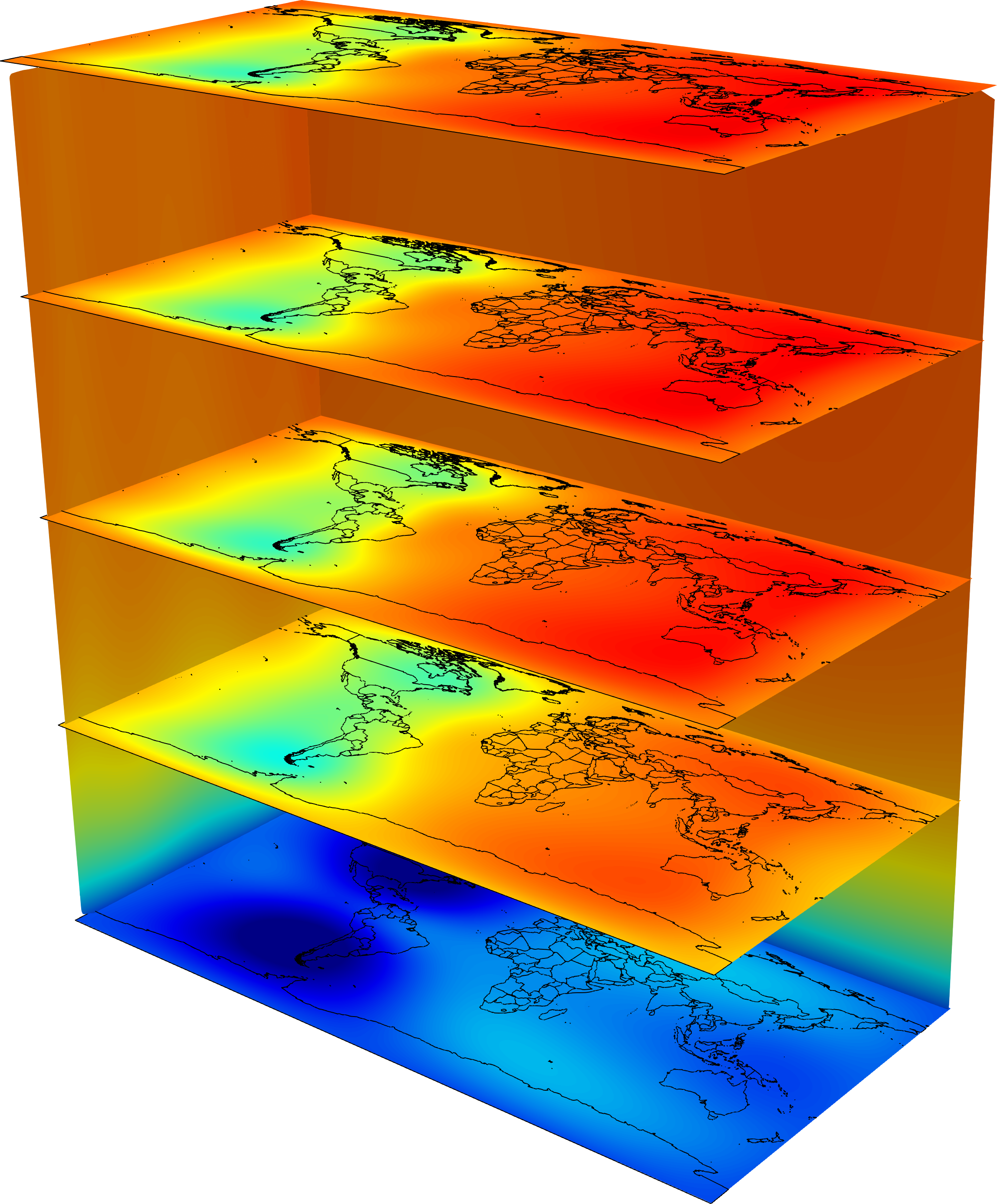}
		\caption{6:00 UTC}
		\label{fig:T_p3H5_t6}
	\end{subfigure} 
	\begin{subfigure}[t]{0.293\textwidth} \centering
		\includegraphics[width=0.9\textwidth]{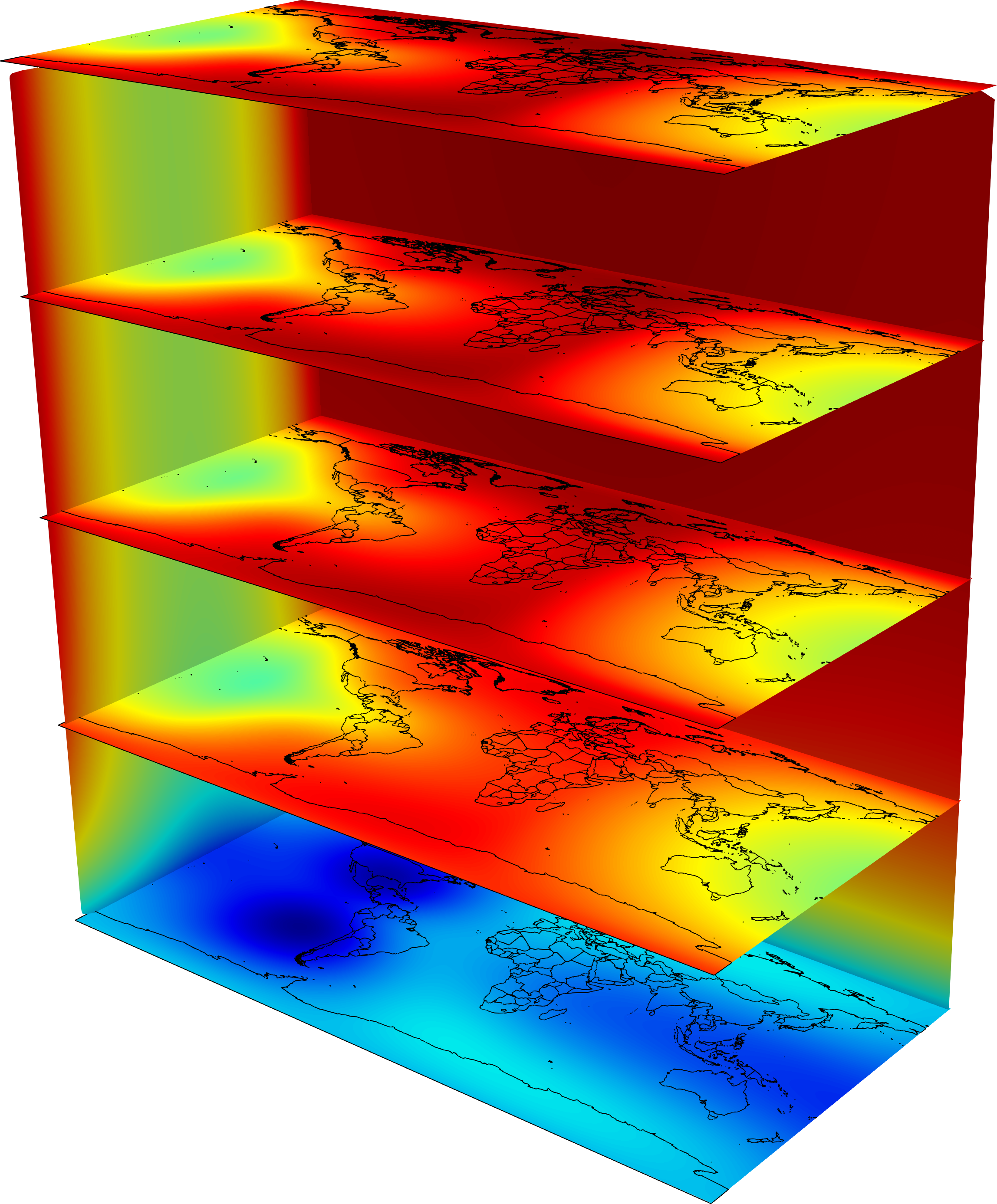}
		\caption{12:00 UTC}
		\label{fig:T_p3H5_t12}
	\end{subfigure} 
	\begin{subfigure}[t]{0.3935\textwidth} \centering
		\includegraphics[width=0.9\textwidth]{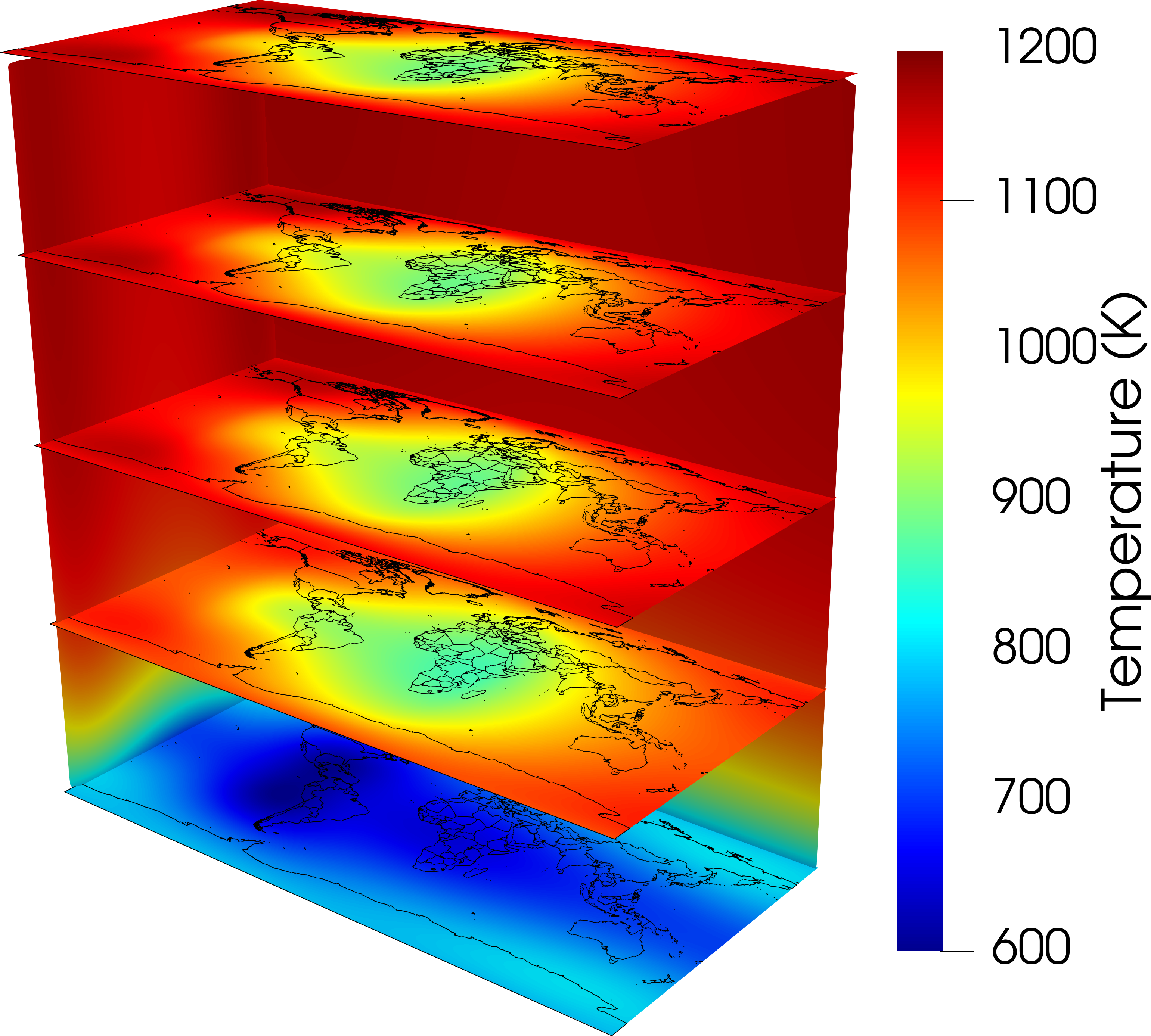}
		\caption{24:00 UTC}
		\label{fig:T_p3H5_t24}
	\end{subfigure} 
	\caption{Temperature estimates at different altitude levels (150, 250, 350, 450 and 550km) during 20/03/2014, obtained on the refined mesh using a cubic approximation.}
	\label{fig:p3H5_12}
\end{figure}

The first computational study presented in this work corresponds to the simulation of the 2014 spring equinox, concretely between 20-23 March 2014.
This case describes a thermosphere under the effect of a moderate-to-high solar activity ($F_{10.7} = 150$) in vernal equinox conditions, that is, with maximum solar incidence along the Equator line and latitudinal symmetry.
The numerical approximation for the temperature, employing cubic polynomials on the refined mesh ($1.25^{\circ}$ of angular resolution) at different instants of time during 20/03/2014, is shown in Figure~\ref{fig:p3H5_12} at various altitudes between 150 and 550km

This example is used to evaluate the accuracy of the current approach and to validate it in a real situation of interest.
In particular, the first section of this study analyzes the convergence of the numerical solution with increasing order of polynomial approximation.
Next, the physics-based thermospheric model is validated against experimental data of the density along a satellite orbit and compared to established empirical and physics-based models of the ionosphere-thermosphere system.
Finally, the results of the radial 1D derivation of the model are presented and compared to the 3D simulations in order to assess the ability of the 1D approximation to replicate the dynamics and the radial structure of the thermosphere. All cases have been run with the an EUV efficiency of $\epsilon=0.25$.

\subsubsection{Convergence analysis} \label{sssc:convergence}
The convergence properties of the computational approach are examined in this example, which is solved using $k=1$, $k=2$ and $k=3$ polynomials on the default mesh (with $2.5^{\circ}$ of angular resolution).
The numerical approximations are analyzed after 12h of simulation (12h UTC on 20/03/2014) and compared to a cubic approximation on the refined mesh (with $1.25^{\circ}$ resolution) which is used as a reference solution, with fixed radial resolution between the two grids.

The simulations are advanced in time using a DIRK scheme. In particular, DIRK(2,2) is used for $k=1$ and $k=2$ simulations, whereas a DIRK(3,3) method is used for $k=3$. In addition, different time step sizes are used for each of the approximations, owing to difficulties in converging the nonlinear systems generated at each time step, when the time step size is too large. Time steps of 10s are used for the $k=1$ computation, whereas time steps of 5s and 4s are employed for the quadratic and cubic calculations, respectively. In addition, these time step sizes are reduced by a factor of two during the first hour of simulation to facilitate a better adaptation of the initial condition to the model dynamics.

\begin{figure}[htbp]
	\centering
	\begin{subfigure}[t]{0.445\textwidth} \centering
		\includegraphics[width=\textwidth]{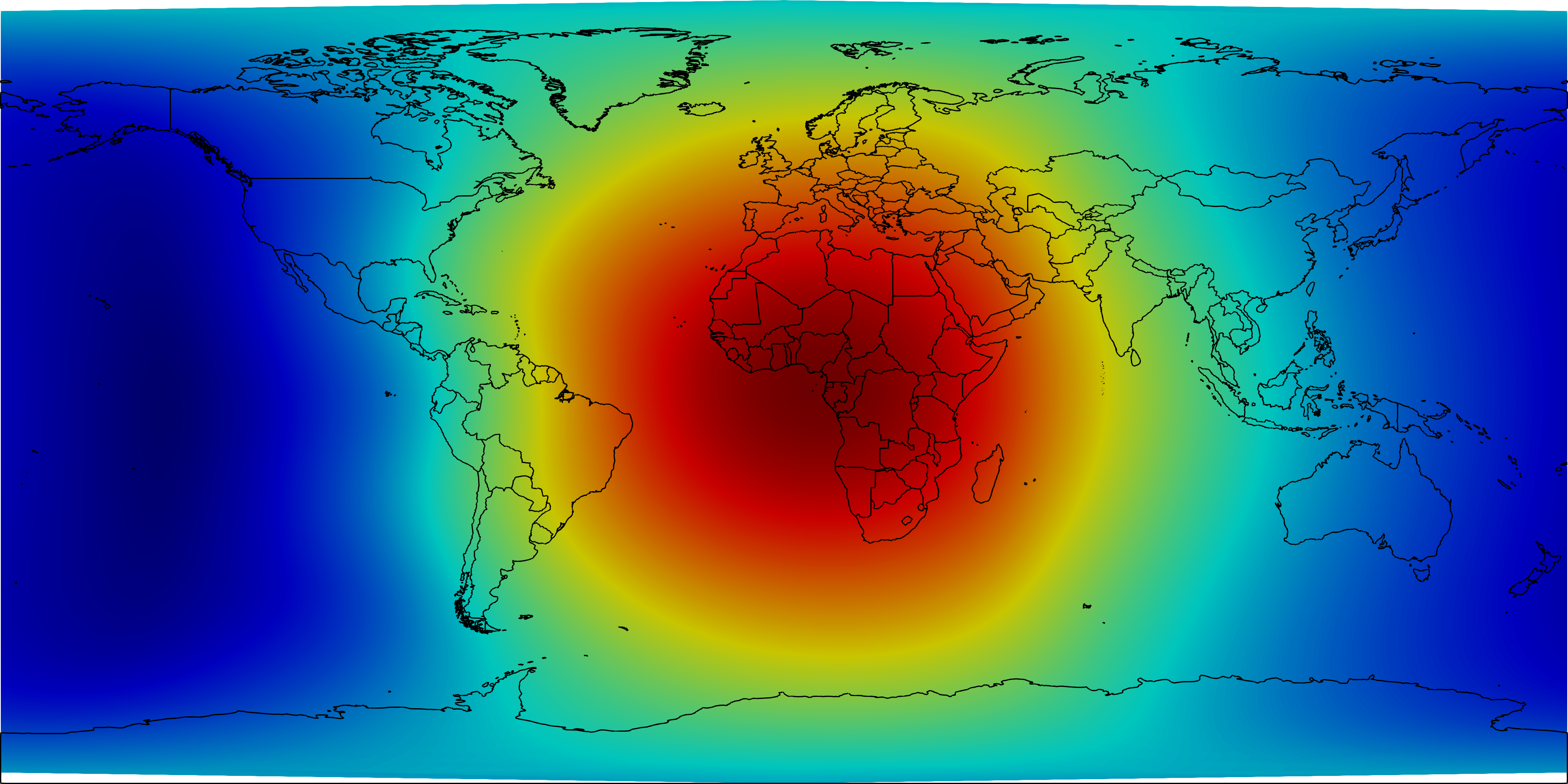}
		\caption{Density, $k=1$}
		\label{fig:rho_p1H3_12_450}
	\end{subfigure}
	\hfill
	\begin{subfigure}[t]{0.535\textwidth} \centering
		\includegraphics[width=\textwidth]{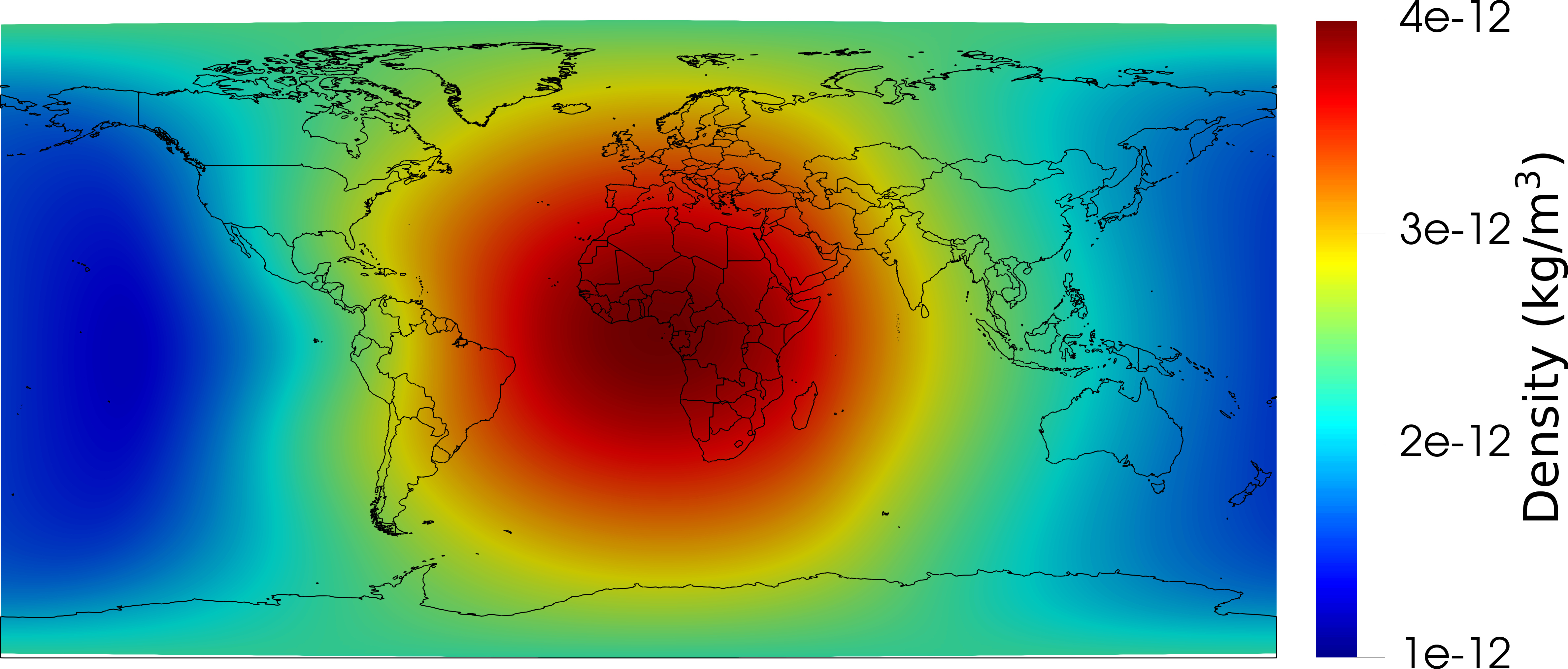}
		\caption{Density, $k=2$}
		\label{fig:rho_p2H3_12_450}
	\end{subfigure}

	\begin{subfigure}[t]{0.4575\textwidth} \centering
		\includegraphics[width=\textwidth]{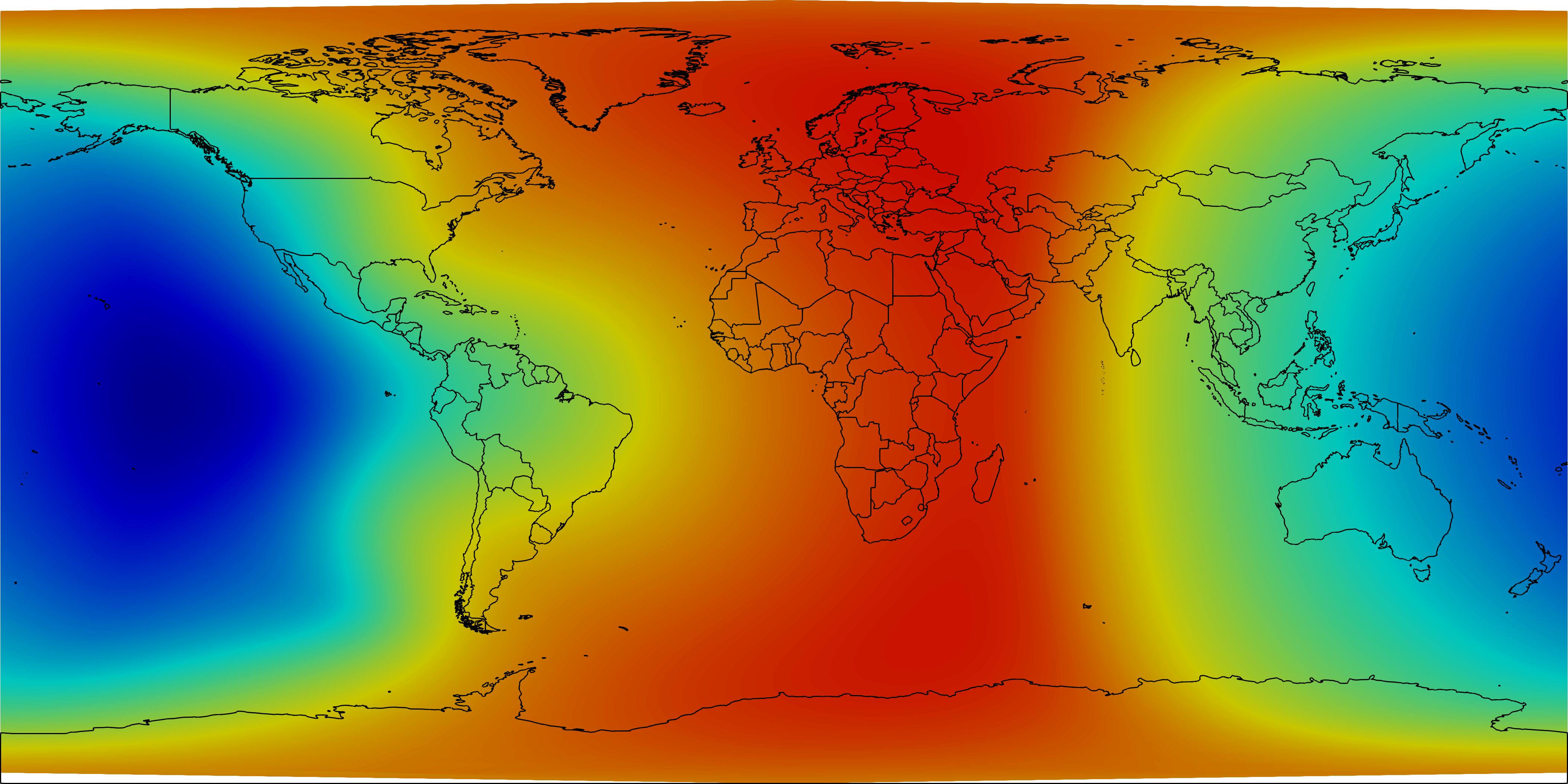}
		\caption{Temperature, $k=1$}
		\label{fig:T_p1H3_12_450}
	\end{subfigure}
	\hfill
	\begin{subfigure}[t]{0.5225\textwidth} \centering
		\includegraphics[width=\textwidth]{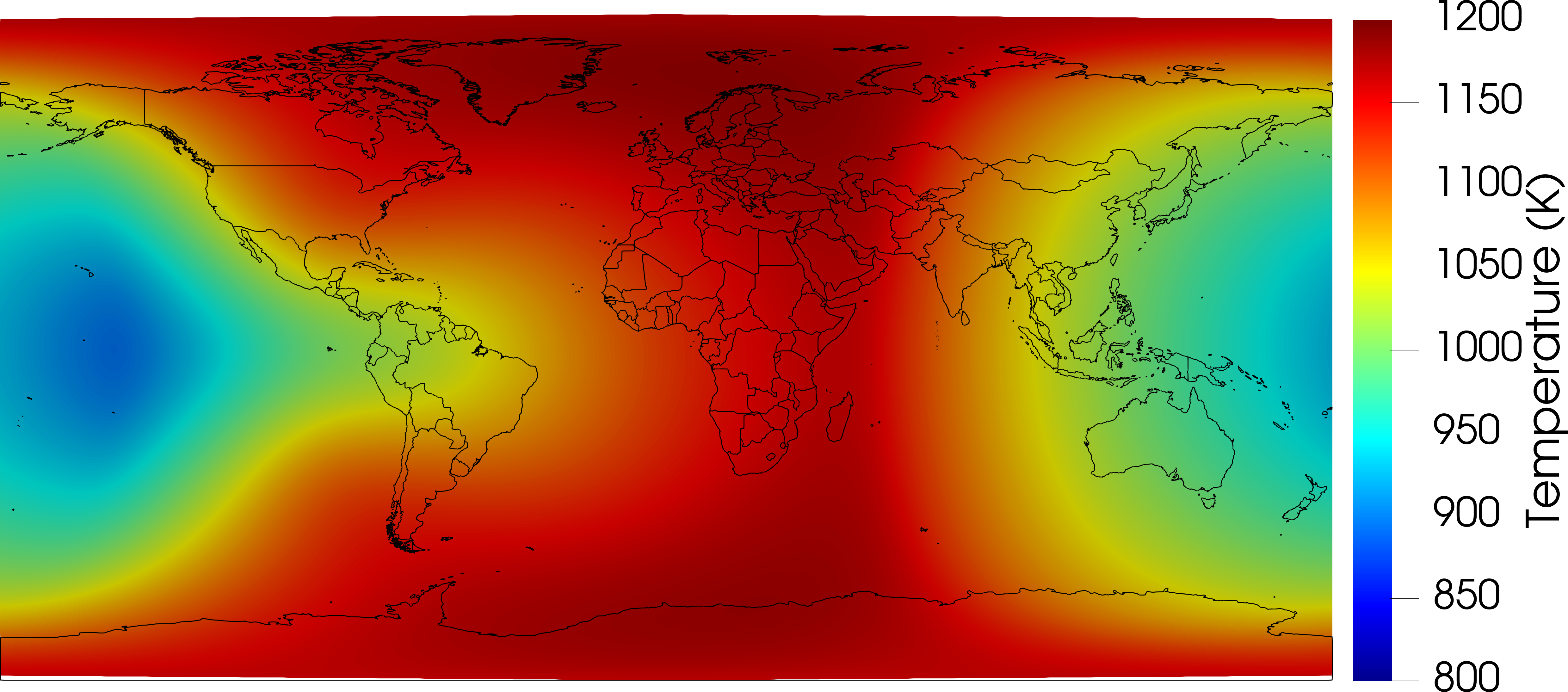}
		\caption{Temperature, $k=2$}
		\label{fig:T_p2H3_12_450}
	\end{subfigure} 
	\caption{Density (top) and temperature (bottom) approximations at 450km of altitude at 12h UTC on 20/03/2014, obtained using different polynomial orders of approximation.}
	\label{fig:case1_solutions}
\end{figure}

Contour maps of the density and temperature fields at 450km of altitude for the linear and quadratic approximations are shown in Figure~\ref{fig:case1_solutions}.
Comparing the two approximations, a more dissipative behavior can be qualitatively observed for the linear case which shows lower temperatures and densities.

\begin{figure}[htbp]
	\centering
	\begin{subfigure}[t]{0.46\textwidth} \centering
		\raisebox{0.75mm}{\includegraphics[width=\textwidth]{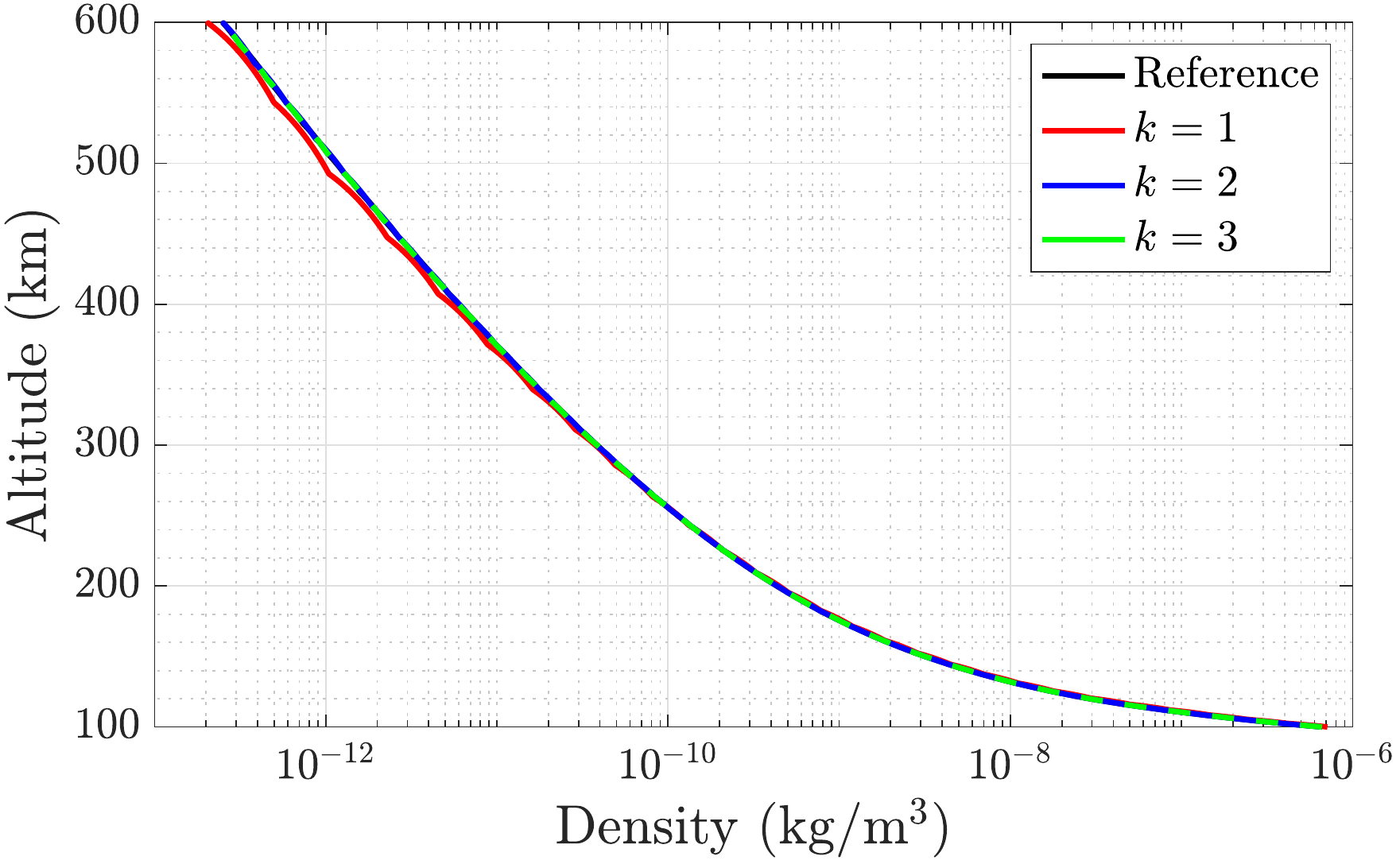}}
	\end{subfigure}
	\quad
	\begin{subfigure}[t]{0.45\textwidth} \centering
		\includegraphics[width=\textwidth]{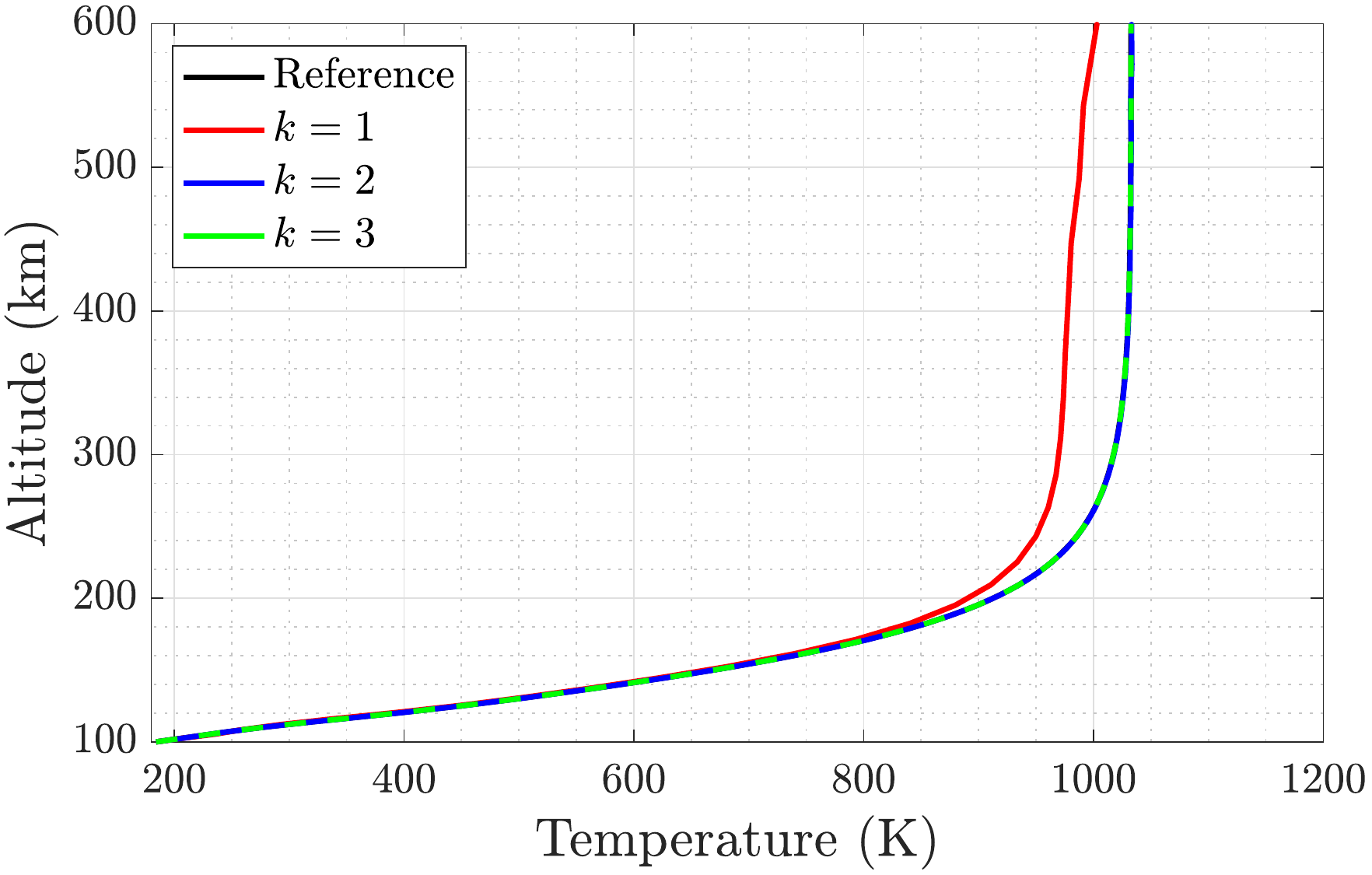}
	\end{subfigure}
	\caption{Radial approximations of the density (left) and temperature (right) above Jicamarca, Per\'u (latitude: -11.9514$^{\circ}$, longitude: -76.8743$^{\circ}$) at 12h UTC on 20/03/2014, using different polynomial orders of approximation.}
	\label{fig:pconv_Jicamarca}
\end{figure}

The dissipative behavior of the linear approximation is also confirmed in the radial 1D analysis over Jicamarca, Per\'u (latitude: -11.9514$^{\circ}$, longitude: -76.8743$^{\circ}$), shown in Figure~\ref{fig:pconv_Jicamarca}. Lower temperatures are obtained for the $k=1$ solution, leading as well to slightly lower levels of density, whereas the quadratic and cubic approximations converge to the reference solution on the finer mesh.

In order to quantify the accuracy of the corresponding approximations, the $\eltwo$ error of the numerical solutions at 12h UTC on 20/03/2014 is computed with respect to the cubic approximation on the refined mesh.
The errors on the density, velocity and temperature approximations are displayed in Table~\ref{tb:pConvergence} and show an exponential decay with the polynomial approximation order.


\begin{table}[htbp]
	\centering
	\begin{tabular}{|c|r|ccc|c r|}
		\hline
		$k$ & $ \ndofs $ & $\norm{E_{\logr}}_{\eltwo}$ & $\norm{E_{\bv}}_{\eltwo}$ & $\norm{E_{T}}_{\eltwo}$ & $\tscheme$ & $\ntime$ \\
		\hline
		1 & 8.71M & $1.35 \times 10^{-2}$ & $3.54 \times 10^{-1}$ & $3.87 \times 10^{-1}$ & DIRK(2,2) & 9000 \\
		2 & 29.39M & $1.20 \times 10^{-4}$ & $1.28 \times 10^{-2}$ & $8.64 \times 10^{-2}$ & DIRK(2,2) & 18000 \\
		3 & 69.67M & $1.46 \times 10^{-7}$ & $1.23 \times 10^{-4}$ & $6.50 \times 10^{-4}$ & DIRK(3,3) & 22500 \\
		\hline
	\end{tabular}
	\caption{Convergence of the $\eltwo$ error of the logarithmic density, velocity and temperature with respect to the polynomial order of approximation, computed at 12h UTC on 20/03/2014. The last two columns indicate the temporal scheme and the number of time steps to reach the final time.}
	\label{tb:pConvergence}
\end{table}

\subsubsection{Density estimation along the Swarm A satellite orbit} \label{sssc:satellite}
\begin{figure}[htbp]
	\centering
	\includegraphics[width=0.46\textwidth]{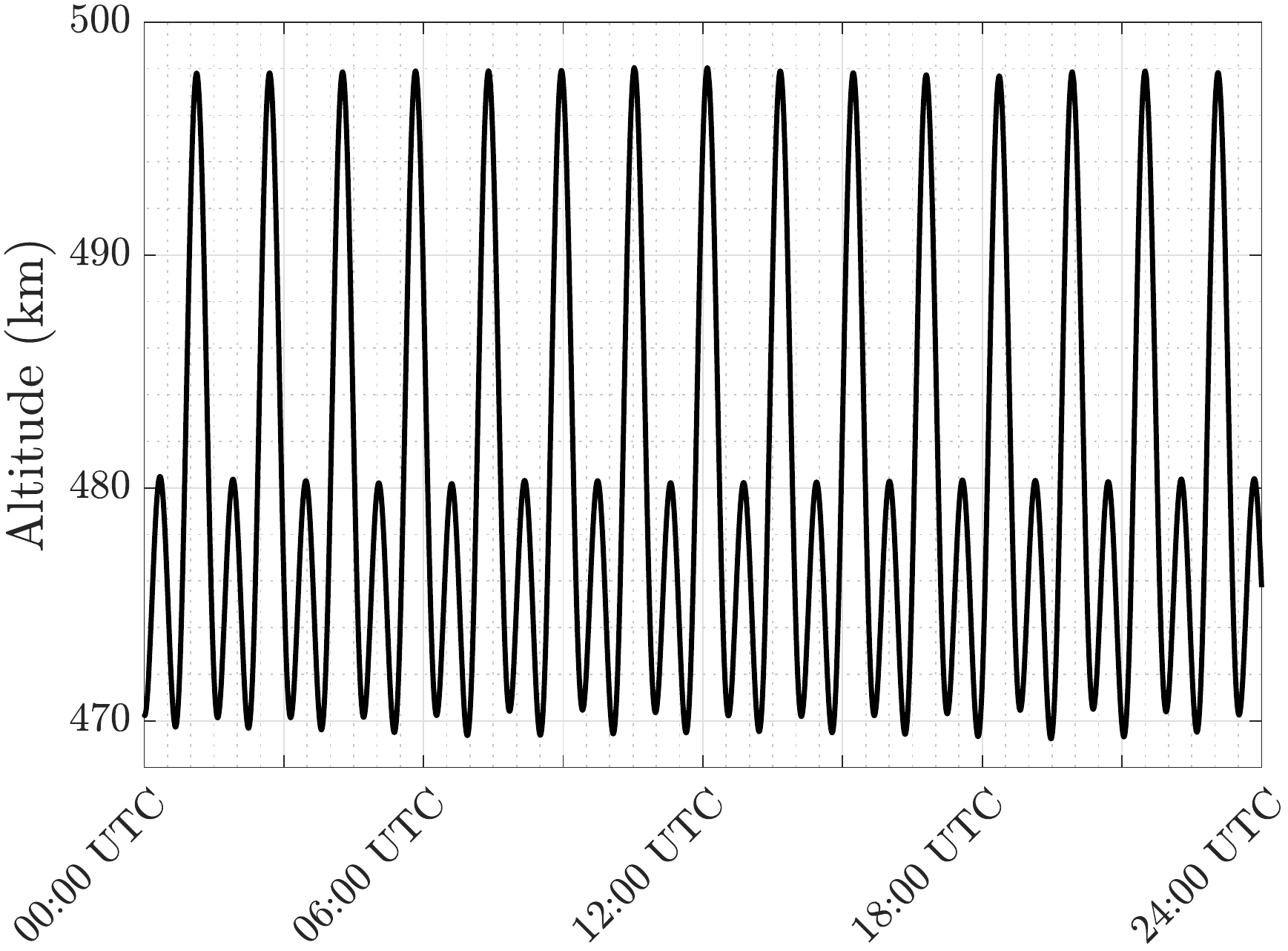}
	\hfill
	\raisebox{1.9mm}{\includegraphics[width=0.49\textwidth]{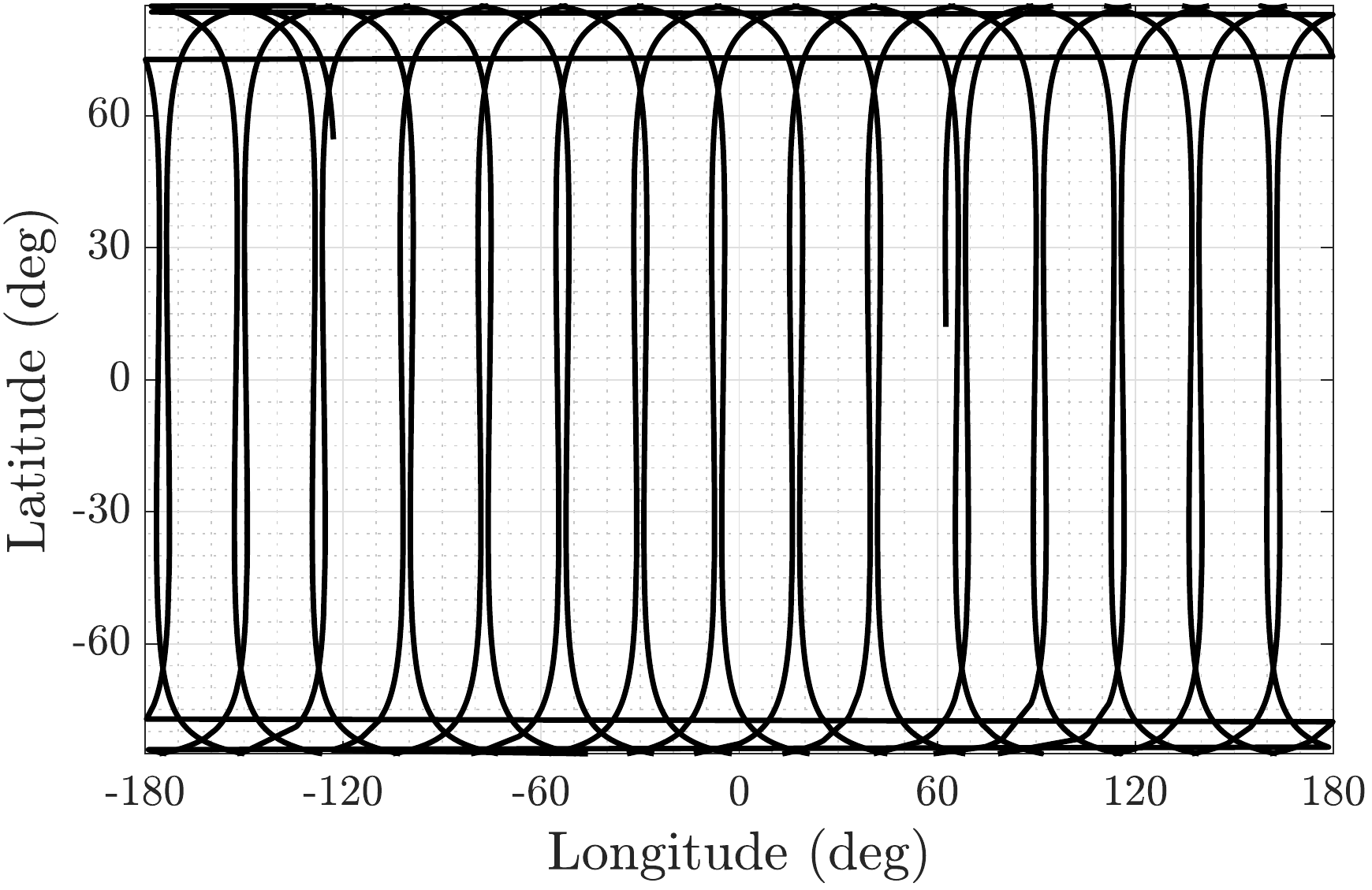}}
	\caption{Schematics of the Swarm A trajectory during 20/03/2014: altitude position in time (left) and latitude-longitude position (right).}
	\label{fig:SwarmAposition}
\end{figure}

The physics-based thermospheric model is validated by comparing the density estimates derived from the numerical approximation with experimental observations gathered along a satellite's orbit.
In particular, the trajectory of the Swarm A satellite during the time of interest, namely on 20/03/2014, is considered.
The satellite performs approximately 15 orbits around the globe and its altitude oscillates between 470 and 500km, as shown in Figure~\ref{fig:SwarmAposition}.

\begin{figure}[htbp]
	\centering
	\includegraphics[width=0.65\textwidth]{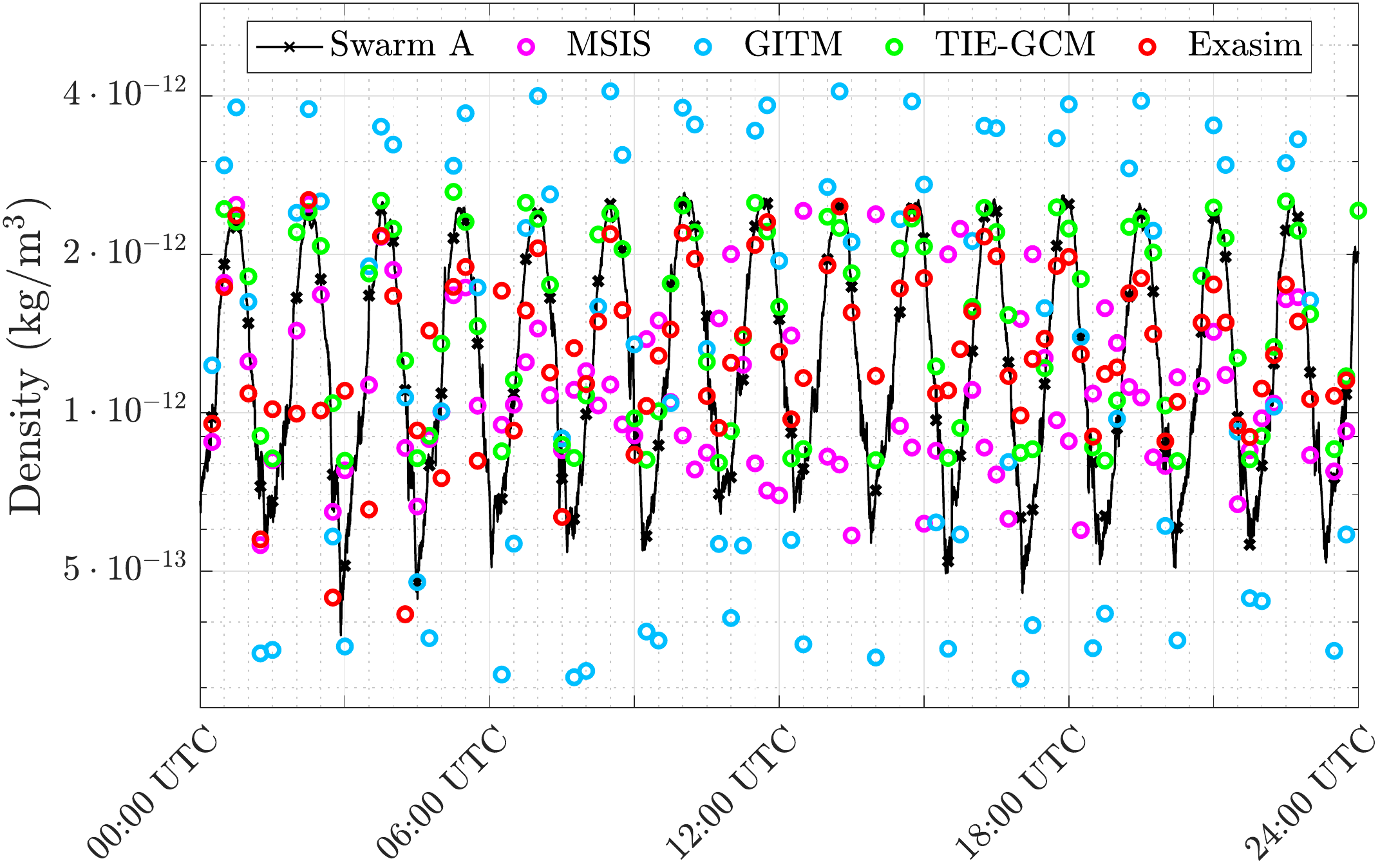}
	\caption{Density predictions of different thermospheric models along the Swarm A satellite trajectory during 20/04/2014, compared to observational data.}
	\label{fig:SwarmA_densities}
\end{figure}

The model's density predictions, interpolated at each point of the satellite's position where the solution is gathered, are contrasted with the density estimates inferred from the drag force recorded by the spacecraft accelerometers. After the relevant corrections and postprocessing~\cite{Siemes2016}, the results are displayed in Figure ~\ref{fig:SwarmA_densities}.
The model and experimental results are also compared with the predictions obtained from the empirical model MSIS, and from the physics-based models GITM and TIE-GCM, which are run at the NASA's CCMC simulation service with the configuration detailed in~\ref{app:CCMC}.

The density predictions made by Exasim align closely with the satellite estimates,  mirroring the behavior of TIE-GCM.
Conversely, GITM exhibits larger oscillation amplitudes in this case, leading to both overpredictions and underpredictions of the density along the orbit. Finally,  for this case, the empirical model MSIS persistently underestimates the density.

It is worth noting that, whereas the satellite density estimates are collected every few seconds (generally 10s for the period of interest), for data management reasons, models collect their solutions every few minutes. In particular, the model solutions are displayed at every 15 minutes. Regarding MSIS, the corresponding evaluations are also performed every 15 minutes for consistency with the physics-based models.

The root mean square error of the different predictions is reported in Table~\ref{tb:SwarmA_RSMerror} for a quantitative assessment of the accuracy of each approximation.
The physics-based ionospheric-thermospheric model TIE-GCM provides the most accurate results, while the proposed physics-based thermospheric model delivers satisfactory density estimates for this case. On the other hand, the predictions from GITM and MSIS display larger degrees of error.

\begin{table}[htbp]
	\centering
	\begin{tabular}{|c|cccc|}
		\hline
		 & MSIS & GITM & TIE-GCM & Exasim \\
		\hline
		RMS error & 0.5325 & 0.4871 & 0.1531 & 0.2528\\
		\hline
	\end{tabular}
	\caption{Root mean square error of the density estimates of different thermospheric models along the Swarm A satellite trajectory with respect to observational data.}
	\label{tb:SwarmA_RSMerror}
\end{table}

This example demonstrates the robust performance of the proposed thermospheric model, providing neutral density predictions along the Swarm A satellite trajectory with error levels comparable to established physics-based models. These established models solve for a broader array of components and describe a considerably more intricate physical system.

Additionally, it's noteworthy that none of the models, including those with the most uncertain parameters, underwent prior calibration to more accurately align with the experimental results. Consequently, a more refined calibration of some of the model parameters could potentially enhance the solutions and minimize discrepancies between various approximations.

\subsubsection{Radial one-dimensional model comparison} \label{sssc:1Dcomparison}
This section evaluates the performance of the radial 1D thermospheric model implemented in Exasim. The 1D model, operated during the March 20-23 time period over Jicamarca, Per\'u (latitude: -11.9514$^{\circ}$, longitude: -76.8743$^{\circ}$), is compared against various empirical and physics-based models
The same resolution as in the cube-sphere meshes is used for the radial direction. Consequently, the simulation employs a 1D stretched mesh consisting of 28 elements with second-order polynomials. A constant growth rate is applied, with the width of the element at the lower boundary being 2.5km.

\begin{figure}[htbp]
	\centering
		\includegraphics[width=0.6\textwidth]{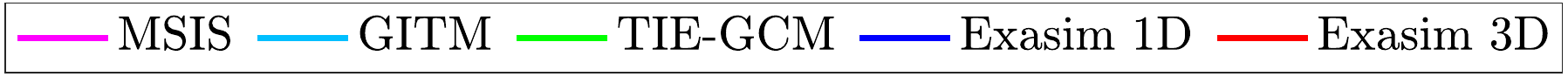} \\
		\includegraphics[width=0.485\textwidth]{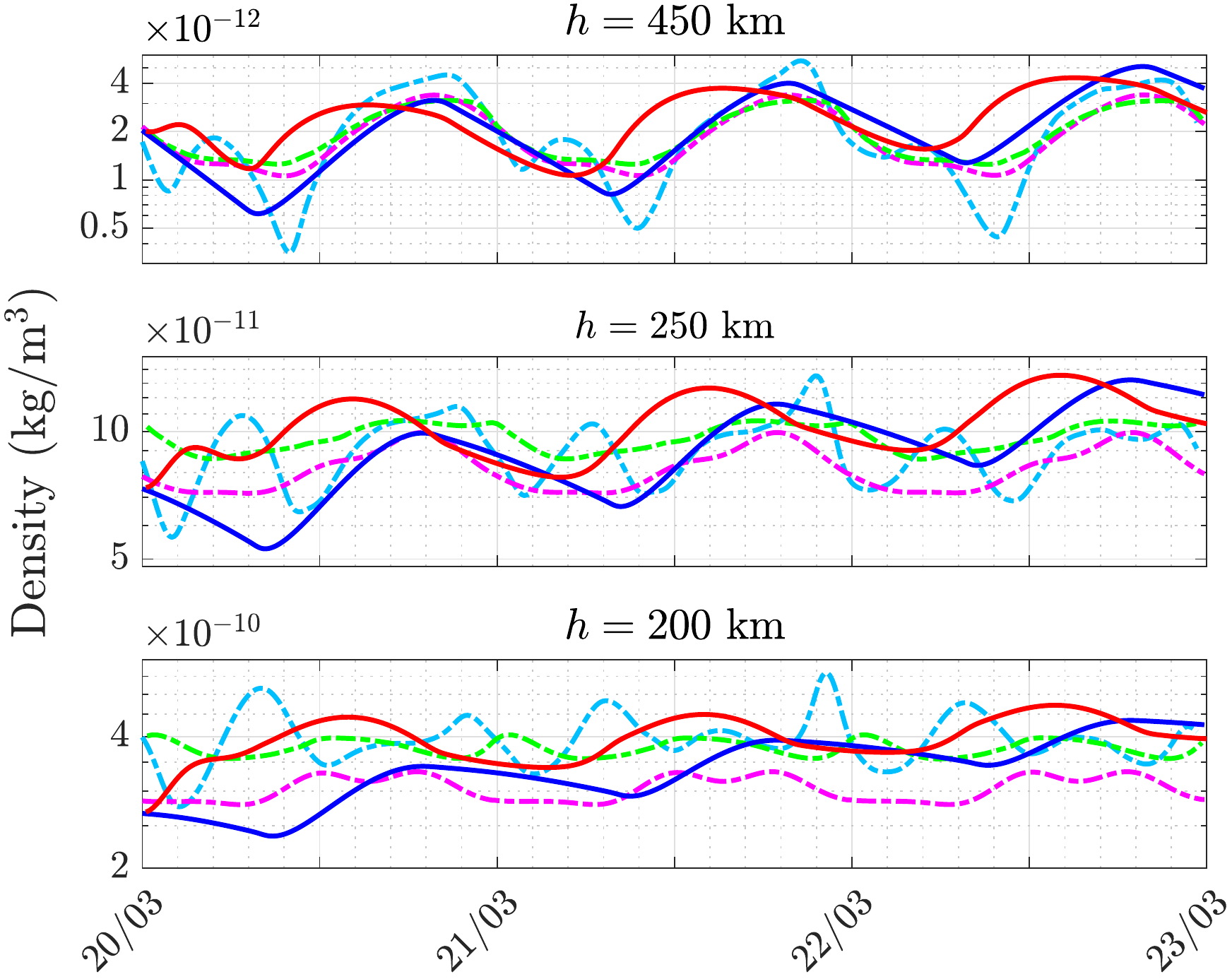} \hfill
		\includegraphics[width=0.485\textwidth]{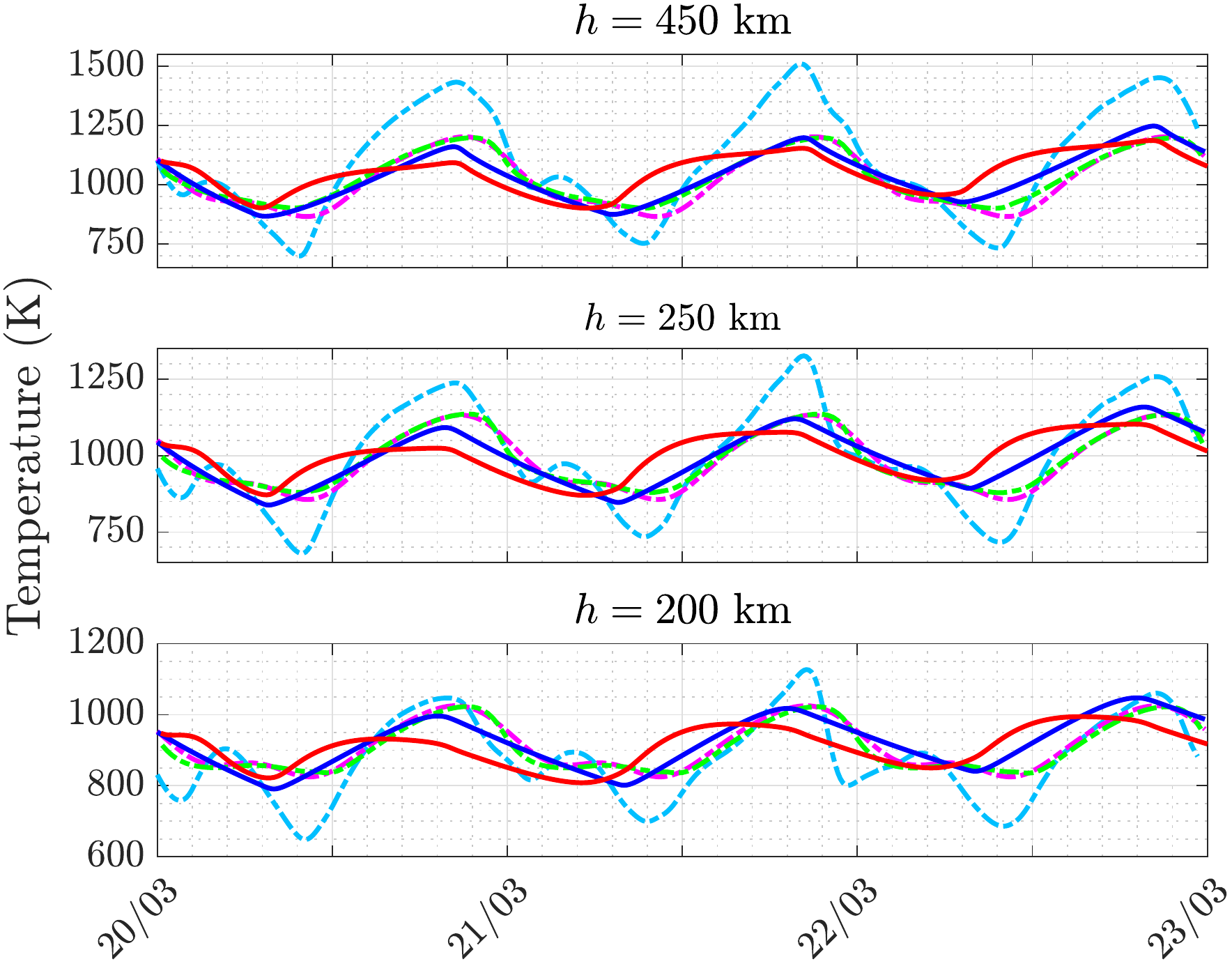}
	\caption{Comparison of the density (left) and temperature (right) temporal evolutions at different altitudes above Jicamarca, Per\'u (latitude: -11.9514, longitude: -76.8743) for different thermospheric models during 20-23 March 2014.}
	\label{fig:1D3D_Jicamarca}
\end{figure}

Figure~\ref{fig:1D3D_Jicamarca} depicts the temporal evolution of the density and temperature fields at different altitudes above Jicamarca  for the different models. For comparison purposes, the results of the 3D simulations were interpolated at the same spatial location.

The temperature results show a strong similarity between the 1D and 3D models. The 1D model also shows a very good agreement with TIE-GCM and MSIS temperature estimates, whereas the GITM approximation displays a larger amplitude.
The presence of 3D effects, such as horizontal winds, Coriolis acceleration or a stronger effect of the inertia terms, produces the slight differences between the 1D and 3D implementations.
These differences are amplified in the density approximation, owing to the nearly exponential behavior of this quantity. Nevertheless, both the 1D and 3D results show a comparable performance to the rest of models, specially at higher altitudes.
Indeed, the simplified description of the lower boundary condition may be responsible for introducing an additional source of uncertainty.

These results serve as further validation of the current approach compared to other existing physics-based models, highlighting the good performance of the 1D model in accurately mirroring the thermospheric dynamics and maintaining consistency with 3D implementations.
Moreover, the 1D implementation facilitates significantly cost-effective runs compared to the 3D counterpart due to a substantial reduction in the number of degrees of freedom, simplified dynamics, the possibility of increasing time steps, and improved problem conditioning. This makes it well-suited for parametric studies or development purposes.


\subsection{Model sensitivity to heating parameters under variable solar activity} \label{ssc:case2}

Lastly, this section presents a parametric study of two principal factors influencing the thermospheric response – the EUV heating source and thermal conductivity. This is done to assess the sensitivity of the physics-based model towards the depiction of the main heating elements.
This analysis investigates the performance of diverse parameter combinations subjected to varying solar conditions, utilizing the 1D model.
This process creates a comprehensive data set that helps understanding the impact of these different terms.  This enables the execution of 3D simulations of the different cases with a single set of parameters that offers the best combined performance and avoids costly nonphysical predictions.
The strategy outlined here, demonstrating the model's versatility in various configurations, can be integrated into a simulation pipeline, which utilizes the 1D model to calibrate the suitability of different parametric configurations before operating the full 3D model.
This methodology can of course be extended to conduct further sensitivity analyses with respect to other model inputs.

In this study, the EUV efficiency is used to modulate the absorption of solar energy into the system. On the other hand, a constant parameter $\alpha_0$ adjusts the amount of thermal dissipation, by modifying the nominal molecular conductivity according to
\begin{equation} 
	\kappa_m  = \alpha_0 \sum_s \left( \frac{n_s}{\sum_k n_k} \alpha_s \right)  T ^{0.69}.
\end{equation}

In particular, the EUV efficiency takes values within a $20\%$ of its nominal value, that is $\epsilon = 0.25 \pm 0.05$, whereas the thermal conductivity is only modulated by decreasing the amount of dissipation up to a maximum of $50\%$.
A uniform distribution is assumed for both quantities, and the discrete values considered for the parametric evaluation are $\epsilon \in \lbrace 0.2, 0.225, 0.25, 0.275, 0.3 \rbrace$ and $\alpha_0 \in \lbrace 0.5, 0.625, 0.75, 0.875, 1 \rbrace$.

All the possible combinations of the heating parameters for three different cases, corresponding to equinox and solstice conditions at various solar irradiance intensities and described in Table~\ref{tb:casesSAdescription}, are considered.
\begin{table}[htbp]
	\centering
	\begin{tabular}{|c|l|l|l|r|r|}
		\hline
		Case & Dates & Description & Solar activity &$F_{10.7}$ & $F_{10.7}^{81}$ \\
		\hline
		1 & 20-23 March 2014 & Spring equinox & Moderate/high & 150 & 150 \\
		\hline
		2 & 20-23 March 2020 & Spring equinox & Low & 70 & 70 \\
		\hline
		3 & 15-18 June 2022 & Summer solstice & Moderate/high & 150 & 120 \\
		\hline
	\end{tabular}
	\caption{Description of the examples considered for the parametric study, corresponding to cases with different solar activity and heating distribution.}
	\label{tb:casesSAdescription}
\end{table}
Note that, in case 2, the solar minimum conditions lead to much lower pressure/density levels at a given altitude than in moderate or high solar conditions. For this reason, the problem is solved up to 500km of altitude, instead of 600km, as done in cases 1 and 3.

\subsubsection{Parametric study on the 1D model} \label{sssc:1Dparametric}
The parametric study leverages the good performance of the 1D model in consistently mimicking thermospheric dynamics and conducts several runs at three different spatial locations. This is done to collect spatially distributed information representative of distinct behavior observed between the Northern and Southern hemispheres, or near the Equator.
The locations of interest are listed as follows:
\begin{itemize}
	\item Millstone Hill, USA: latitude 42.6347$^{\circ}$, longitude -71.4883$^{\circ}$,
	\item Quito, Ecuador: latitude -0.1859$^{\circ}$, longitude -78.4663$^{\circ}$,
	\item Santiago, Chile: latitude -33.3458$^{\circ}$, longitude -70.6786$^{\circ}$.
\end{itemize}

The problems are solved using a quadratic approximation in the radial 1D mesh described in previous sections, consisting of 28 elements (26 elements for case 2), with geometric growth with altitude and a first element with a height of 2.5km.
The time integration uses a DIRK(2,2) scheme, with time steps of 30s.

\begin{figure}[htbp]
	\centering
	\begin{subfigure}[t]{0.5\textwidth} \centering
		\includegraphics[width=0.8\textwidth]{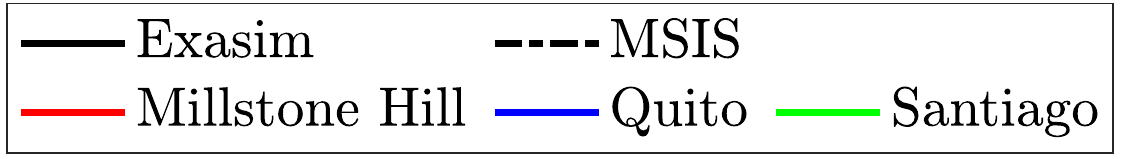}
	\end{subfigure} \\
	\begin{subfigure}[t]{0.325\textwidth} \centering
		\includegraphics[width=\textwidth]{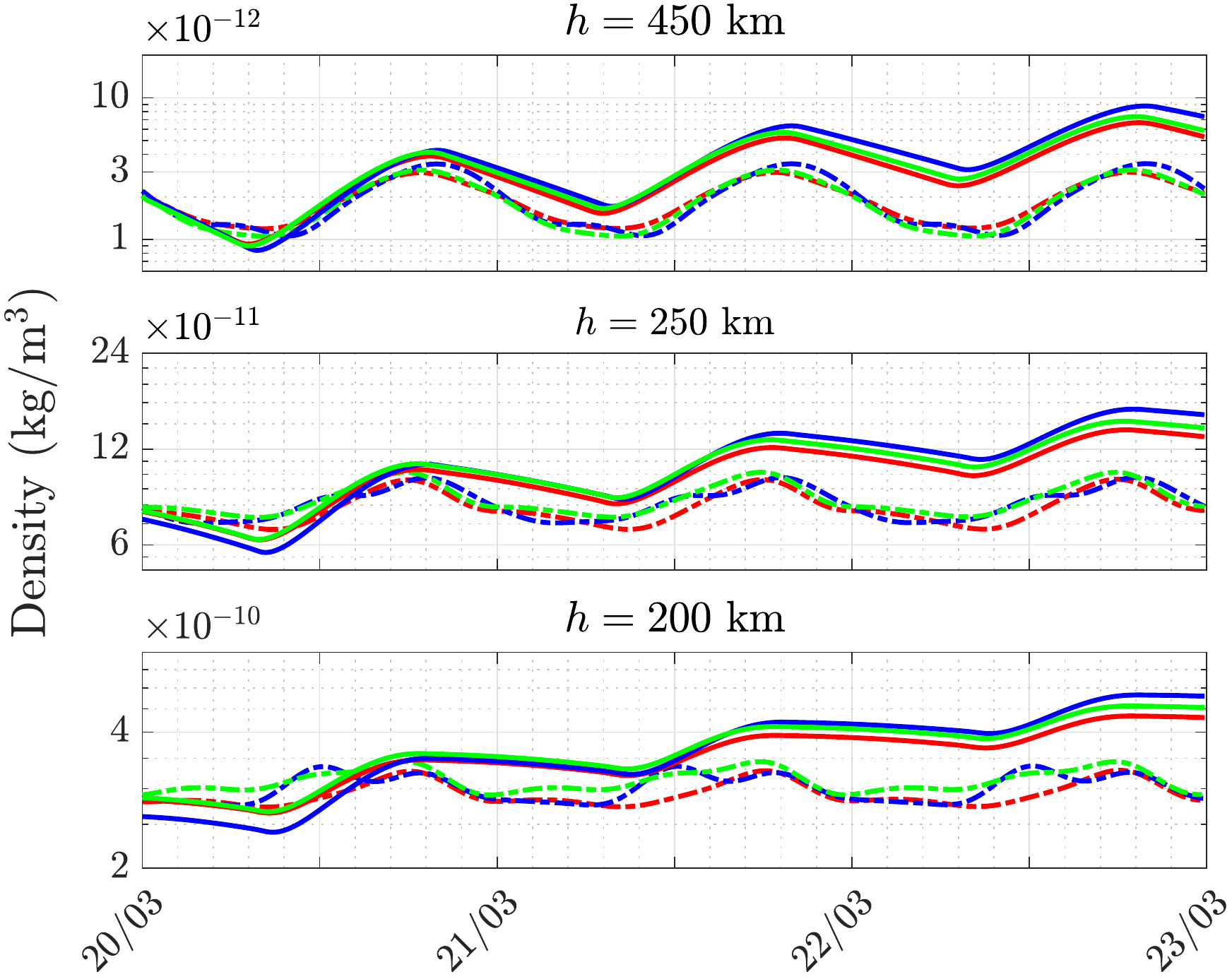}
		\caption{Case 1, $\epsilon=0.25, \alpha_0 = 0.75$}
		\label{fig:density1D_ex1_3}
	\end{subfigure}
	\hfill
	\begin{subfigure}[t]{0.325\textwidth} \centering
		\includegraphics[width=\textwidth]{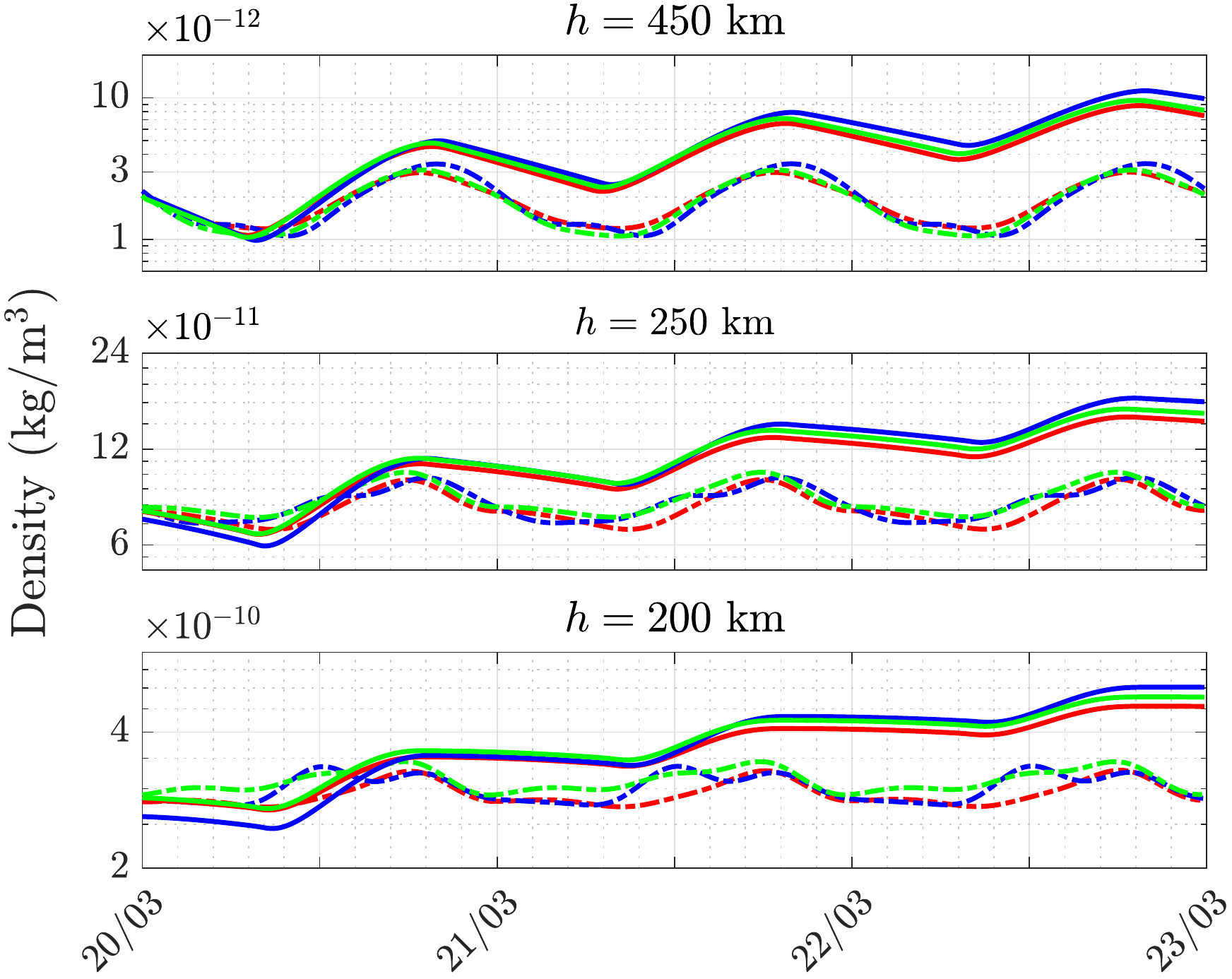}
		\caption{Case 1, $\epsilon=0.25, \alpha_0 = 0.625$}
		\label{fig:density1D_ex1_1}
	\end{subfigure} 
	\hfill
	\begin{subfigure}[t]{0.325\textwidth} \centering
		\includegraphics[width=\textwidth]{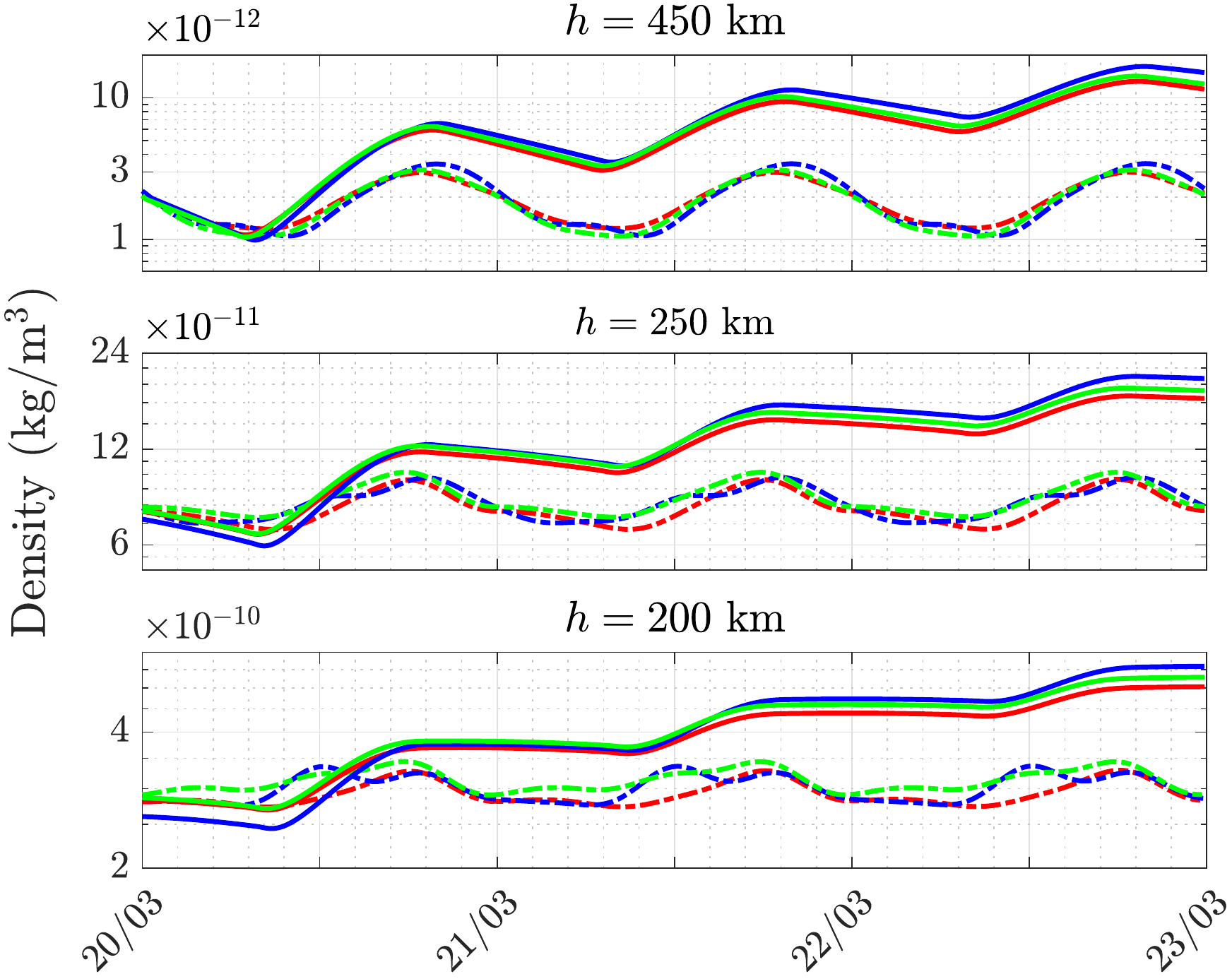}
		\caption{Case 1, $\epsilon=0.3, \alpha_0 = 0.625$}
		\label{fig:density1D_ex1_2}
	\end{subfigure} 
	\vspace{3mm}
	\begin{subfigure}[t]{0.325\textwidth} \centering
		\includegraphics[width=\textwidth]{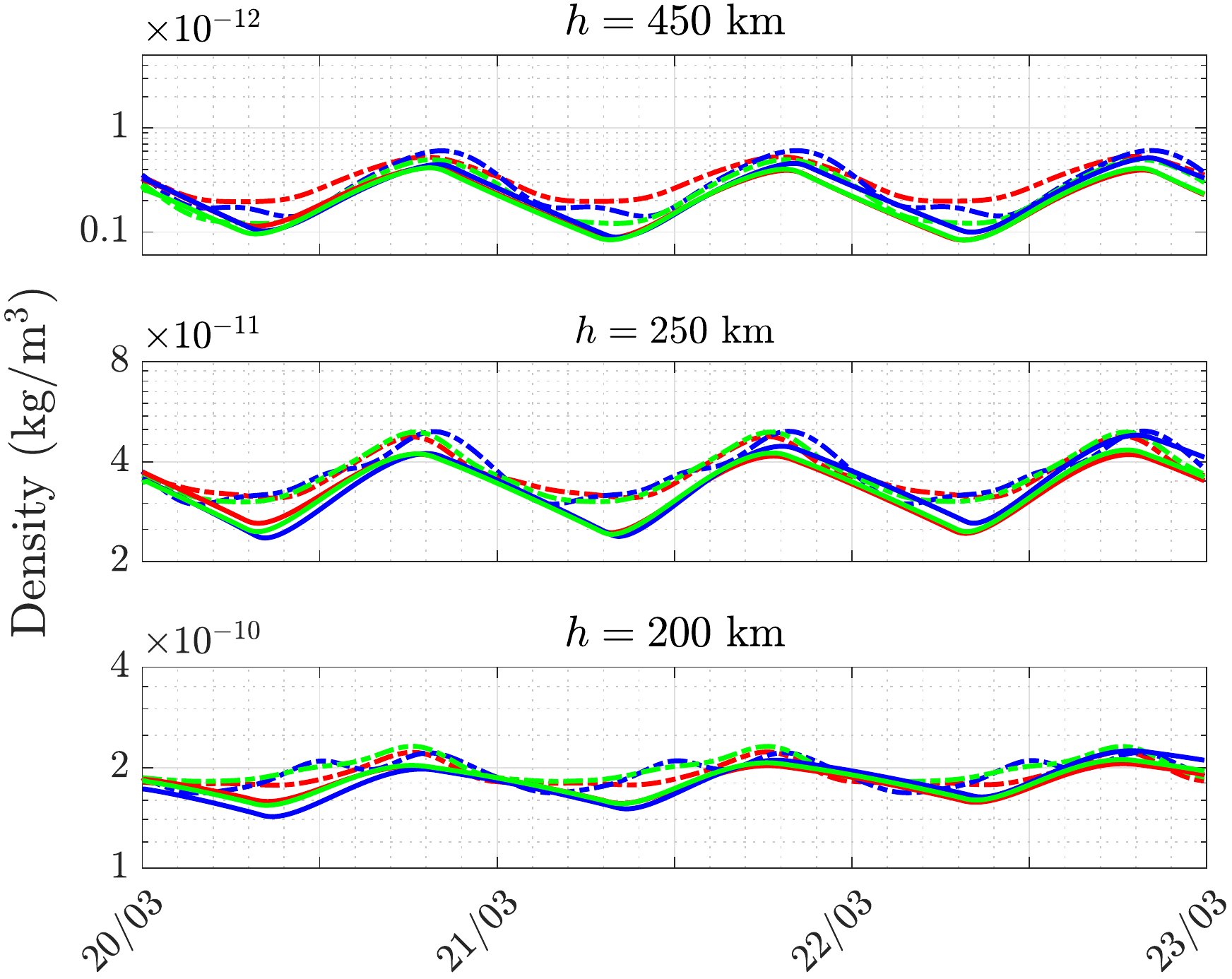}
		\caption{Case 2, $\epsilon=0.25, \alpha_0 = 0.75$}
		\label{fig:density1D_ex2_3}
	\end{subfigure}
	\hfill
	\begin{subfigure}[t]{0.325\textwidth} \centering
		\includegraphics[width=\textwidth]{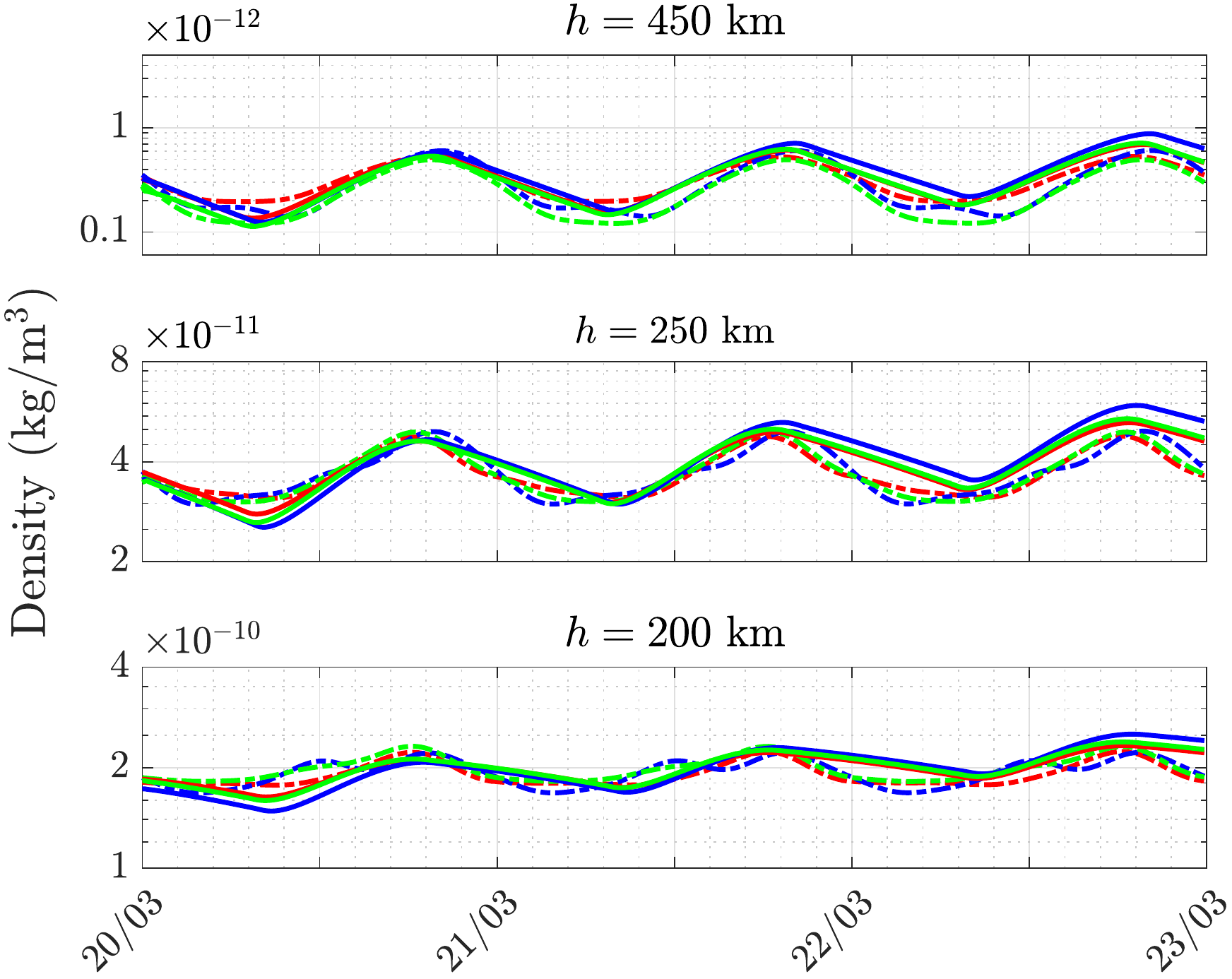}
		\caption{Case 2, $\epsilon=0.25, \alpha_0 = 0.625$}
		\label{fig:density1D_ex2_1}
	\end{subfigure} 
	\hfill
	\begin{subfigure}[t]{0.325\textwidth} \centering
		\includegraphics[width=\textwidth]{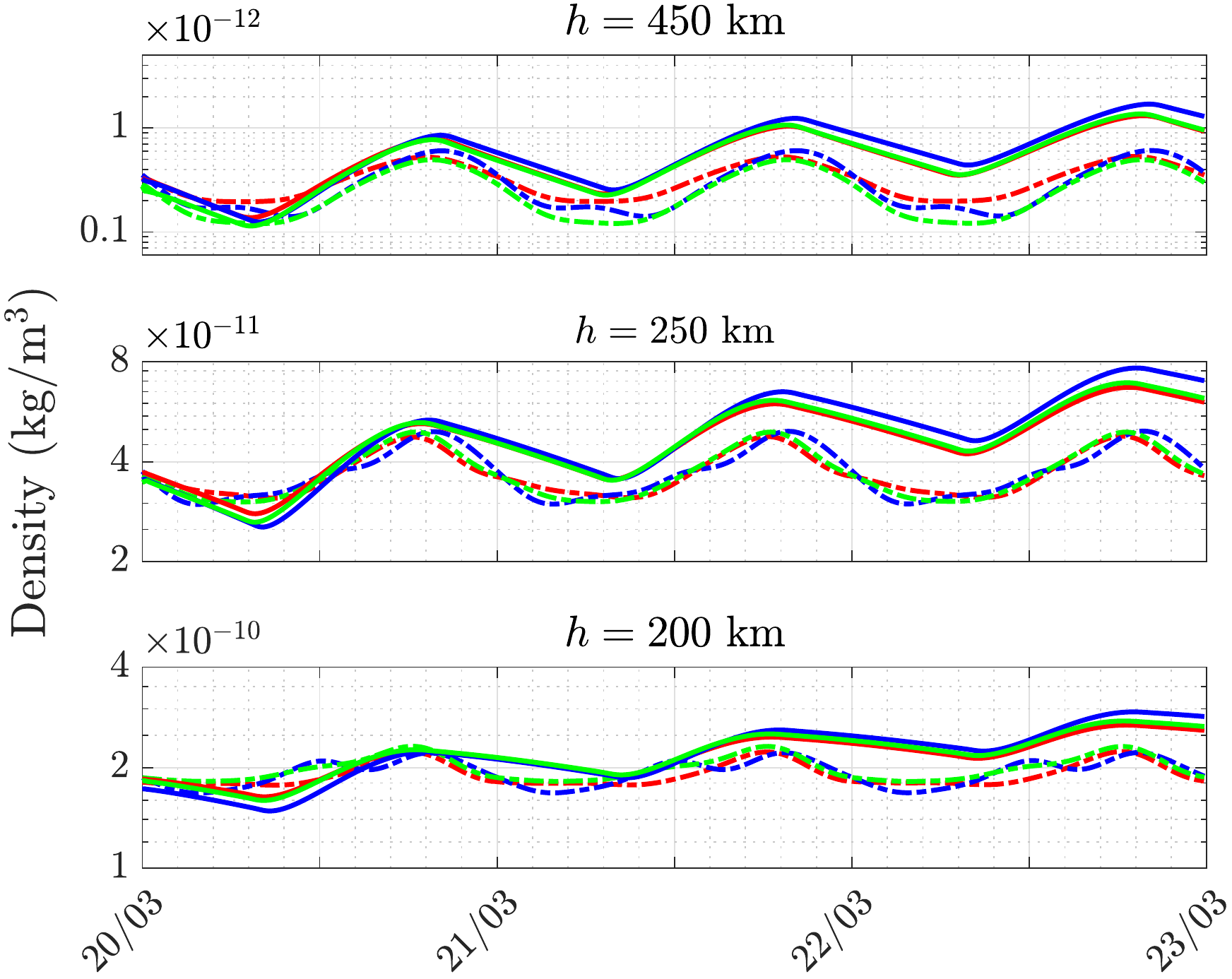}
		\caption{Case 2, $\epsilon=0.3, \alpha_0 = 0.625$}
		\label{fig:density1D_ex2_2}
	\end{subfigure} 
	\vspace{3mm}	
	\begin{subfigure}[t]{0.325\textwidth} \centering
		\includegraphics[width=\textwidth]{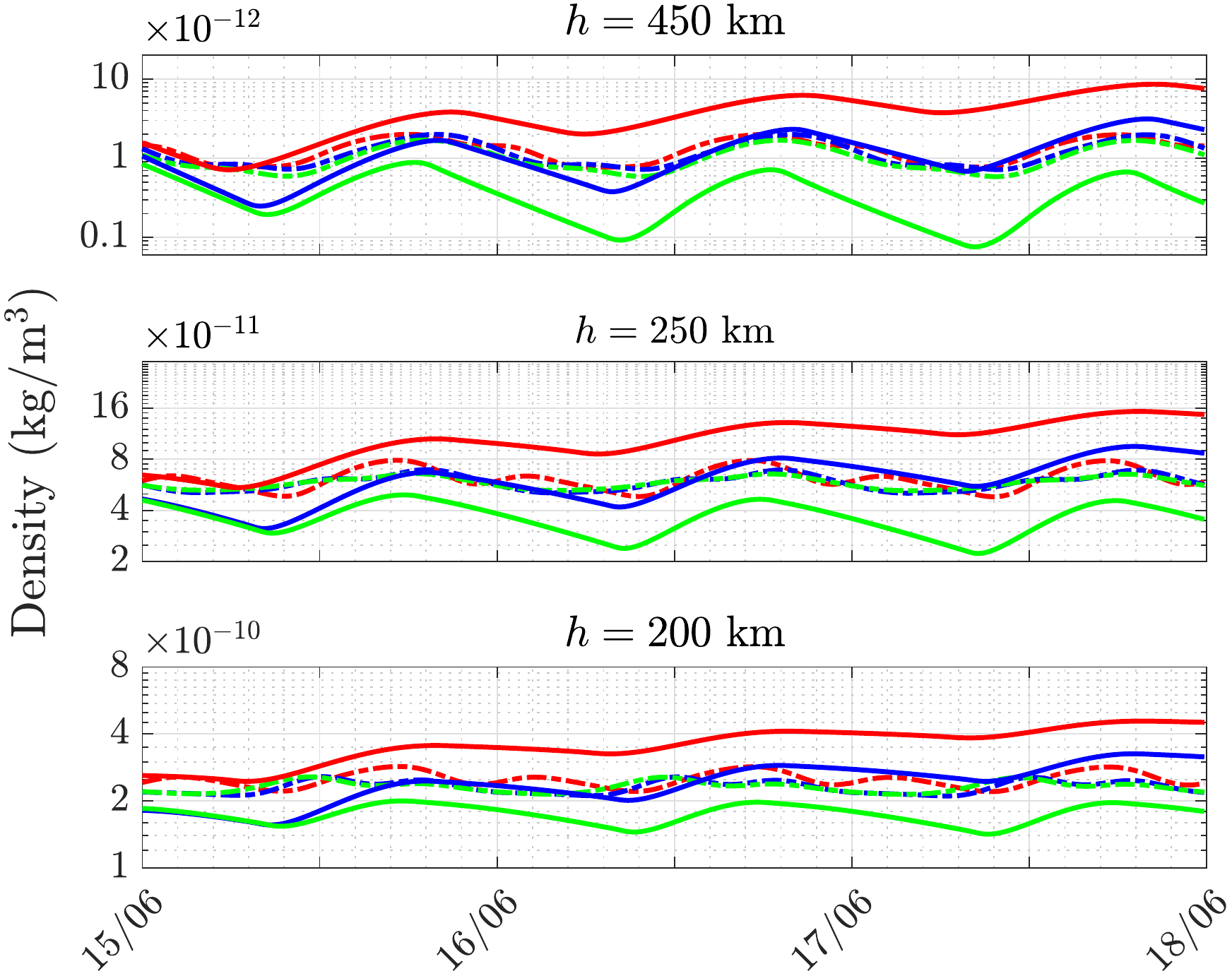}
		\caption{Case 3, $\epsilon=0.25, \alpha_0 = 0.75$}
		\label{fig:density1D_ex3_3}
	\end{subfigure}
	\hfill
	\begin{subfigure}[t]{0.325\textwidth} \centering
		\includegraphics[width=\textwidth]{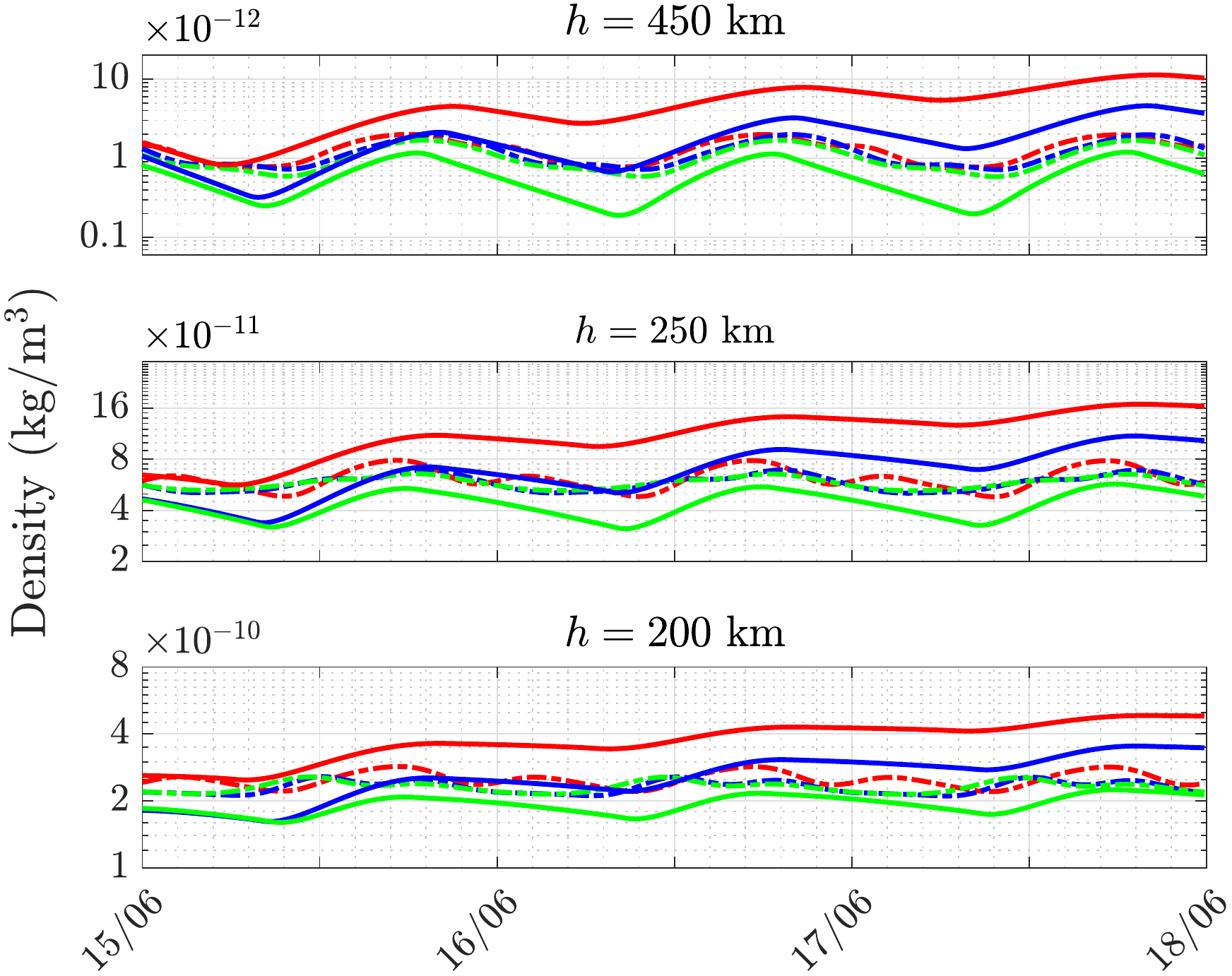}
		\caption{Case 3, $\epsilon=0.25, \alpha_0 = 0.625$}
		\label{fig:density1D_ex3_1}
	\end{subfigure} 
	\hfill
	\begin{subfigure}[t]{0.325\textwidth} \centering
		\includegraphics[width=\textwidth]{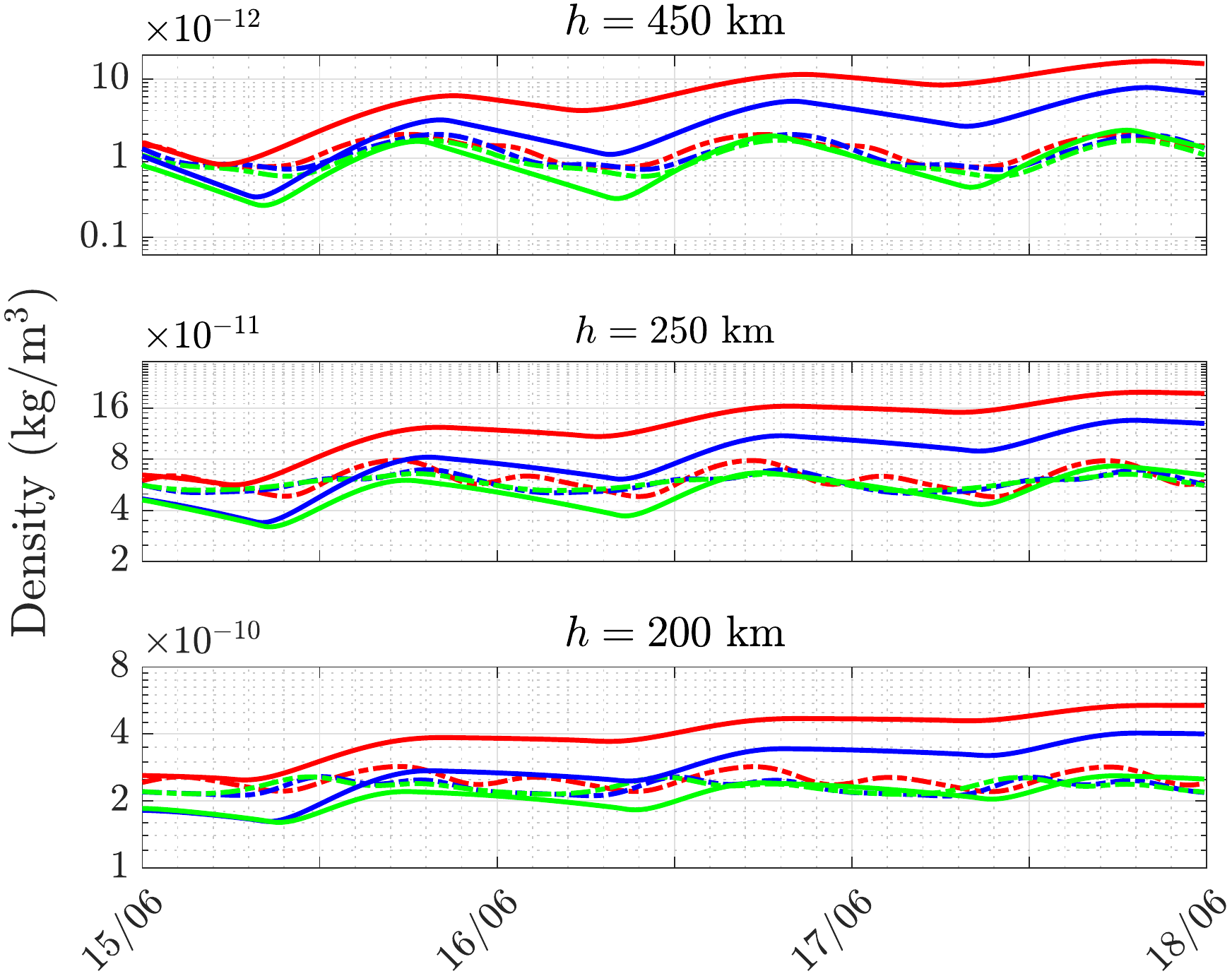}
		\caption{Case 3, $\epsilon=0.3, \alpha_0 = 0.625$}
		\label{fig:density1D_ex3_2}
	\end{subfigure} 
	\caption{Temporal evolution of the density in different cases of interest (top to bottom) at various altitudes above Millstone Hill, Quito and Santiago, using three different combinations of parameters (left to right). The results correspond to the 1D model using a quadratic approximation. MSIS estimates are included for reference.}
	\label{fig:case2_density1D}
\end{figure}

Figure~\ref{fig:case2_density1D} displays the temporal evolution of the predicted density at the different locations, for each of the cases of study, subject to three combinations of parameters, namely $(\epsilon,\alpha_0) = \lbrace (0.25,0.75),(0.25,0.625),(0.3,0.625)\rbrace$
The results are compared to the MSIS estimates, which offers a reference prediction for each case.

The results show the effect of the thermal conductivity and EUV efficiency in the dynamics of the solution and their role in the different cases.
On the one hand, higher EUV efficiencies translate into a higher input of energy to the system. Likewise, smaller thermal conductivities imply a lower capacity to dissipate heat. In both cases, we observe a similar trend of increasing density (and temperature).
On the other hand, we can observe how, for the same choice of parameters, the external solar activity can lead to a different dynamic response. Whereas for equinox conditions, a choice of parameters can work very well for solar minima (for instance, $\epsilon=0.25$ and $\alpha_0 = 0.625$ in case 2), it may lead to a more inaccurate behavior in solar maxima (case 1). Similarly, a certain set of parameters can offer accurate predictions at a given location, but an increasing or decreasing tendency at other locations. For instance, $\epsilon=0.25$ and $\alpha_0 = 0.75$ in case 3 displays a stable tendency for Quito (which follows MSIS predictions), but show an underdissipative  behavior in the North hemisphere and overdissipative in the South hemisphere.

In order to evaluate the suitability of a particular candidate in the parametric space, the $\ell^2$ relative error, with respect to the MSIS estimates, of the density predictions at four different altitudes (150, 200, 250 and 350km) is computed for each case and location of interest. The maximum error value computed at these different altitudes is taken into account.
Given that some combinations of parameters may lead to nonphysical or unstable solutions, these cases are penalized with a large value of the error metric (error $= 10^2$).

\begin{figure}[htbp]
	\centering
	\begin{subfigure}[t]{0.65\textwidth} \centering
		\includegraphics[width=0.6\textwidth]{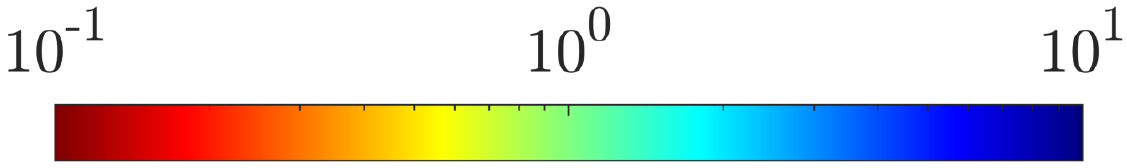}
	\end{subfigure} \\
	\begin{subfigure}[t]{0.325\textwidth} \centering
		\includegraphics[width=0.9\textwidth]{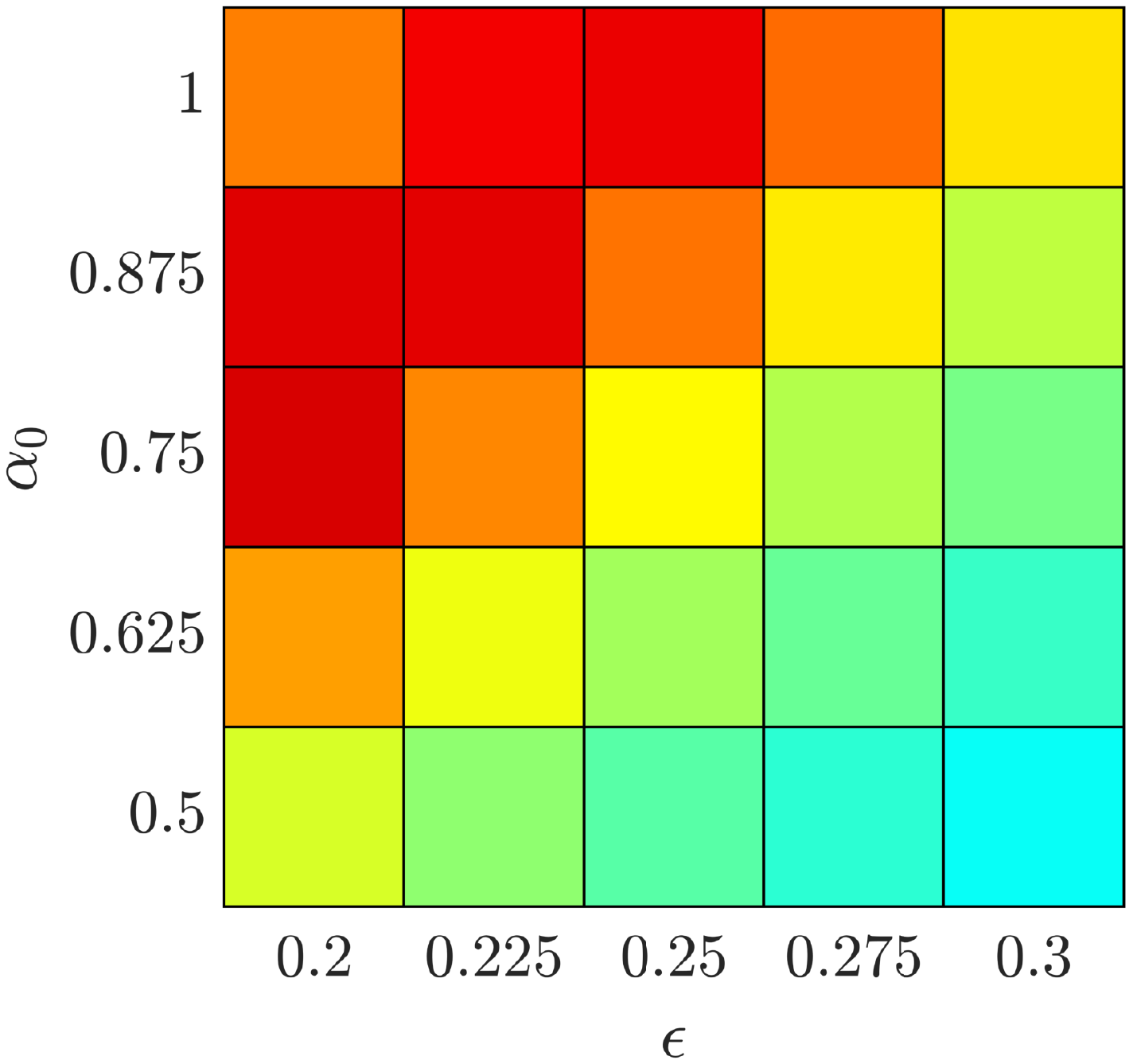}
		\caption{Case 1, Millstone Hill}
		\label{fig:SA_eRho_Millstone_case1}
	\end{subfigure}
	\hfill
	\begin{subfigure}[t]{0.325\textwidth} \centering
		\includegraphics[width=0.9\textwidth]{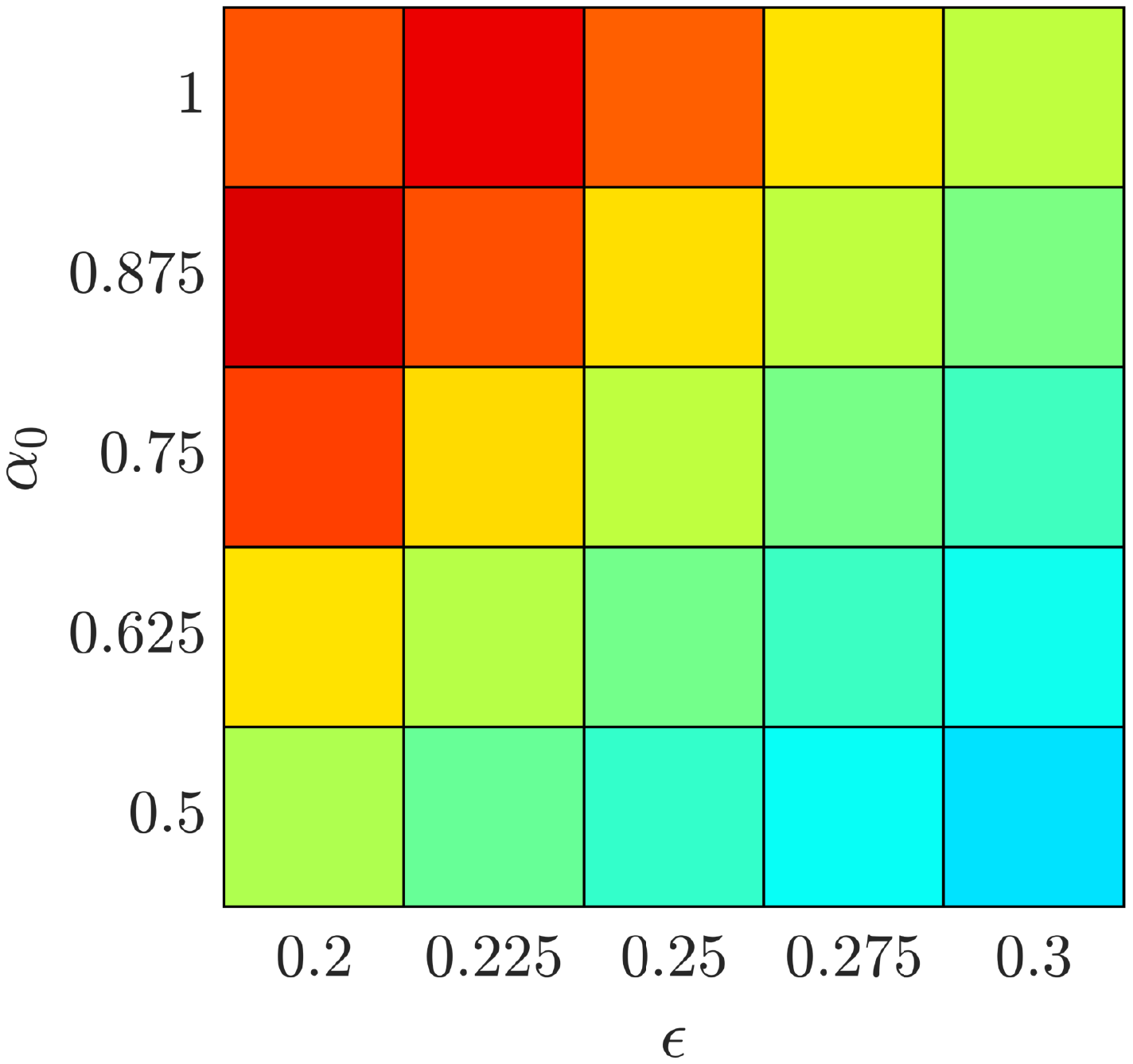}
		\caption{Case 1, Quito}
		\label{fig:SA_eRho_Quito_case1}
	\end{subfigure} 
	\hfill
	\begin{subfigure}[t]{0.325\textwidth} \centering
		\includegraphics[width=0.9\textwidth]{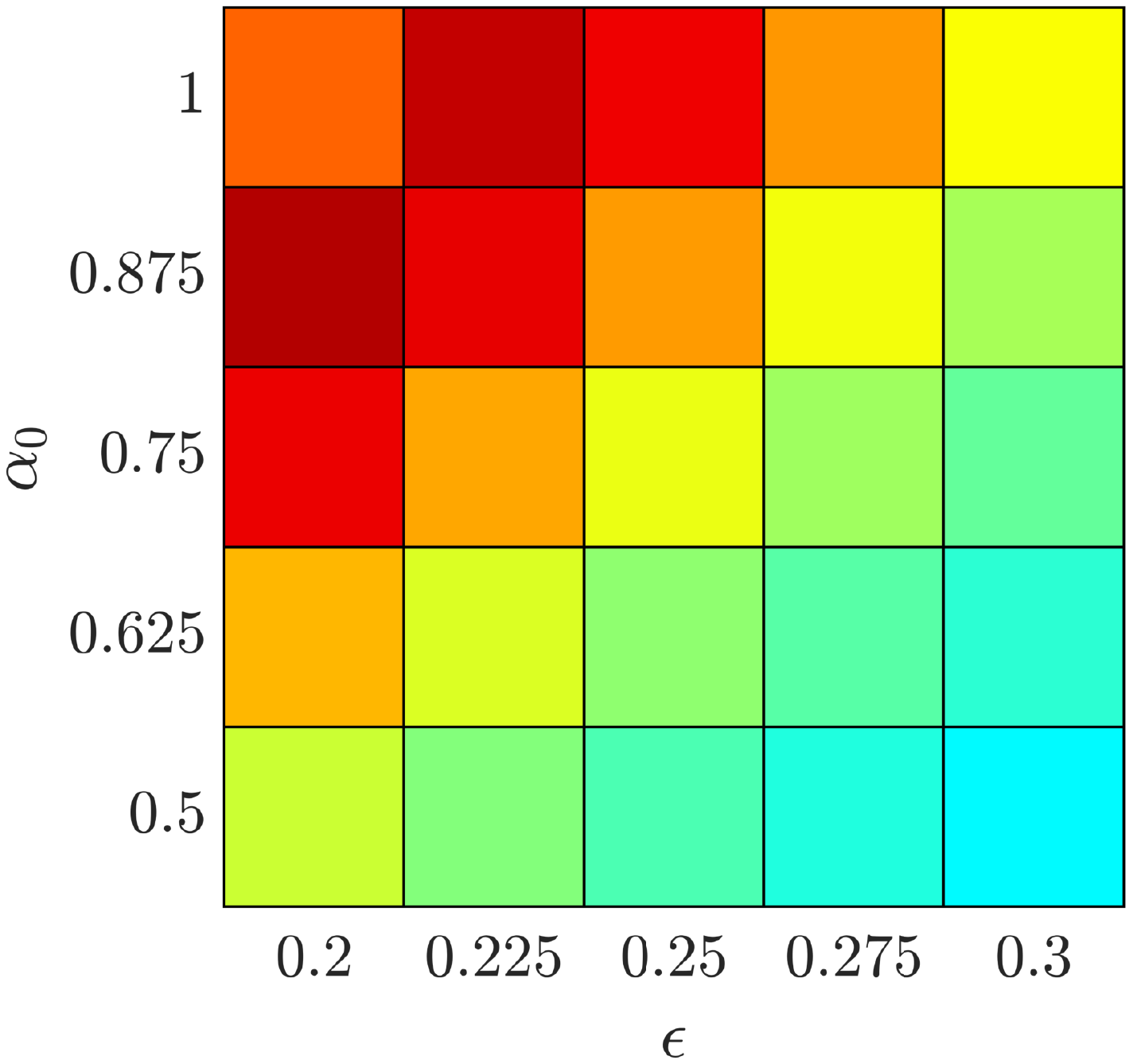}
		\caption{Case 1, Santiago}
		\label{fig:SA_eRho_Santiago_case1}
	\end{subfigure} 
\vspace{3mm}	
	\begin{subfigure}[t]{0.325\textwidth} \centering
		\includegraphics[width=0.9\textwidth]{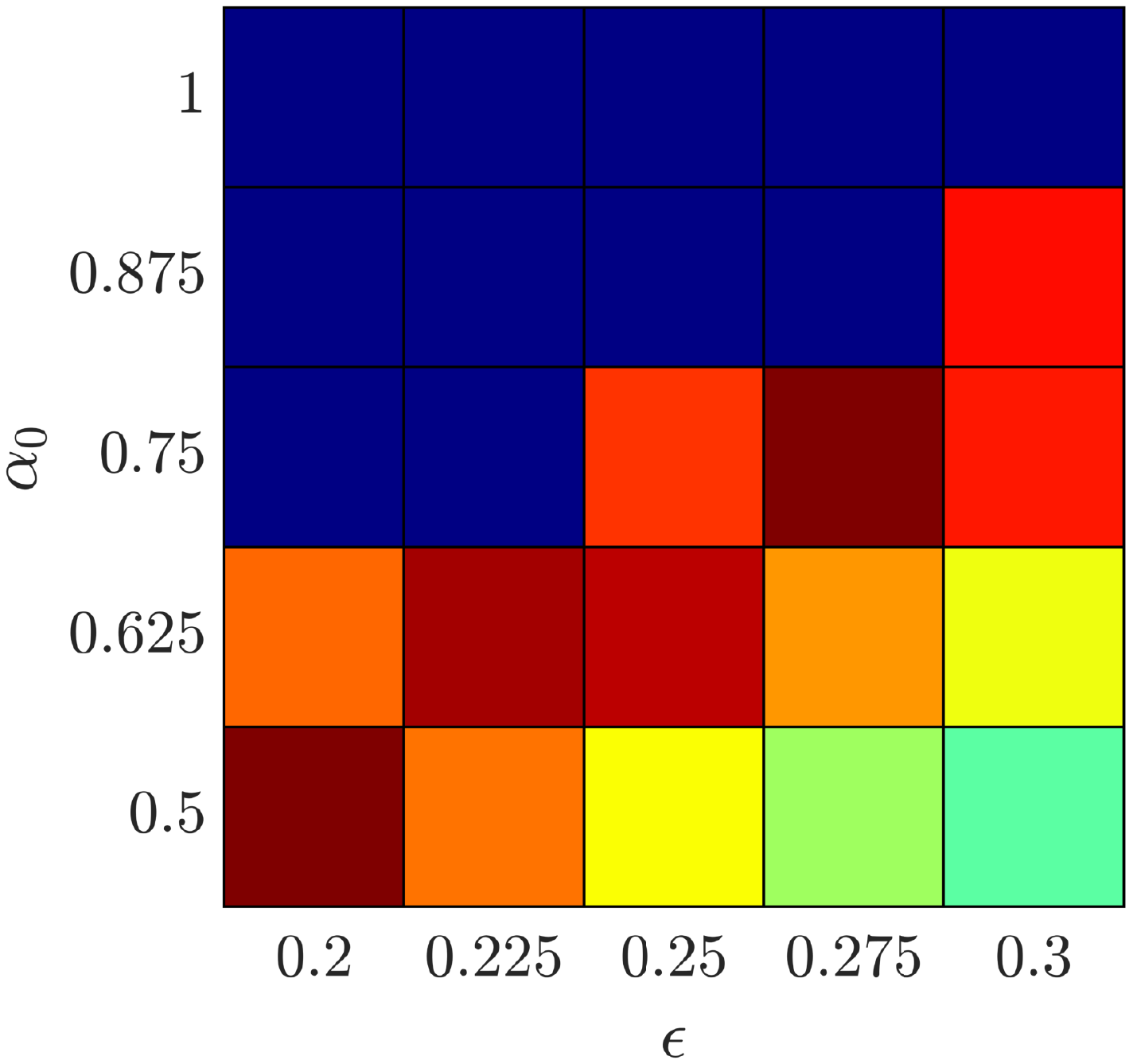}
		\caption{Case 2, Millstone Hill}
		\label{fig:SA_eRho_Millstone_case2}
	\end{subfigure}
	\hfill
	\begin{subfigure}[t]{0.325\textwidth} \centering
		\includegraphics[width=0.9\textwidth]{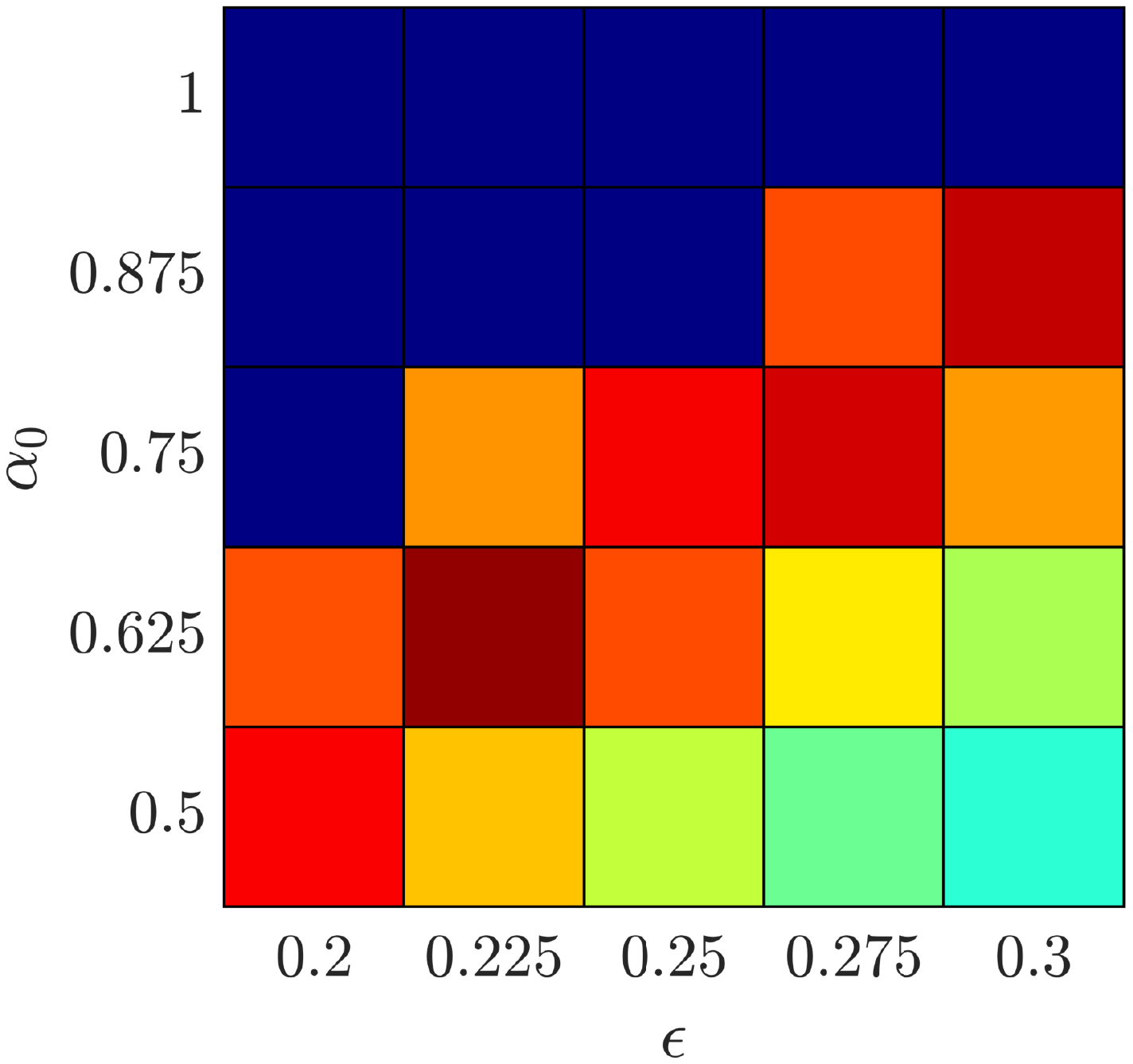}
		\caption{Case 2, Quito}
		\label{fig:SA_eRho_Quito_case2}
	\end{subfigure} 
	\hfill
	\begin{subfigure}[t]{0.325\textwidth} \centering
		\includegraphics[width=0.9\textwidth]{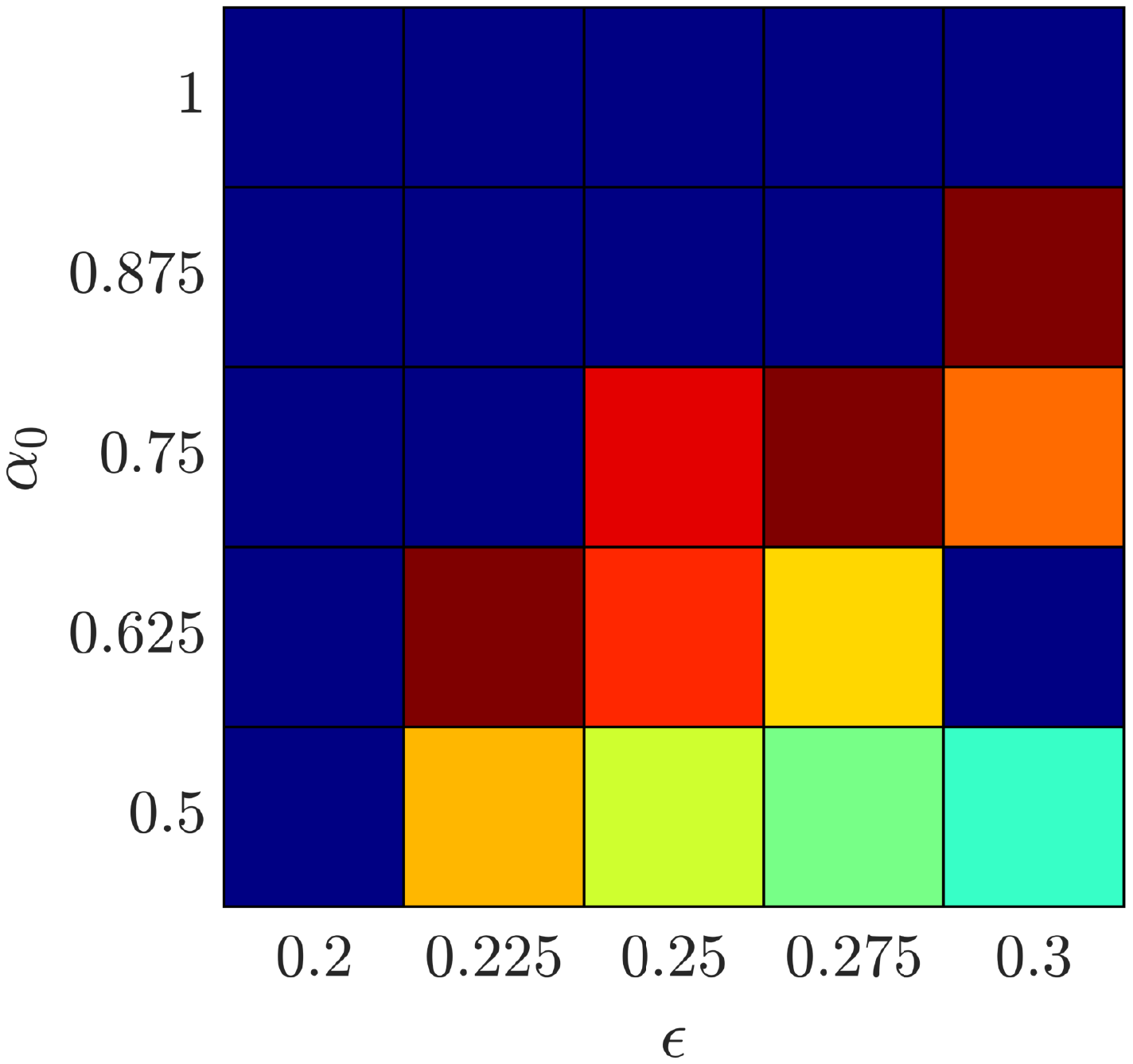}
		\caption{Case 2, Santiago}
		\label{fig:SA_eRho_Santiago_case2}
	\end{subfigure} 
\vspace{3mm}	
	\begin{subfigure}[t]{0.325\textwidth} \centering
		\includegraphics[width=0.9\textwidth]{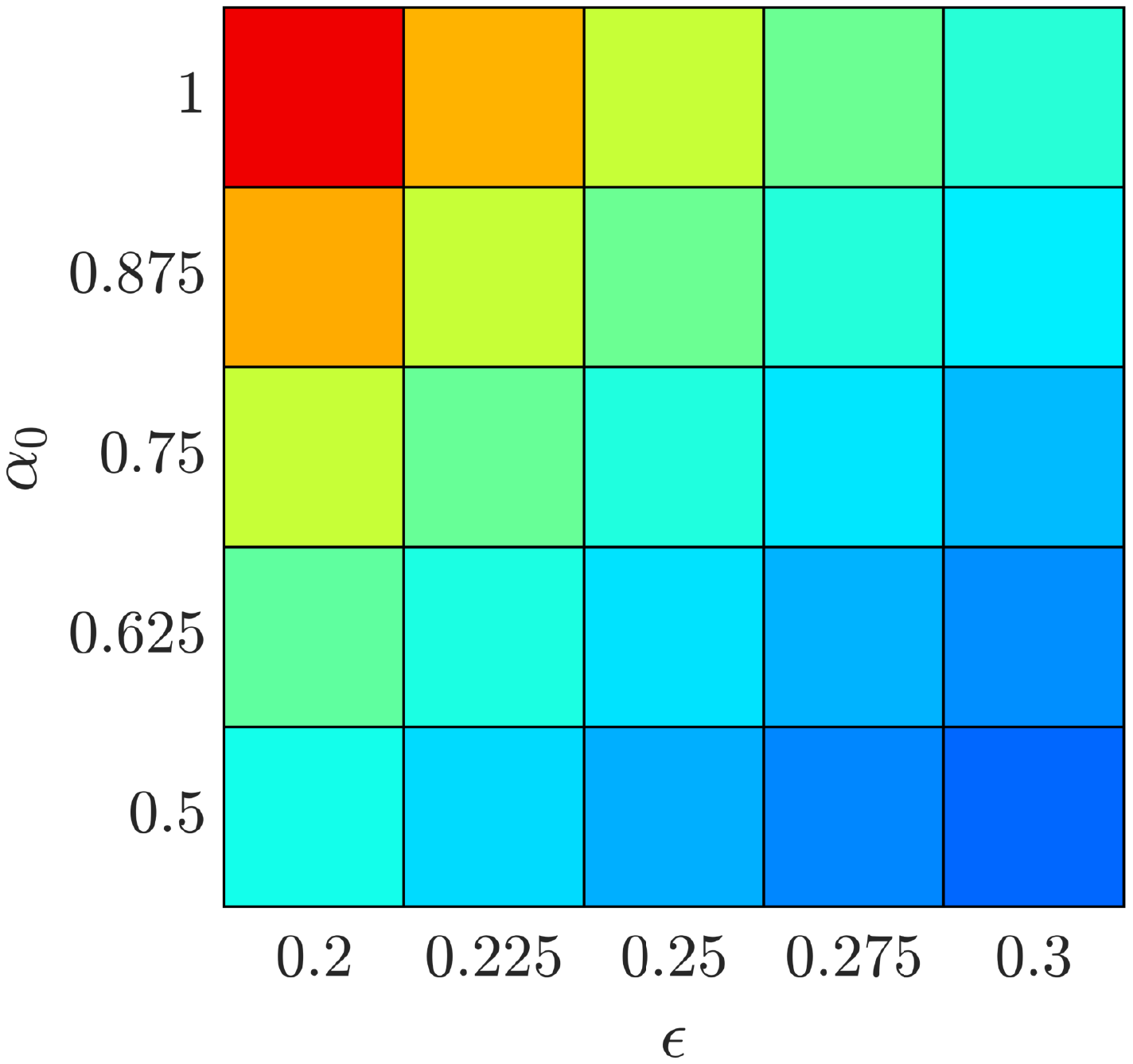}
		\caption{Case 3, Millstone Hill}
		\label{fig:SA_eRho_Millstone_case3}
	\end{subfigure}
	\hfill
	\begin{subfigure}[t]{0.325\textwidth} \centering
		\includegraphics[width=0.9\textwidth]{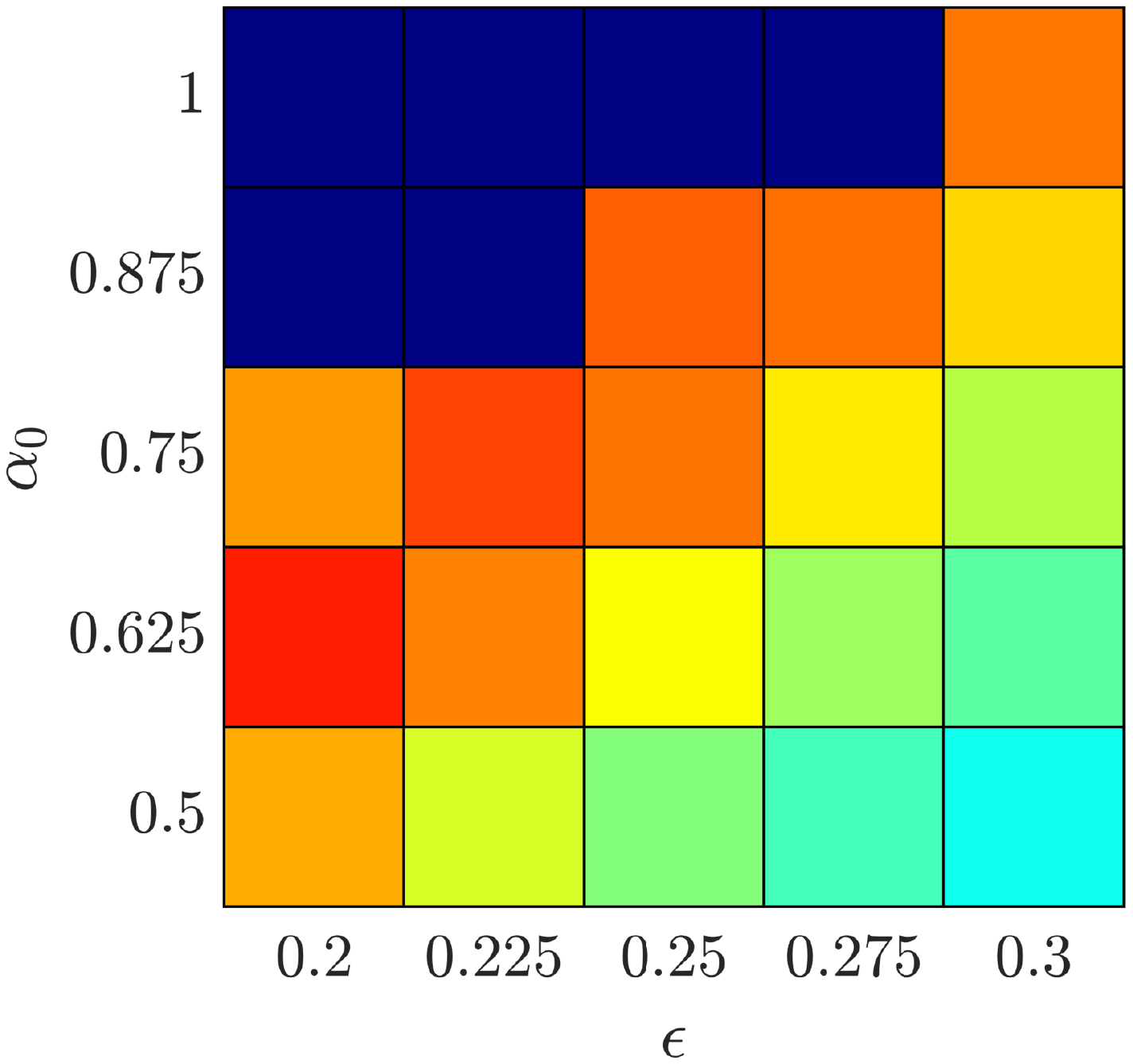}
		\caption{Case 3, Quito}
		\label{fig:SA_eRho_Quito_case3}
	\end{subfigure} 
	\hfill
	\begin{subfigure}[t]{0.325\textwidth} \centering
		\includegraphics[width=0.9\textwidth]{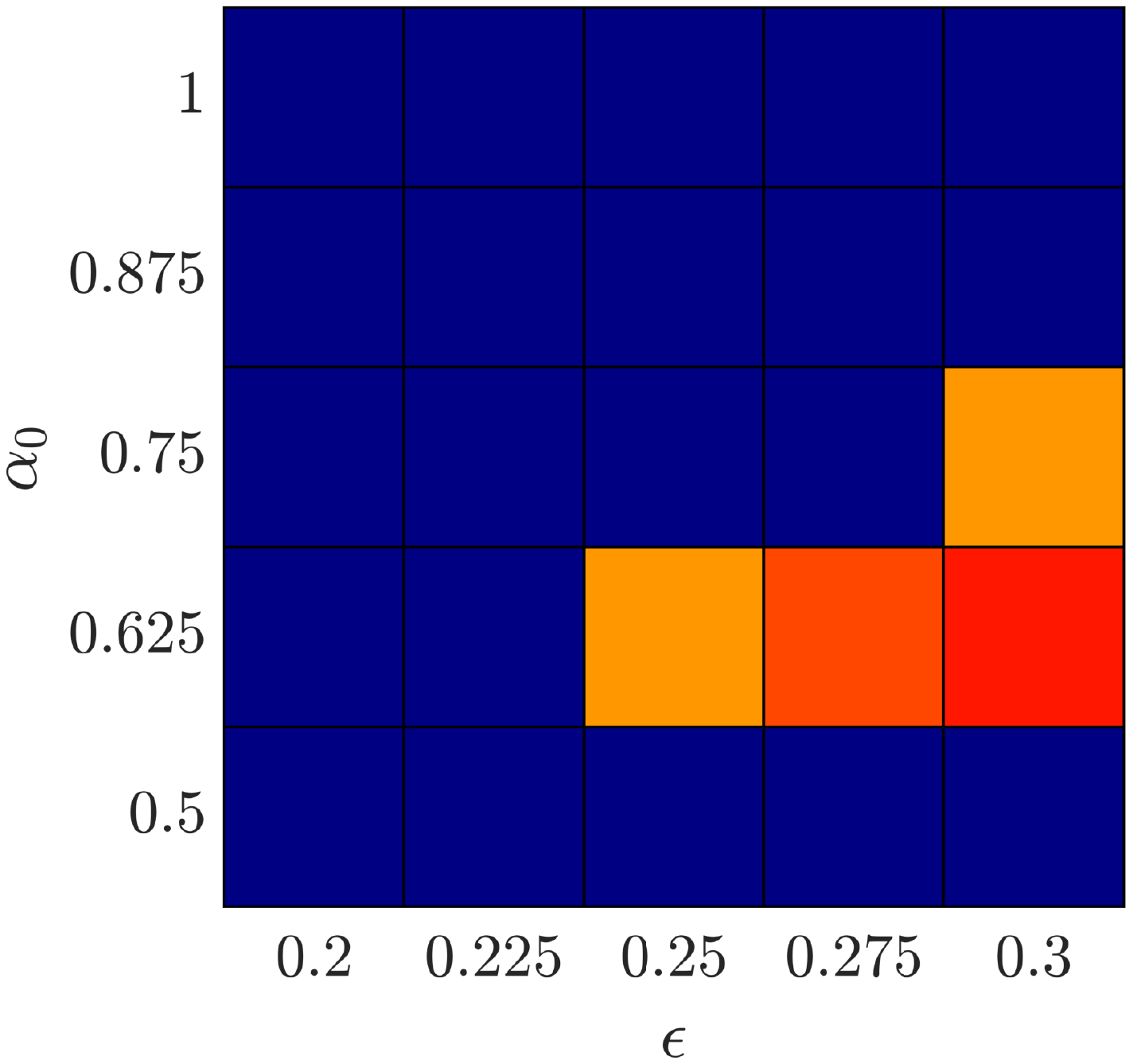}
		\caption{Case 3, Santiago}
		\label{fig:SA_eRho_Santiago_case3}
	\end{subfigure} 
	\caption{Parametric study on the thermal conductivity factor, $\alpha_0$, and EUV efficiency, $\epsilon$, for different cases of study and locations: $\ell^2$ error of the density prediction with respect to the MSIS estimates.}
	\label{fig:SA_eRho}
\end{figure}

Figure~\ref{fig:SA_eRho} displays the error metric for each case and location of interest for all possible combinations of EUV efficiencies and thermal conductivities considered in this study.
An error metric approaching 0 (in maroon) defines a good candidate on the parametric space, whereas inaccurate solutions have a large value associated to them (navy blue).

The results confirm the tendencies observed in Figure~\ref{fig:case2_density1D} and quantify the effect of the choice of parameters for each case and location.
The plots display a very similar performance between the different locations in cases 1 and 2, corresponding to solar equinox in moderate-to-high and low solar activity, owing to a symmetric heating distribution with respect to the Equator line, whereas a very differentiated behavior can be observed under solar solstice conditions.

Case 1 displays the best results for choices of parameters leading to higher dissipation and smaller input of energy, given the higher solar activity associated to this case.
On the other hand, case 2 shows its best performance for combinations of parameters with a compromise between thermal dissipation and energy input: that is, low levels of thermal conductivity and low EUV efficiency, or higher conductivity and higher EUV efficiency. Similarly, this case shows how a combination of parameters with an associated higher dissipation of energy or smaller photoionization leads to poor or non-physical results.
Finally, case 3 shows a very distinct behavior for the different locations. Whereas, in the North hemisphere, the best performance is achieved for $\epsilon = 0.2$ and $\alpha_0=1$, that is, the least energetic and most dissipative configuration. This choice of parameters leads to unstable results in the South hemisphere.

\subsubsection{Model performance under variable solar activity} \label{sssc:3Dcases}
Finally, the three cases of interest, detailed in Table~\ref{tb:casesSAdescription}, are reproduced using the 3D model, employing the set of parameters that offer the best combined results for all cases and locations, which corresponds to an EUV efficiency of $\epsilon = 0.25$ and a thermal conductivity factor of $\alpha_0 = 0.625$.

\begin{figure}[htbp]
	\centering
	\begin{subfigure}[t]{0.286\textwidth} \centering
		\includegraphics[width=0.9\textwidth]{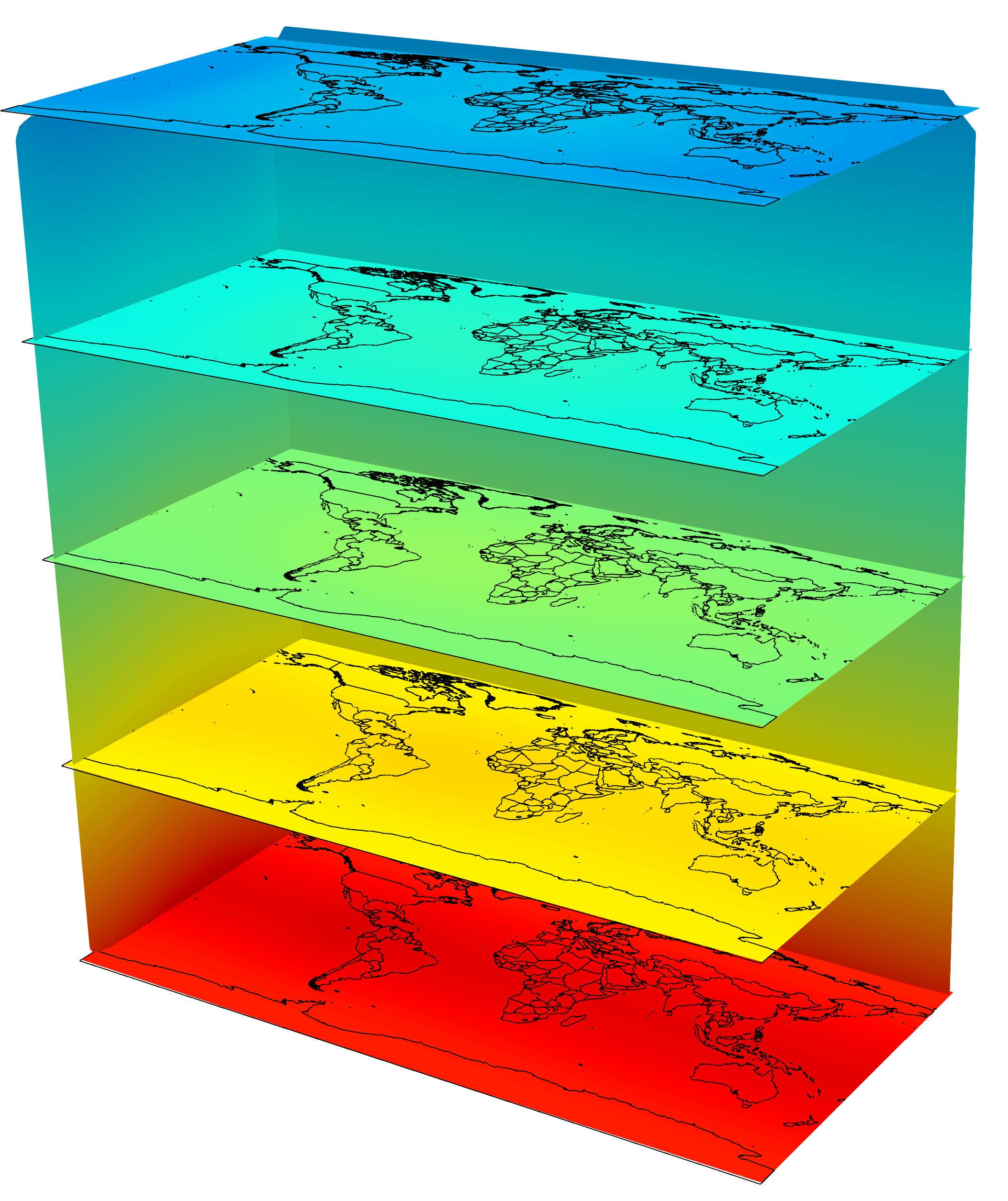}
		\caption{Case 1, 22/03/2014}
		\label{fig:ex1_rho_60}
	\end{subfigure}
	\hfill
	\begin{subfigure}[t]{0.276\textwidth} \centering
		\includegraphics[width=0.9\textwidth]{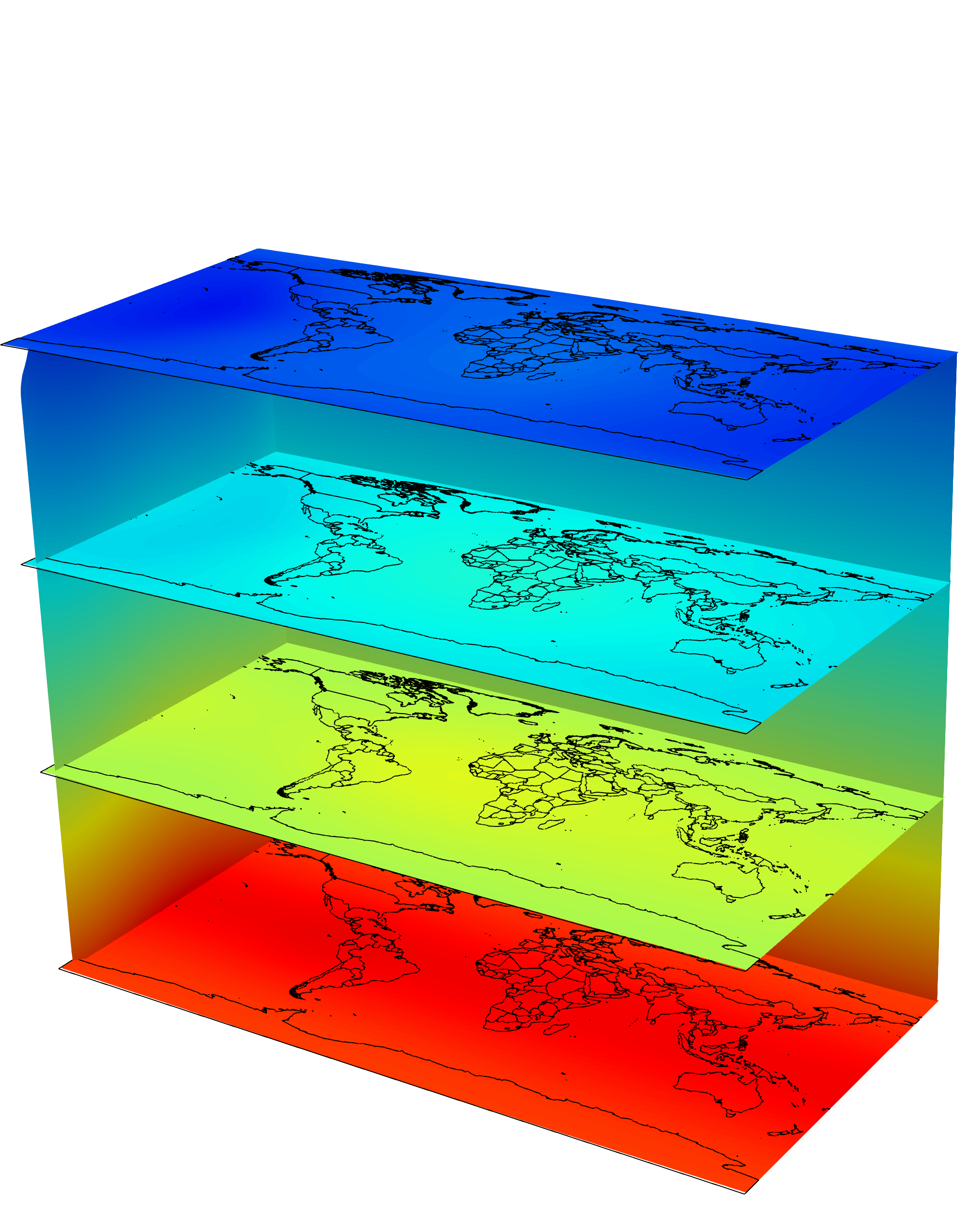}
		\caption{Case 2, 22/03/2020}
		\label{fig:ex2_rho_60}
	\end{subfigure} 
	\hfill
	\begin{subfigure}[t]{0.418\textwidth} \centering
		\includegraphics[width=0.9\textwidth]{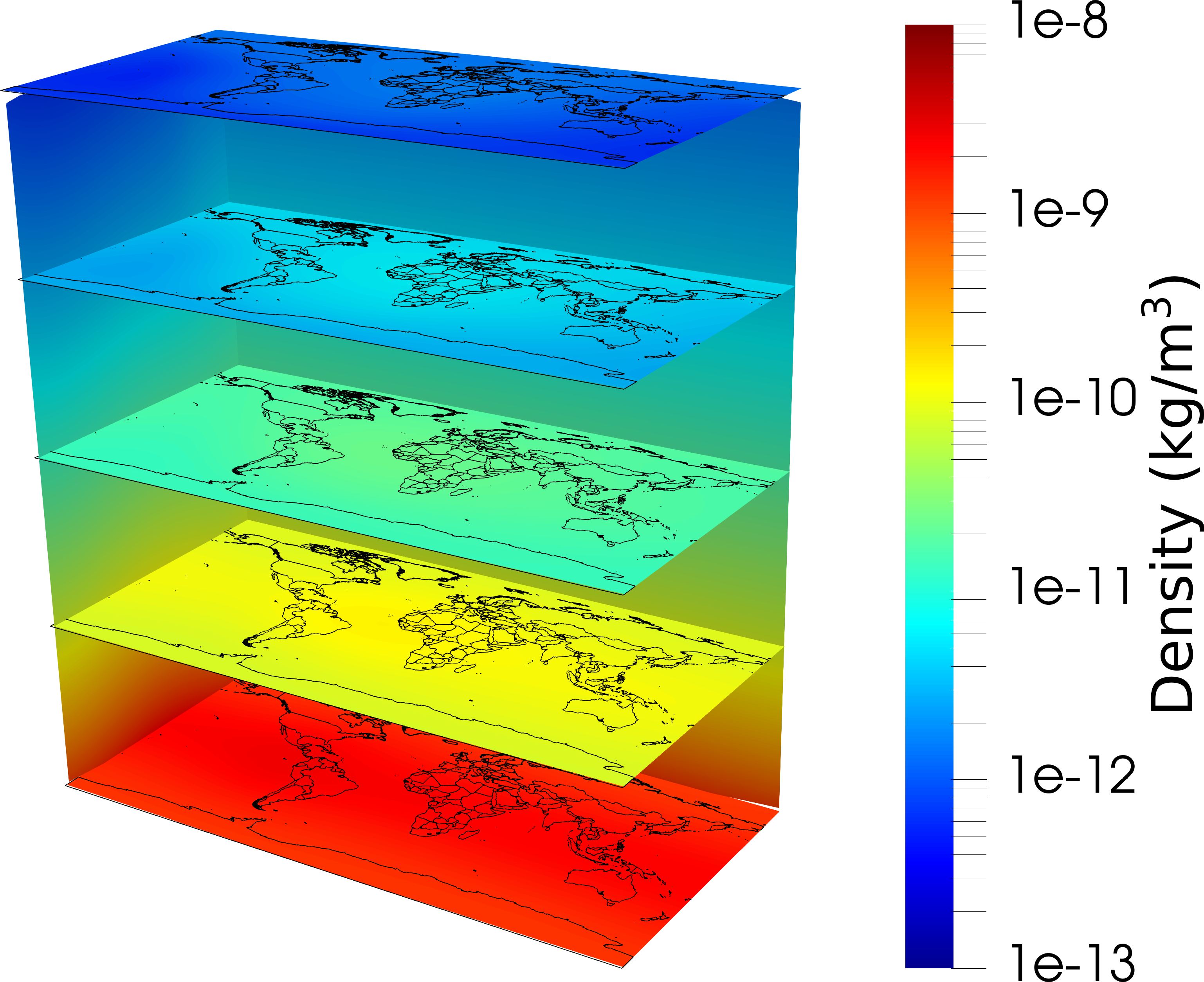}
		\caption{Case 3, 17/06/2022}
		\label{fig:ex3_rho_60}
	\end{subfigure} 
	
	\begin{subfigure}[t]{0.289\textwidth} \centering
		\includegraphics[width=0.9\textwidth]{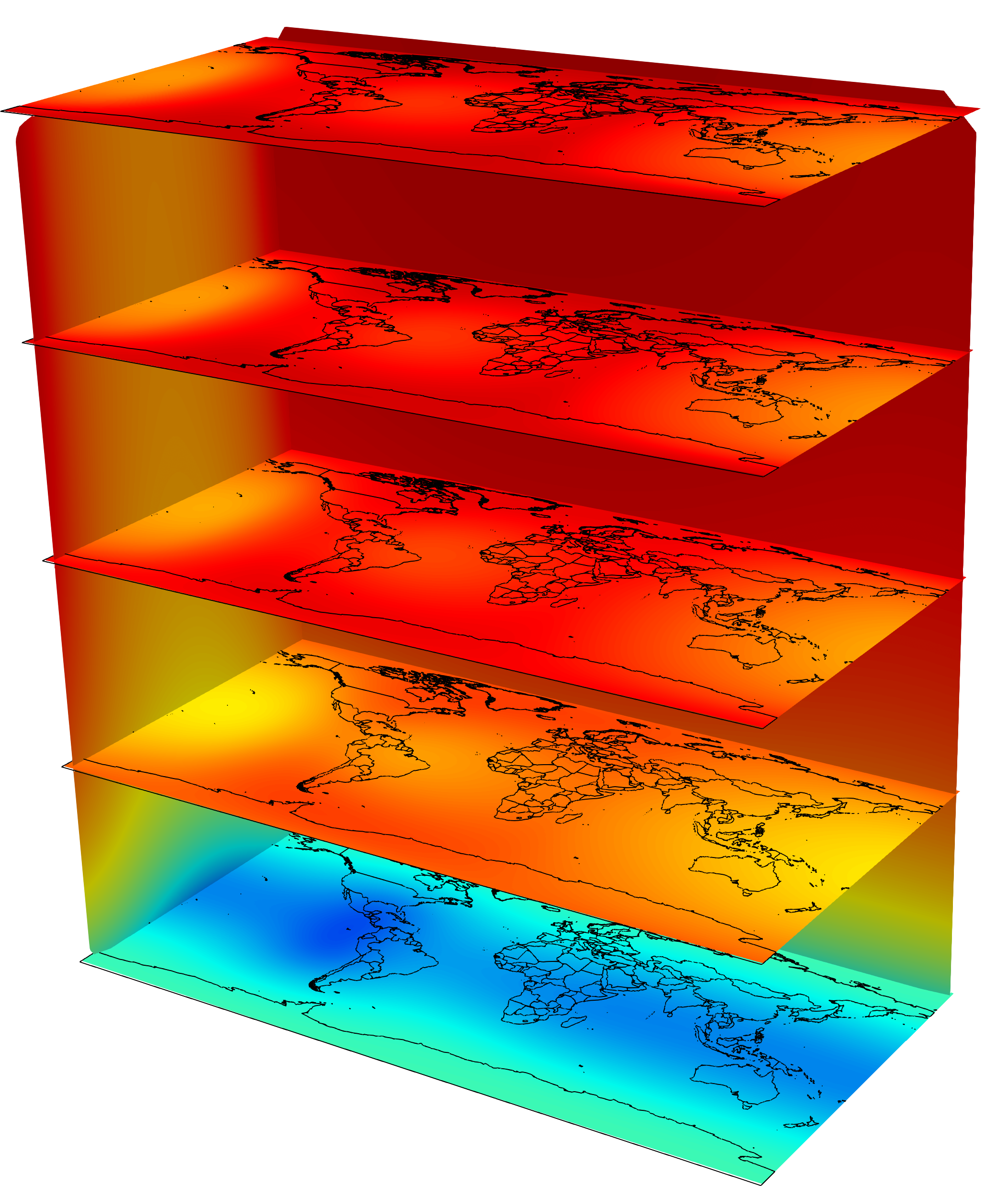}
		\caption{Case 1, 22/03/2014}
		\label{fig:ex1_T_60}
	\end{subfigure}
	\hfill
	\begin{subfigure}[t]{0.28\textwidth} \centering
		\includegraphics[width=0.9\textwidth]{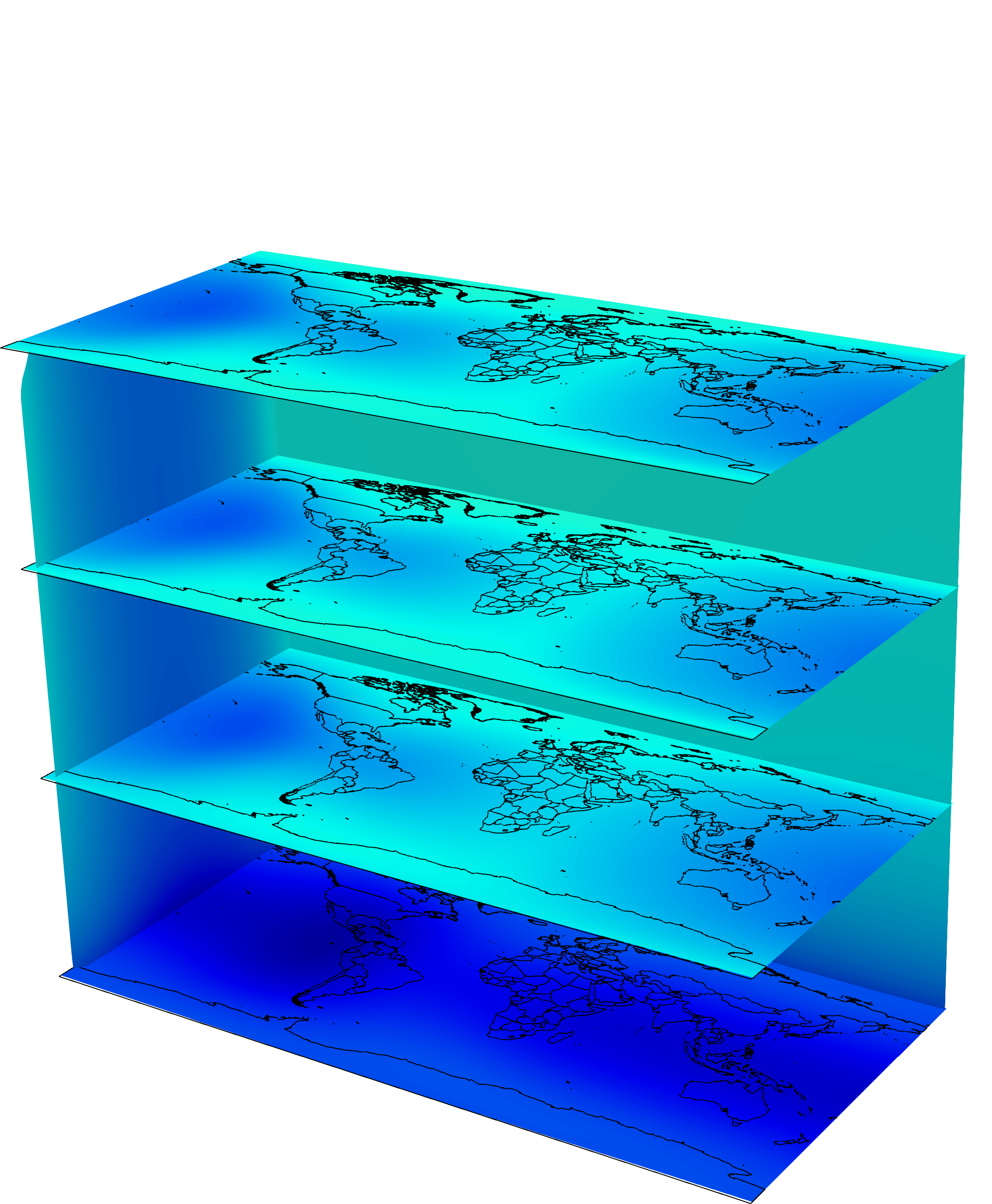}
		\caption{Case 2, 22/03/2020}
		\label{fig:ex2_T_60}
	\end{subfigure} 
	\hfill
	\begin{subfigure}[t]{0.411\textwidth} \centering
		\includegraphics[width=0.9\textwidth]{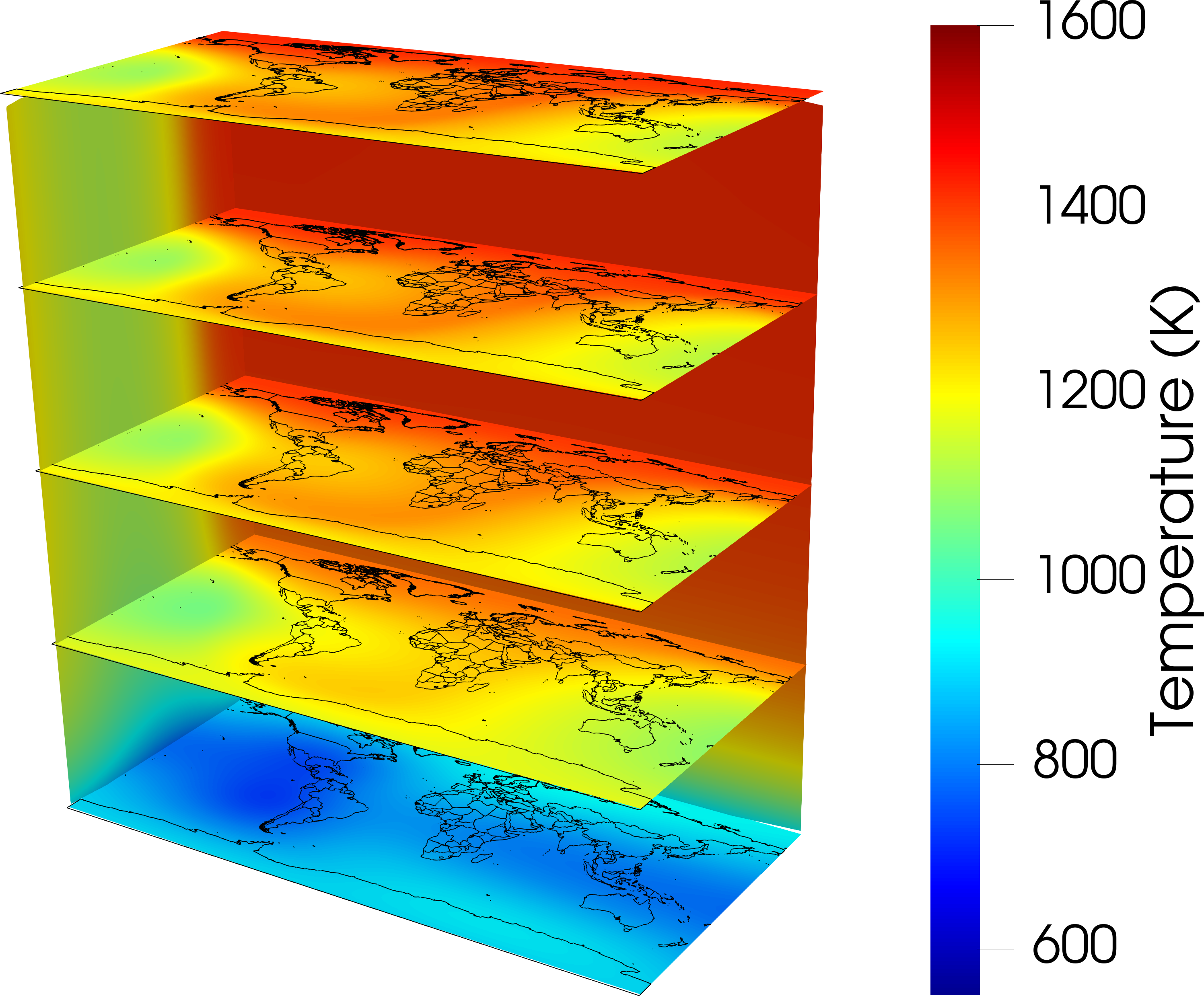}
		\caption{Case 3, 17/06/2022}
		\label{fig:ex3_T_60}
	\end{subfigure} 
	\caption{Numerical approximations of the density (top) and temperature (bottom) fields at different altitudes (150, 250, 350, 450 and 550km --if applicable--), computed with $k=2$ polynomials for the different cases of study at 12:00 UTC on the specified days.}
	\label{fig:sol_60}
\end{figure}

The problems are solved using a quadratic approximation on a cube-sphere mesh with $2.5^{\circ}$ of angular resolution. The simulations are advanced in time using a DIRK(2,2) scheme with time steps of 5s.
The model predictions of density and temperature for each of these cases are displayed in Figure~\ref{fig:sol_60} at various altitude levels between 150 and 550km after 2.5 days of simulation.

The plots show the effects of the different solar conditions for each of the cases.
In particular, case 1 and 3 show higher temperature and density predictions, due to the higher solar irradiance.
In addition, case 1 displays more uniformly distributed temperatures and density fields because of the symmetric heating distribution under equinox conditions, whereas case 3 presents an asymmetric distribution with higher temperatures in the North hemisphere owing to the Summer solstice conditions.
On the other hand, case 2 displays the effect of low solar conditions, with much lower temperatures and densities, which are reached at an even lower altitude.

\begin{figure}[htbp]
	\centering
	\begin{subfigure}[t]{\textwidth} \centering
		\includegraphics[width=0.485\textwidth]{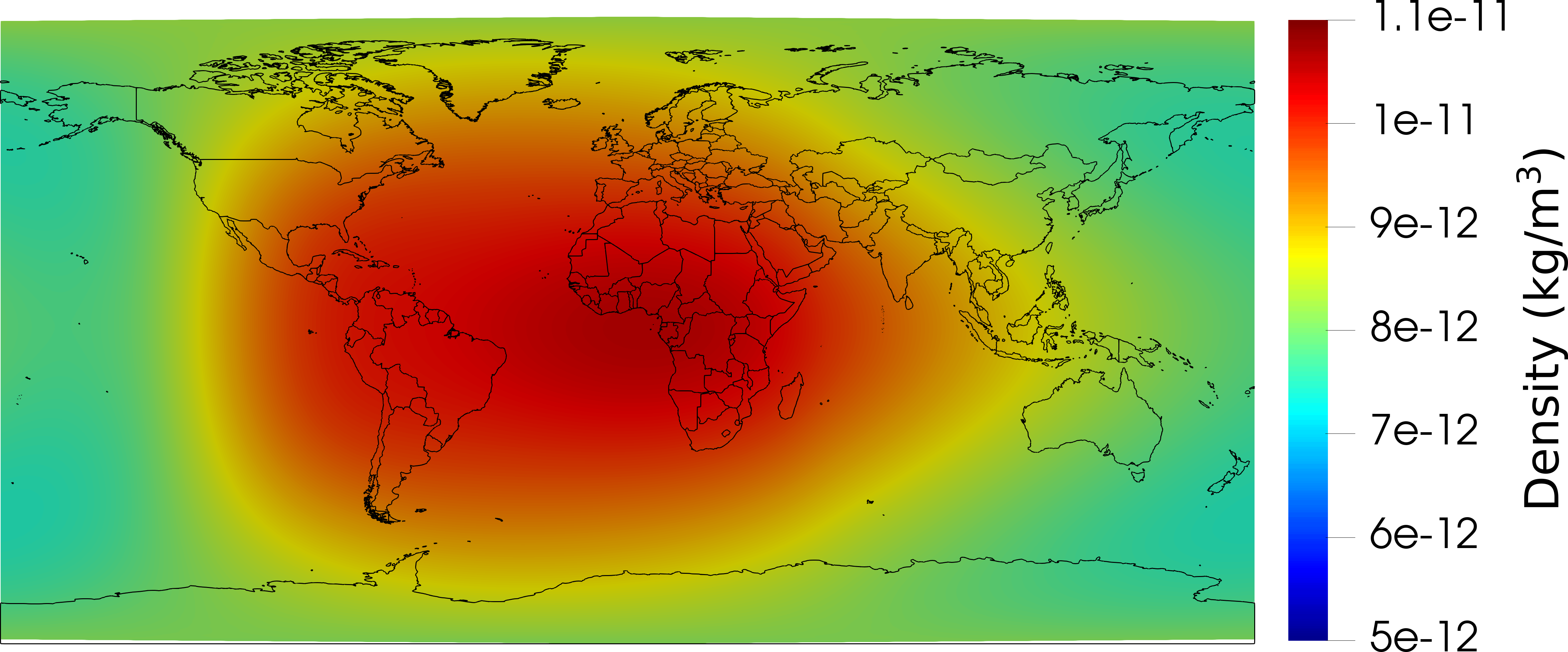} \hfill
		\includegraphics[width=0.485\textwidth]{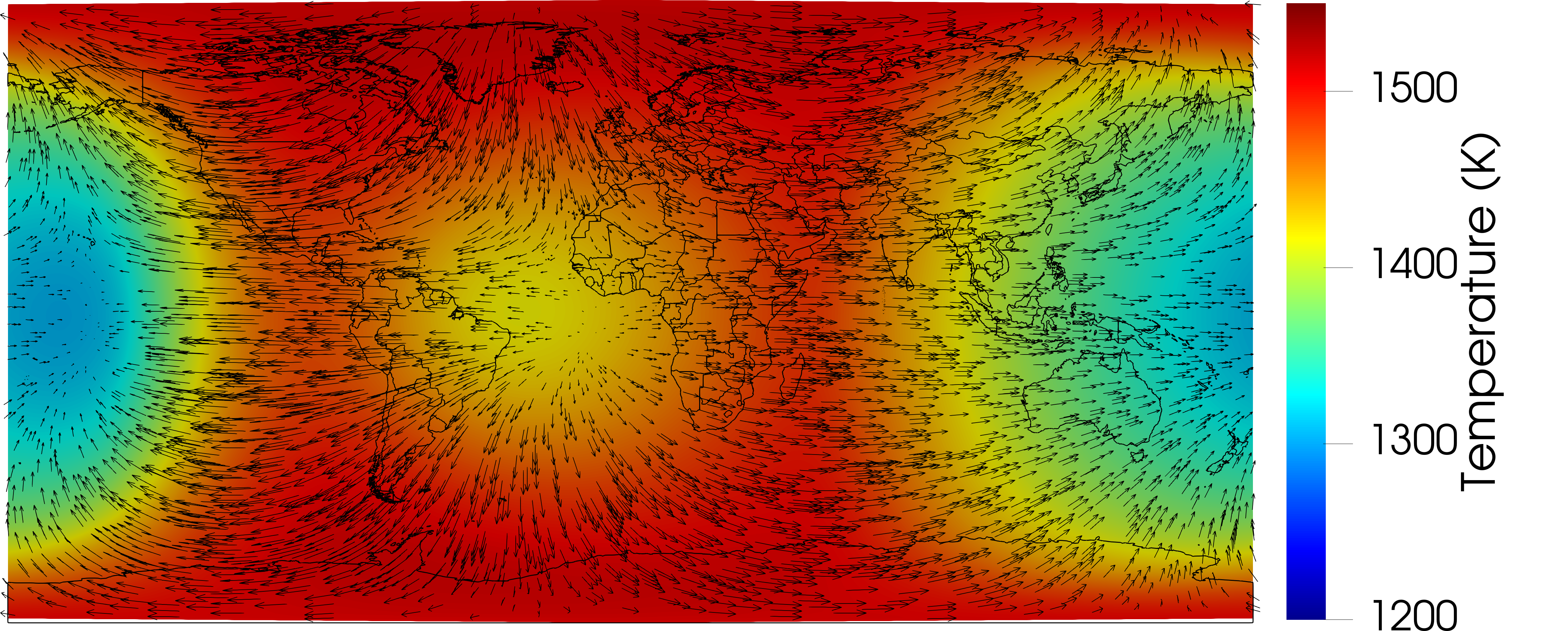}
		\caption{Case 1: 12:00 UTC on 22/03/2014.}
		\label{fig:ex1_60_450}
	\end{subfigure} 
	\begin{subfigure}[t]{\textwidth} \centering
		\includegraphics[width=0.49\textwidth]{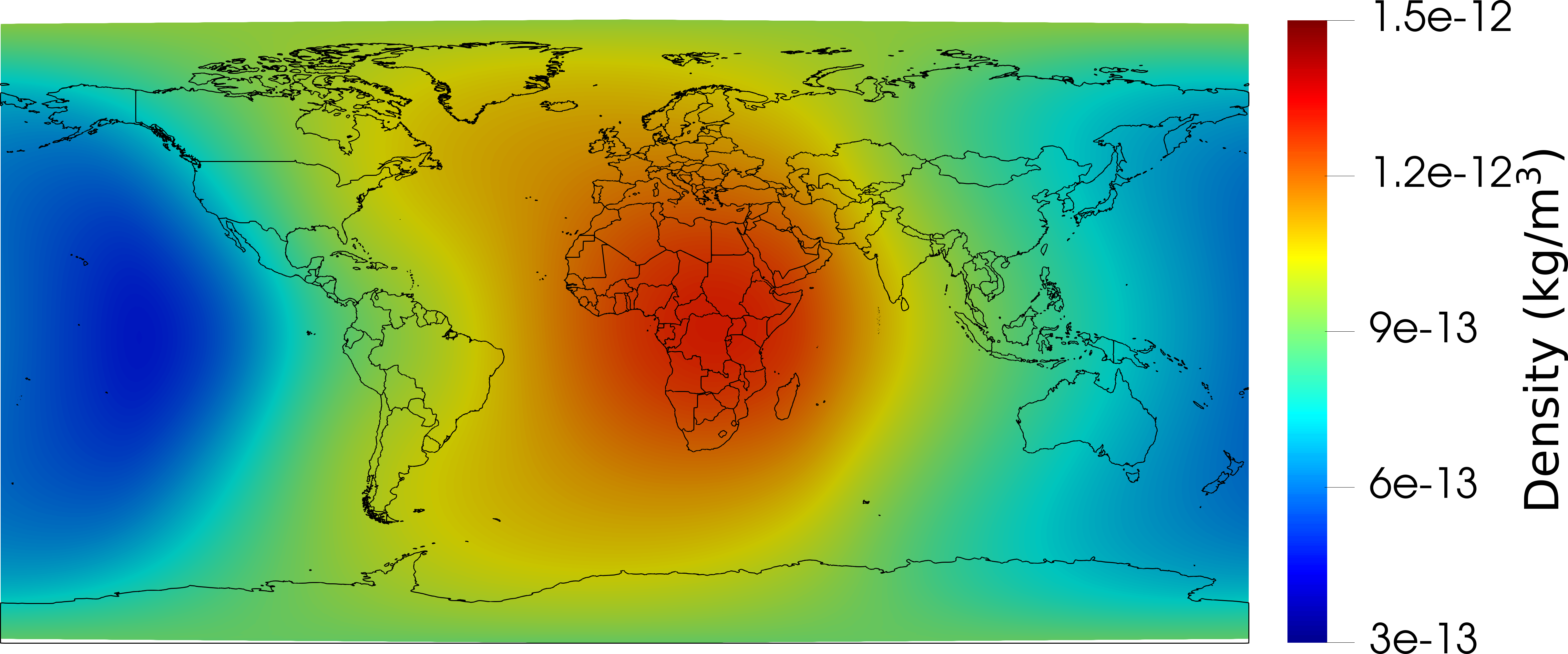} \hfill
		\includegraphics[width=0.48\textwidth]{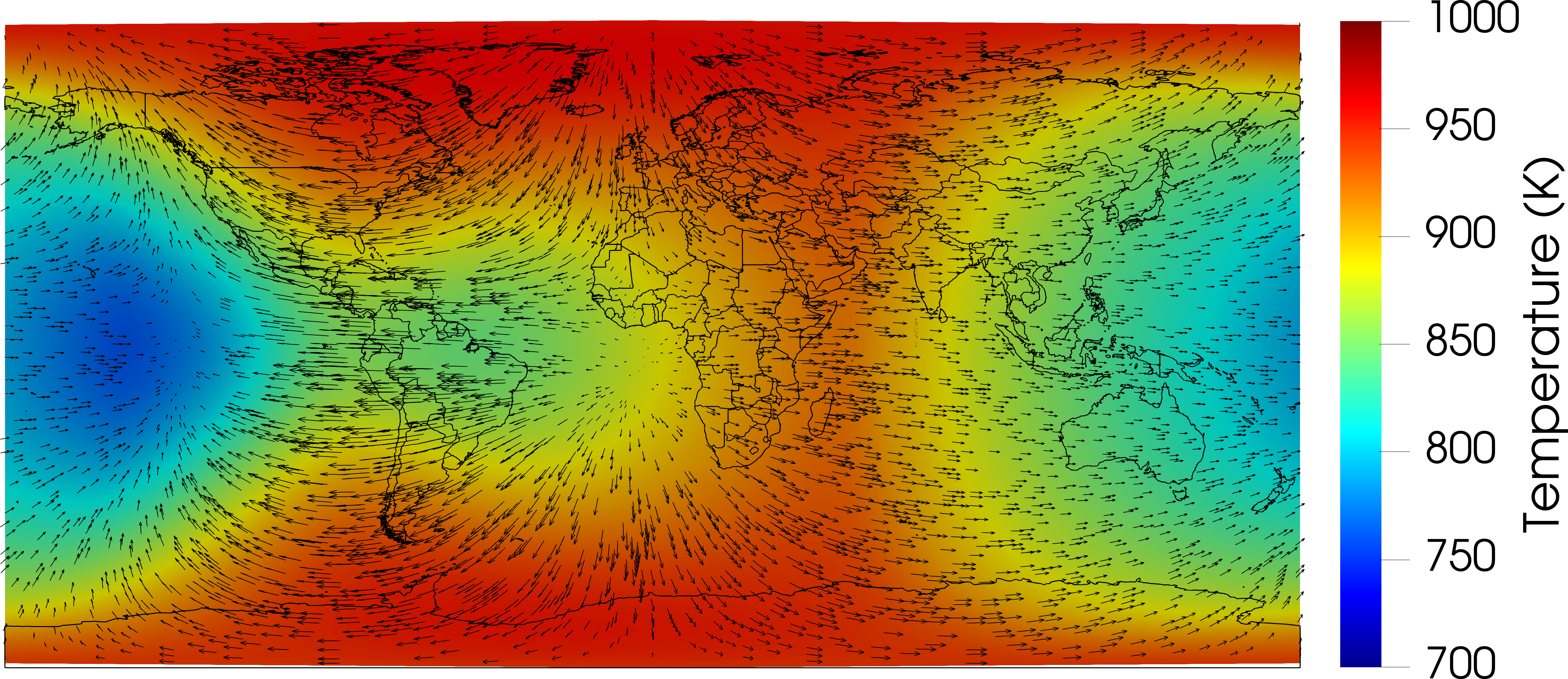}
		\caption{Case 2: 12:00 UTC on 22/03/2020.}
		\label{fig:ex2_60_450}
	\end{subfigure} 
	\begin{subfigure}[t]{\textwidth} \centering
		\includegraphics[width=0.485\textwidth]{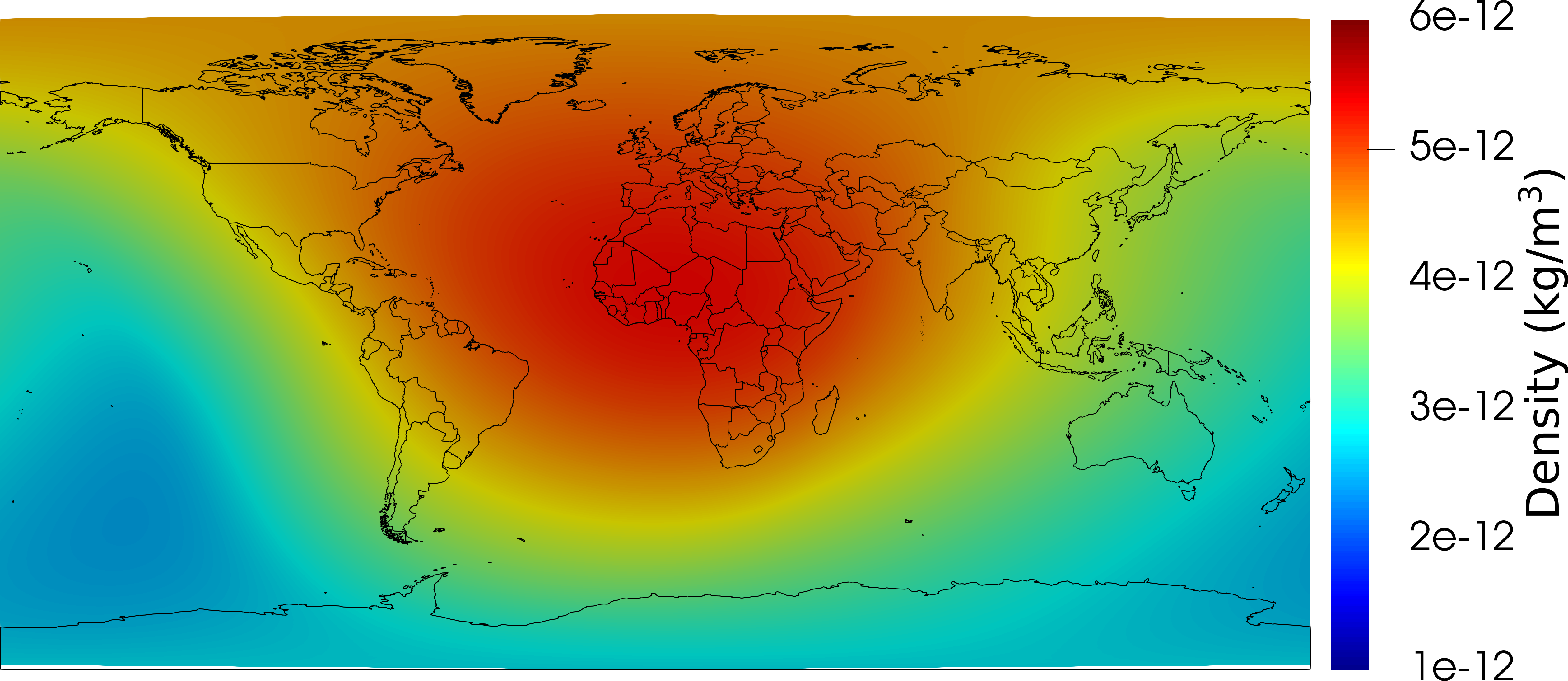} \hfill
		\includegraphics[width=0.485\textwidth]{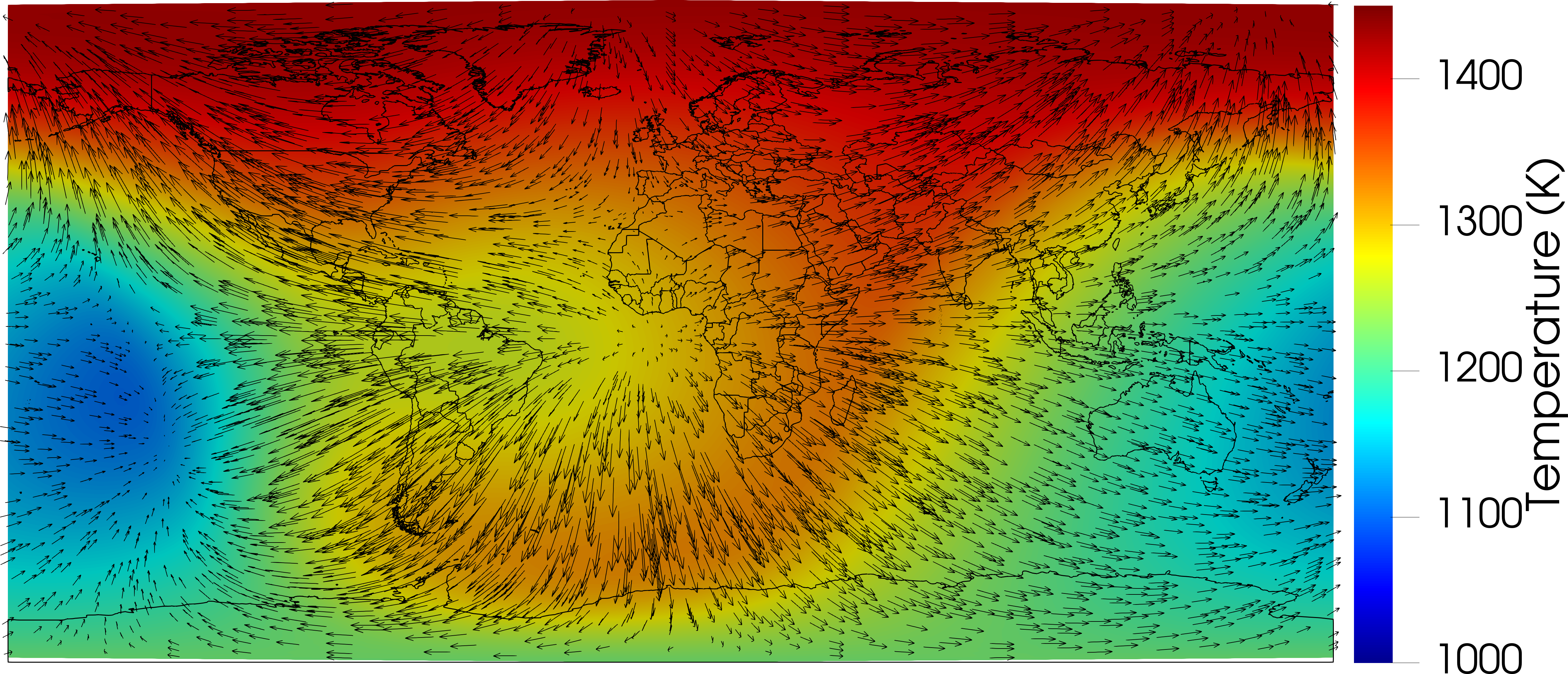}
		\caption{Case 3: 12:00 UTC on 17/06/2022.}
		\label{fig:ex3_60_450}
	\end{subfigure} 
	\caption{Density (left) and temperature (right) estimates at 450km of altitude for the three cases of interest, obtained with a quadratic approximation. Note that the color bars are adapted to each case.}
	\label{fig:case2_60}
\end{figure}

All these effects can be better observed in Figure~\ref{fig:case2_60}, which shows the density and temperature predictions at 450km of altitude after 2.5 days of simulation.

The similar patterns in cases 1 and 2, attributable to the same declination angle between the Sun and the Earth, can be clearly observed, despite of the significant differences in magnitude of both density and temperature fields caused by the intensity of the solar irradiance.
This contrasts with case 3, which shows the effect of the solar rays impinging more obliquely with respect to the Equator, which results in higher temperatures and densities over the North hemisphere. This is consistent with the northern summer solstice conditions.
For all these cases, the velocity field, which is shown superimposed to the temperature plots, displays the logical pattern of velocities moving from hotter and denser areas to colder and less dense regions.

\begin{figure}[htbp]
	\centering
	\begin{subfigure}[t]{0.5\textwidth} \centering
		\includegraphics[width=0.8\textwidth]{SA_legend}
	\end{subfigure} \\
	\begin{subfigure}[t]{\textwidth} \centering
		\includegraphics[width=0.45\textwidth]{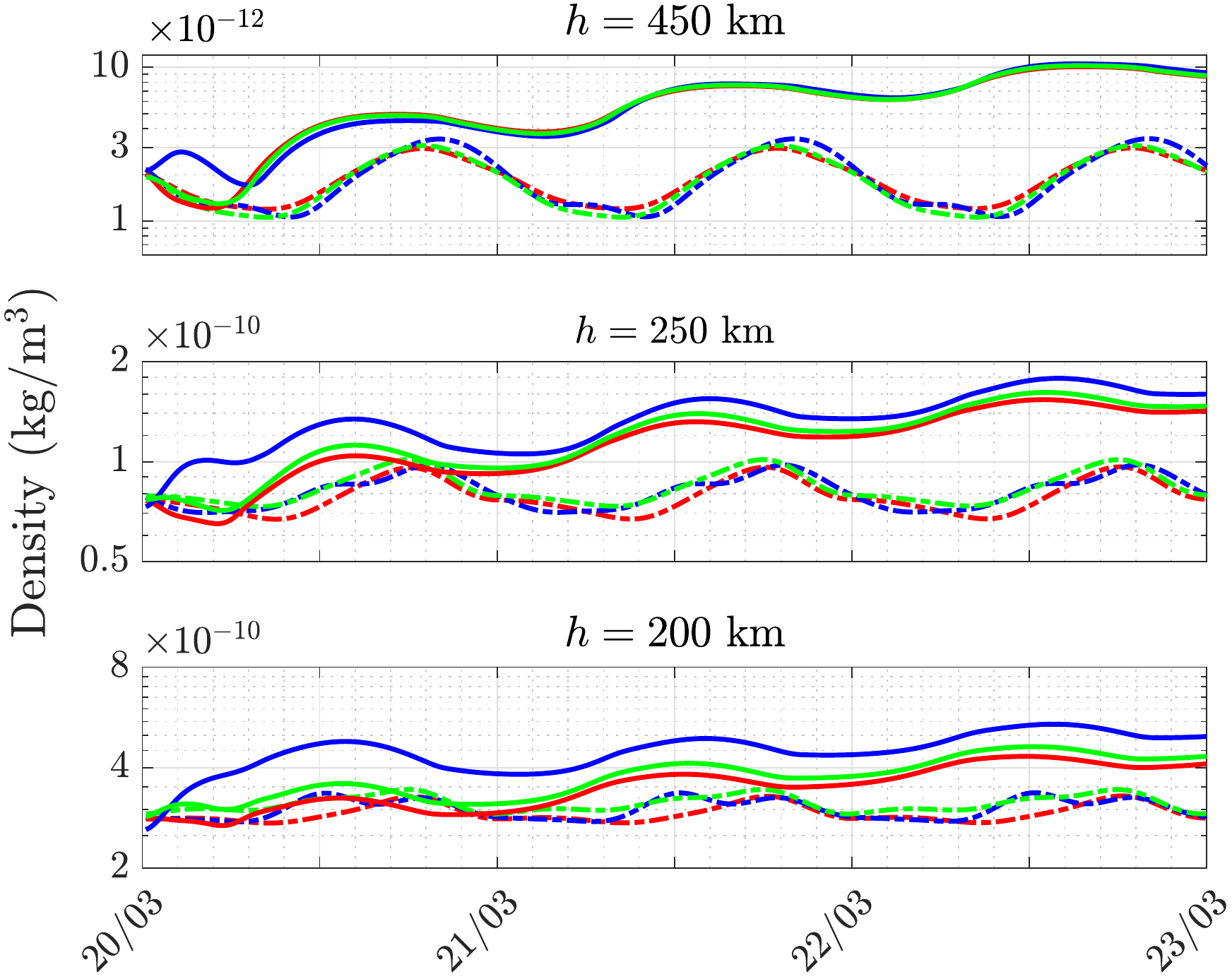}
		\qquad
		\includegraphics[width=0.45\textwidth]{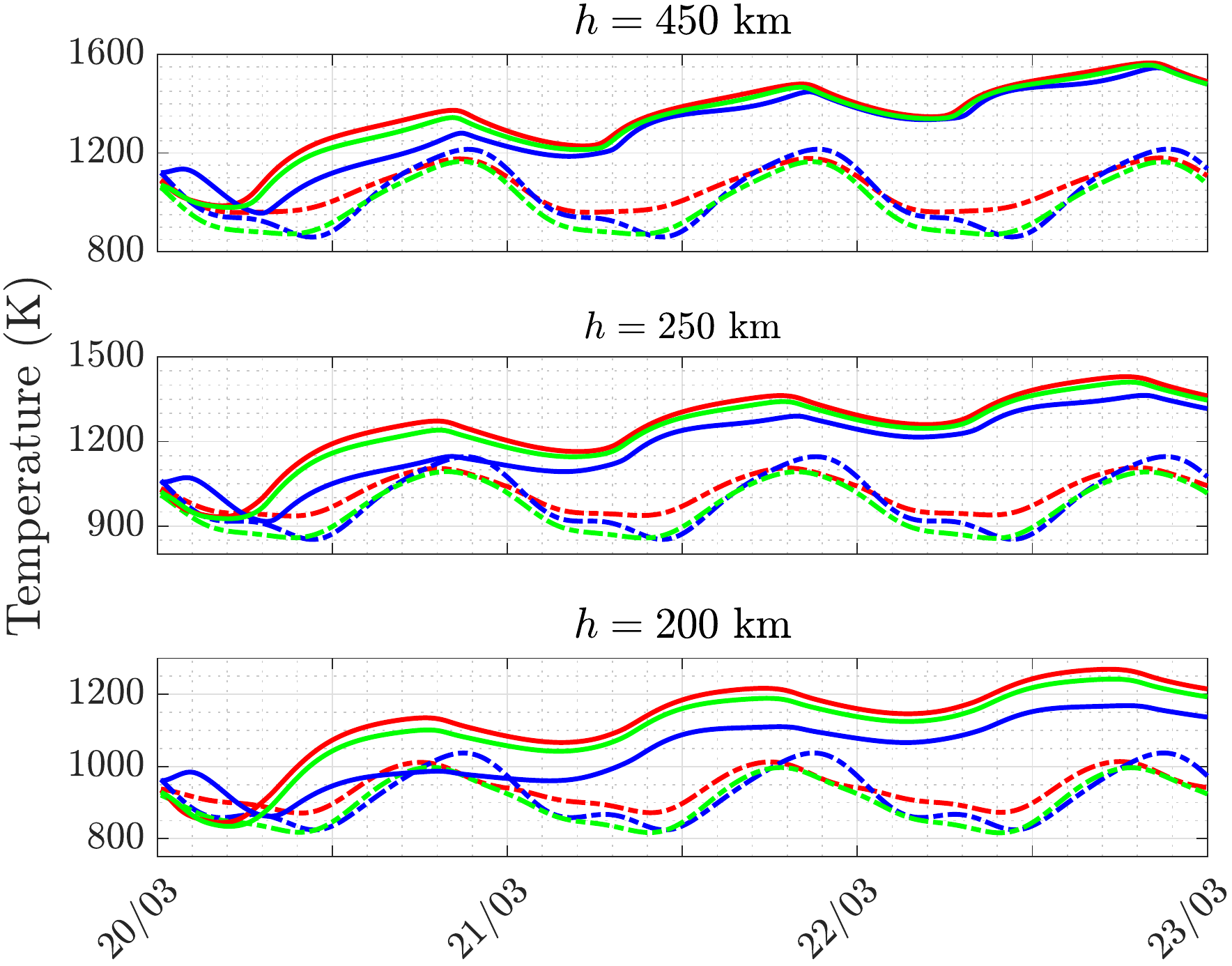}
		\caption{Case 1: 2014 spring equinox, 20-23 March}
		\label{fig:ex1}
	\end{subfigure} 
	\begin{subfigure}[t]{\textwidth} \centering
		\includegraphics[width=0.45\textwidth]{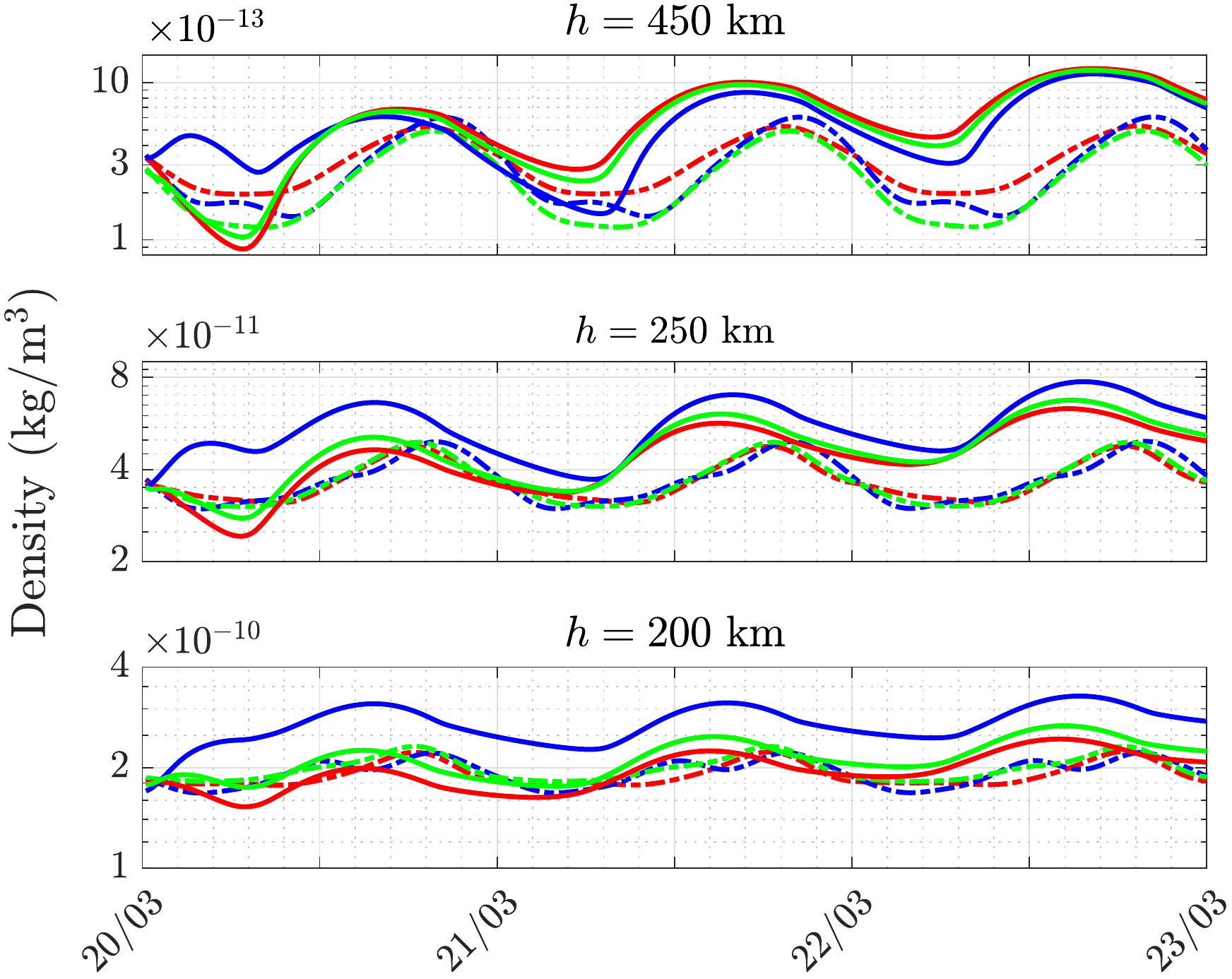}
		\qquad
		\includegraphics[width=0.45\textwidth]{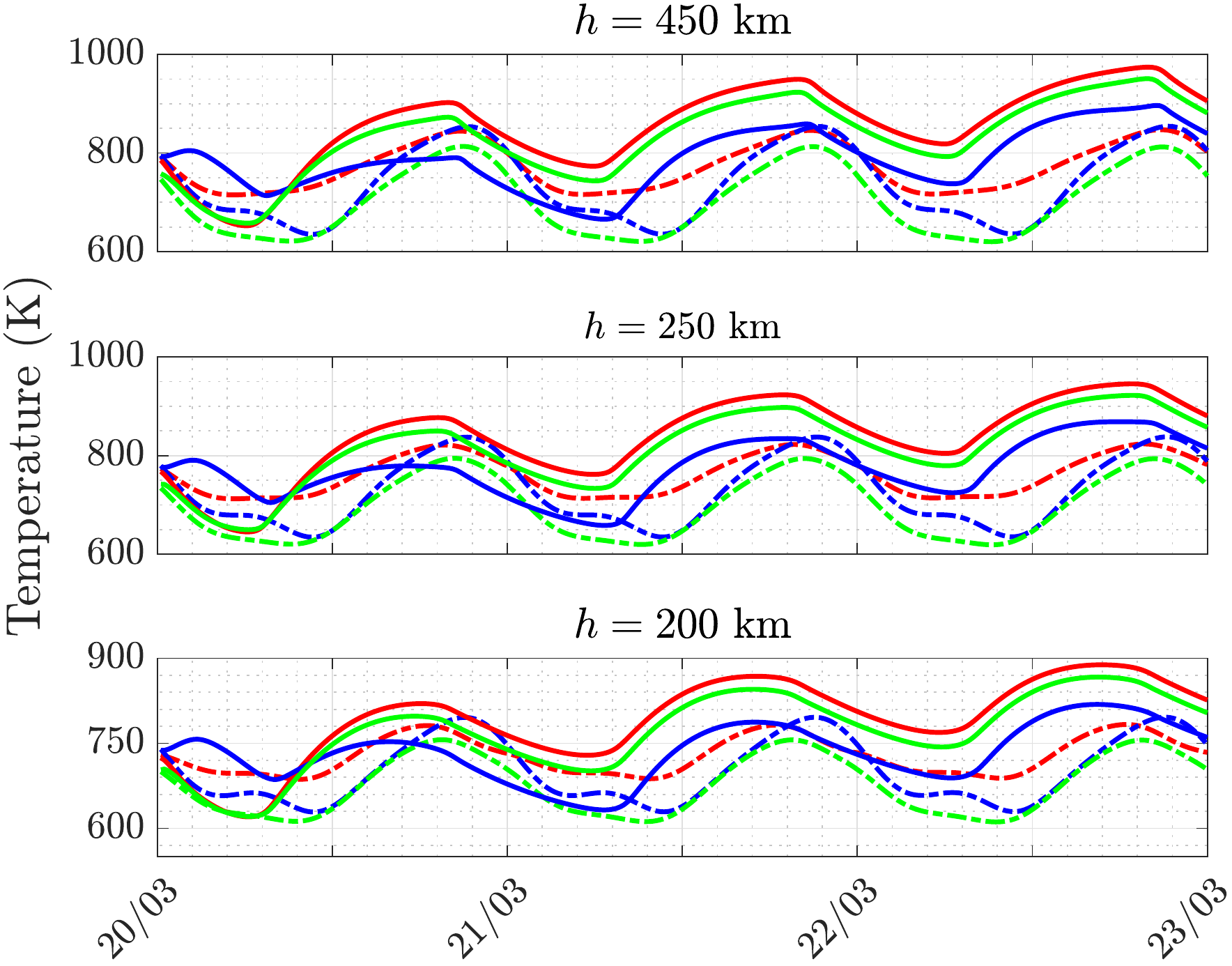}
		\caption{Case 2: 2020 spring equinox, 20-23 March}
		\label{fig:ex2}
	\end{subfigure} 
	\begin{subfigure}[t]{\textwidth} \centering
		\includegraphics[width=0.45\textwidth]{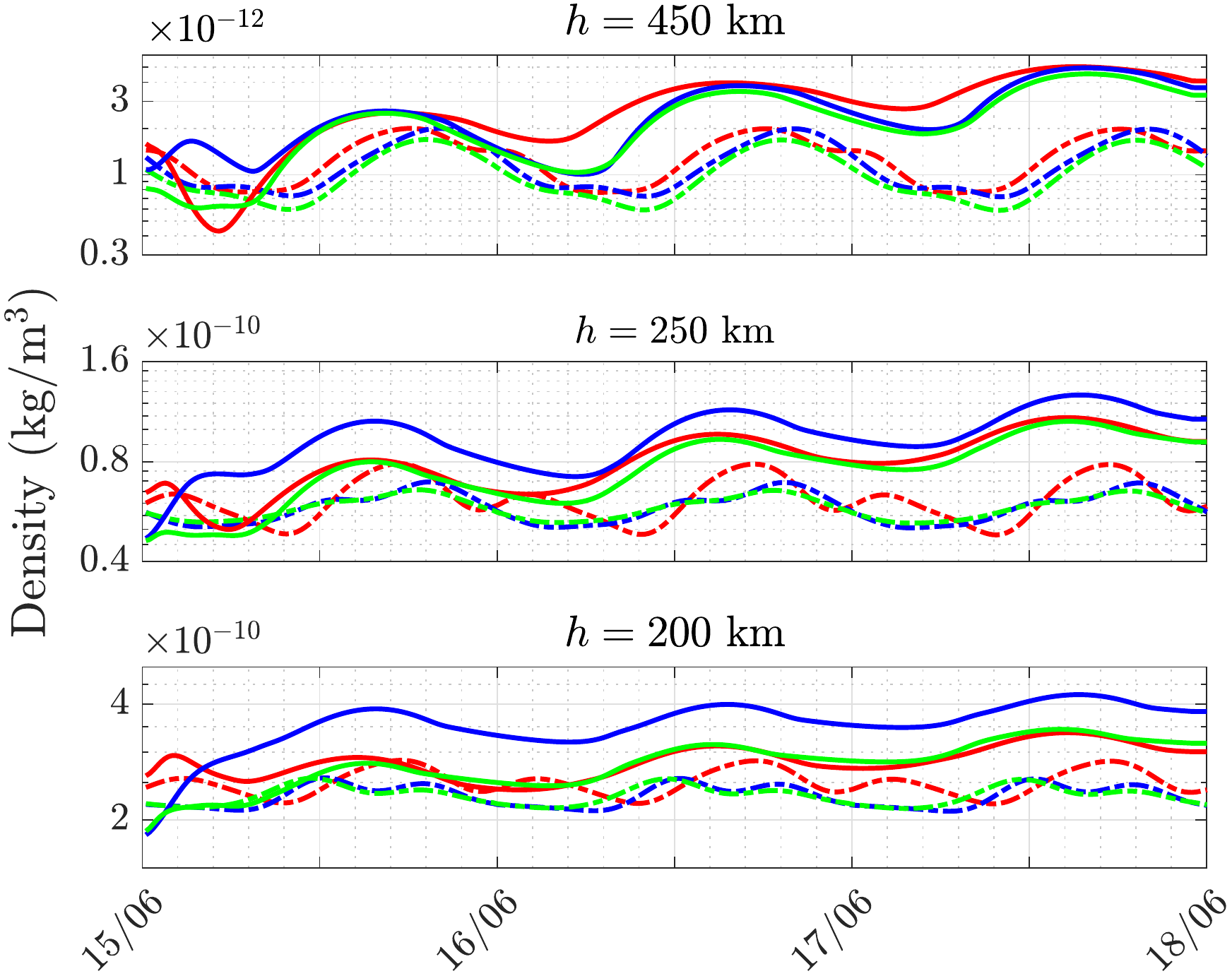}
		\qquad
		\includegraphics[width=0.45\textwidth]{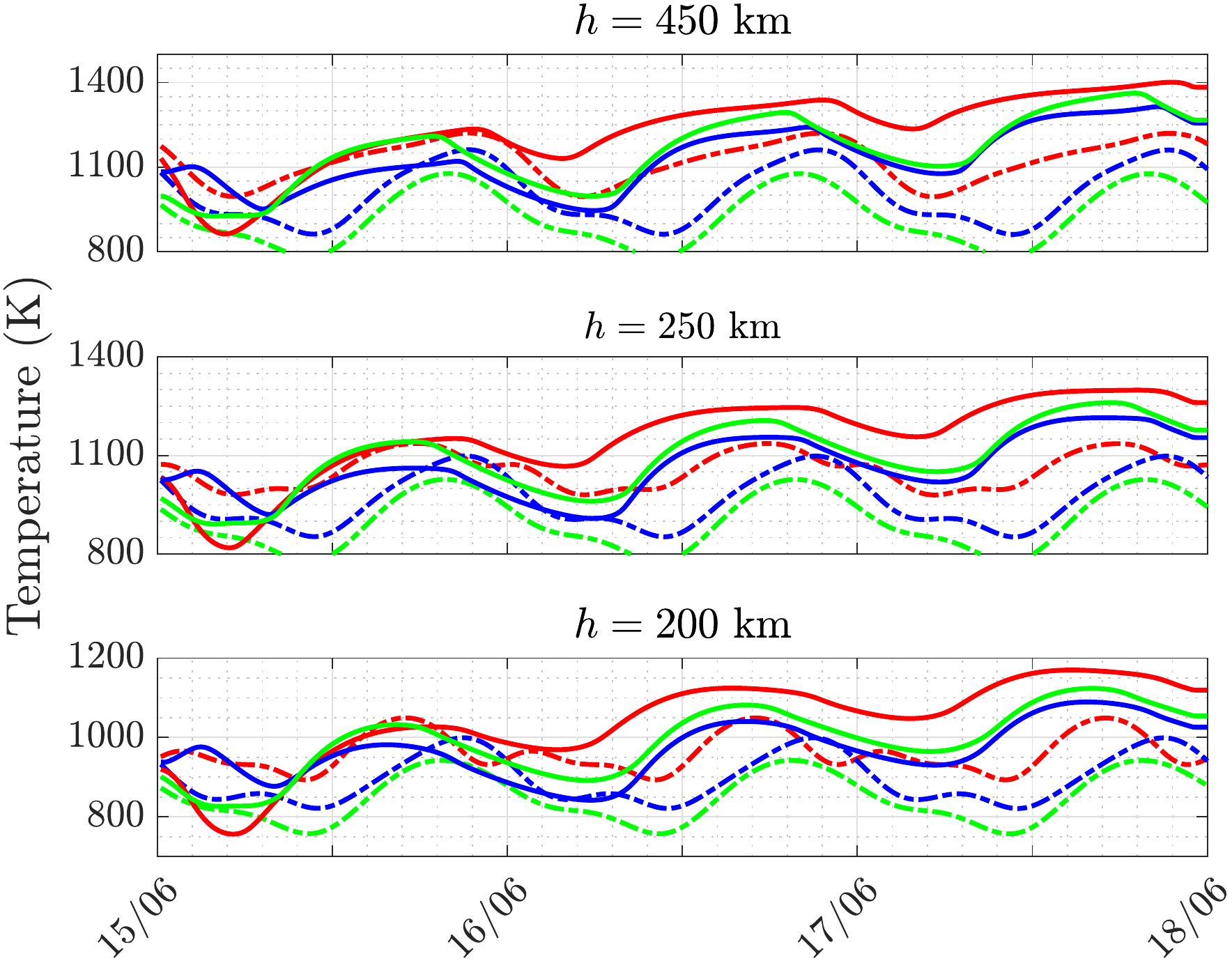}
		\caption{Case 3: 2022 summer solstice, 15-18 June}
		\label{fig:ex3}
	\end{subfigure} 
	\caption{Temporal history of the density (left) and temperature (right) at different altitudes above Millstone Hill, Quito and Santiago for the three cases of interest. The results correspond to the 3D model using $k=2$ polynomials. MSIS estimates are included for reference.}
	\label{fig:case2_3D}
\end{figure}

Finally, in order to evaluate the effect of the single choice of parameters on the different cases, the evolution of the density and temperature fields at different altitudes above Millstone Hill, Quito and Santiago is presented for the three cases of interest in Figure~\ref{fig:case2_3D}.

The results show an underdissipative behavior in case 1 motivated by an insufficient amount of thermal conductivity, which gives rise to an increasing tendency of temperatures and densities in time.
Note that this behavior could be expected from this choice of parameters since the best performance in the parametric study (Figure~\ref{fig:SA_eRho}) was achieved for higher thermal conductivities and lower EUV efficiencies.
The selection of parameters produces a more adequate fit in case 2, whose density and temperature estimates show a more accurate representation without significant increasing or decreasing tendencies over time.
Finally, the results in case 3 offer a stable approximation of densities, especially at lower altitudes, with increasing tendency at higher altitudes. This behavior is also observed for the temperature predictions.
More significant differences between the three locations are also observed at lower altitudes.

\section{Conclusions} \label{sc:conclusion}
This study has presented a high-order DG formulation of a physics-based space weather model for the dynamics of the upper atmosphere.
A model describing a neutral thermosphere in non-hydrostatic equilibrium has been introduced.
The approach is based on the compressible Navier-Stokes equations in the rotating Earth, formulated in terms of a novel set of variables to accommodate the large range of densities present in the problem and to reduce the stiffness in the system of equations.
The thermospheric dynamics is driven by a heat source term that models the energy absorption from solar EUV radiation.
The proposed numerical approach is based on a matrix-free DG formulation for the numerical solution of the model, making it suitable for GPU architectures and enabling the solution of large systems of equations.

A numerical example simulating the thermosphere in equinox conditions has been presented to validate the proposed methodology, both in 1D and 3D.
The 3D model shows a good agreement with experimental data along the Swarm A satellite orbit, offering neutral density predictions within the typical ranges of well established physics-based models.
Both 1D and 3D models show similar behavior and agree well with the other established empirical and physics-based models.

A set of cases with variable solar activity has been also presented to illustrate the ability of the model to describe a coherent dynamics and produce estimates of neutral quantities of interest in different scenarios.
These cases have also been used to study the model's sensitivity to some thermal parameters, such as the EUV heating source and the molecular thermal conductivity.
This computational study has showcased the 1D model's potential for integration into parametric studies. These studies facilitate capturing the physics-based model's sensitivities to factors like thermal conductivity or EUV efficiency.
While the analysis has concentrated on a very limited set of constant parameters, it could be broadened and formalized into global sensitivity analyses or incorporated into data assimilation studies to enhance the approach's prediction capabilities. Furthermore, these methods could be employed to derive more complex expressions for model quantities with larger associated uncertainty.

\section*{Acknowledgments}
The authors gratefully acknowledge the National Science Foundation for supporting this work (under grant number NSF-PHY-2028125).
In addition, the authors thank the Oak Ridge Leadership Computing Facility at the Oak Ridge National Laboratory for providing access to their GPU computational resources, supported by the Office of Science of the U.S. Department of Energy under Contract No. DE-AC05-00OR22725. The authors are also grateful to Aaron J. Ridley from the Univeristy of Michigan and Richard Linares from MIT for many technical discussions and advice.

Simulation results for GITM~\cite{Ridley2006}, developed by A. J. Ridley and collaborators at the University of Michigan, and TIE-GCM~\cite{Roble1982,Roble1988,Richmond1992,Qian2014}, developed by R. G. Roble and collaborators at the NCAR High Altitude Observatory, have been obtained through the Community Coordinated Modeling Center (CCMC) simulation services (\url{https://ccmc.gsfc.nasa.gov}), at NASA Goddard Space Flight Center.
Finally, Swarm A satellite data is available through Technical University Delft (\url{http://thermosphere.tudelft.nl}).

\appendix
\section{Absorption cross sections of the EUV spectrum} \label{app:EUVcrosssections}
As reported in Section~\ref{sc:EUV}, the computation of the EUV source term is performed  over a number of discrete wavelength values.
Table~\ref{tb:EUVcrosssections} summarizes the absorption cross sections for each species $\phi_s$, the EUV flux spectrum $F_{74113}$, and the scaling factor $A$ at each discrete wavelength bin $\lambda$. These values are obtained from~\cite{Richards1994, Torr1979}.

\begin{table}[htbp]
	\centering
	\begin{tabular}{|r|r|r|r|r|r|r|}
		\hline
		\multicolumn{1}{|r|}{$\lambda$ ({\AA})} & \multicolumn{1}{c|}{$F_{74113}$ (m$^{-2}$ s$^{-1}$)} &\multicolumn{1}{c|}{$A$} & \multicolumn{1}{c|}{$\phi_{\ch{N2}}$ (m$^2$)} & \multicolumn{1}{c|}{$\phi_{\ch{O2}}$ (m$^2$)} &\multicolumn{1}{c|}{$\phi_{\ch{O}}$ (m$^2$)} & \multicolumn{1}{c|}{$\phi_{\ch{He}}$ (m$^2$)} \\
		\hline
		50-100 & 1.200 $\times 10^{-13}$ & 0.0100 & 0.720 $\times 10^{-22}$ & 1.316 $\times 10^{-22}$ & 0.730 $\times 10^{-22}$ & 0.144 $\times 10^{-22}$ \\ 
		100-150 & 0.450 $\times 10^{-13}$ & 0.0071 & 2.261 $\times 10^{-22}$ & 3.806 $\times 10^{-22}$ & 1.839 $\times 10^{-22}$ & 0.479 $\times 10^{-22}$ \\ 
		150-200 & 4.800 $\times 10^{-13}$ & 0.0134 & 4.958 $\times 10^{-22}$ & 7.509 $\times 10^{-22}$ & 3.732 $\times 10^{-22}$ & 1.157 $\times 10^{-22}$ \\ 
		200-250 & 3.100 $\times 10^{-13}$ & 0.0195 & 8.392 $\times 10^{-22}$ & 10.900 $\times 10^{-22}$ & 5.202 $\times 10^{-22}$ & 1.601 $\times 10^{-22}$ \\ 
		256.3 & 0.460 $\times 10^{-13}$ & 0.0028 & 10.210 $\times 10^{-22}$ & 13.370 $\times 10^{-22}$ & 6.050 $\times 10^{-22}$ & 2.121 $\times 10^{-22}$ \\ 
		284.15 & 0.210 $\times 10^{-13}$ & 0.1380 & 10.900 $\times 10^{-22}$ & 15.790 $\times 10^{-22}$ & 7.080 $\times 10^{-22}$ & 2.595 $\times 10^{-22}$ \\ 
		250-300 & 1.679 $\times 10^{-13}$ & 0.0265 & 10.493 $\times 10^{-22}$ & 14.387 $\times 10^{-22}$ & 6.461 $\times 10^{-22}$ & 2.321 $\times 10^{-22}$ \\ 
		303.31 & 0.800 $\times 10^{-13}$ & 0.0250 & 11.670 $\times 10^{-22}$ & 16.800 $\times 10^{-22}$ & 7.680 $\times 10^{-22}$ & 2.953 $\times 10^{-22}$ \\ 
		303.78 & 6.900 $\times 10^{-13}$ & 0.0033 & 11.700 $\times 10^{-22}$ & 16.810 $\times 10^{-22}$ & 7.700 $\times 10^{-22}$ & 2.962 $\times 10^{-22}$ \\ 
		300-350 & 0.965 $\times 10^{-13}$ & 0.0225 & 13.857 $\times 10^{-22}$ & 17.438 $\times 10^{-22}$ & 8.693 $\times 10^{-22}$ & 3.544 $\times 10^{-22}$ \\ 
		368.07 & 0.650 $\times 10^{-13}$ & 0.0066 & 16.910 $\times 10^{-22}$ & 18.320 $\times 10^{-22}$ & 9.840 $\times 10^{-22}$ & 4.268 $\times 10^{-22}$ \\ 
		350-400 & 0.314 $\times 10^{-13}$ & 0.0365 & 16.395 $\times 10^{-22}$ & 18.118 $\times 10^{-22}$ & 9.687 $\times 10^{-22}$ & 4.142 $\times 10^{-22}$ \\ 
		400-450 & 0.383 $\times 10^{-13}$ & 0.0074 & 21.675 $\times 10^{-22}$ & 20.310 $\times 10^{-22}$ & 11.496 $\times 10^{-22}$ & 5.447 $\times 10^{-22}$ \\ 
		465.22 & 0.290 $\times 10^{-13}$ & 0.0075 & 23.160 $\times 10^{-22}$ & 21.910 $\times 10^{-22}$ & 11.930 $\times 10^{-22}$ & 6.563 $\times 10^{-22}$ \\ 
		450-500 & 0.285 $\times 10^{-13}$ & 0.0202 & 23.471 $\times 10^{-22}$ & 23.101 $\times 10^{-22}$ & 12.127 $\times 10^{-22}$ & 7.208 $\times 10^{-22}$ \\ 
		500-550 & 0.452 $\times 10^{-13}$ & 0.0088 & 24.501 $\times 10^{-22}$ & 24.606 $\times 10^{-22}$ & 12.059 $\times 10^{-22}$ & 0.958 $\times 10^{-22}$ \\ 
		554.37 & 0.720 $\times 10^{-13}$ & 0.0033 & 24.130 $\times 10^{-22}$ & 26.040 $\times 10^{-22}$ & 12.590 $\times 10^{-22}$ & 0 \\ 
		584.33 & 1.270 $\times 10^{-13}$ & 0.0052 & 22.400 $\times 10^{-22}$ & 22.720 $\times 10^{-22}$ & 13.090 $\times 10^{-22}$ & 0 \\ 
		550-600 & 0.357 $\times 10^{-13}$ & 0.0037 & 22.787 $\times 10^{-22}$ & 26.610 $\times 10^{-22}$ & 13.024 $\times 10^{-22}$ & 0 \\ 
		609.76 & 0.530 $\times 10^{-13}$ & 0.0162 & 22.790 $\times 10^{-22}$ & 28.070 $\times 10^{-22}$ & 13.400 $\times 10^{-22}$ & 0 \\ 
		629.73 & 1.590 $\times 10^{-13}$ & 0.0033 & 23.370 $\times 10^{-22}$ & 32.060 $\times 10^{-22}$ & 13.400 $\times 10^{-22}$ & 0 \\ 
		600-650 & 0.342 $\times 10^{-13}$ & 0.0118 & 23.339 $\times 10^{-22}$ & 26.017 $\times 10^{-22}$ & 13.365 $\times 10^{-22}$ & 0 \\ 
		650-700 & 0.230 $\times 10^{-13}$ & 0.0043 & 31.755 $\times 10^{-22}$ & 21.919 $\times 10^{-22}$ & 17.245 $\times 10^{-22}$ & 0 \\ 
		703.31 & 0.360 $\times 10^{-13}$ & 0.0030 & 26.540 $\times 10^{-22}$ & 27.440 $\times 10^{-22}$ & 11.460 $\times 10^{-22}$ & 0 \\ 
		700-750 & 0.141 $\times 10^{-13}$ & 0.0047 & 24.662 $\times 10^{-22}$ & 28.535 $\times 10^{-22}$ & 10.736 $\times 10^{-22}$ & 0 \\ 
		765.15 & 0.170 $\times 10^{-13}$ & 0.0039 & 120.490 $\times 10^{-22}$ & 20.800 $\times 10^{-22}$ & 4.000 $\times 10^{-22}$ & 0 \\ 
		770.41 & 0.260 $\times 10^{-13}$ & 0.0128 & 14.180 $\times 10^{-22}$ & 18.910 $\times 10^{-22}$ & 3.890 $\times 10^{-22}$ & 0 \\ 
		789.36 & 0.702 $\times 10^{-13}$ & 0.0033 & 16.487 $\times 10^{-22}$ & 26.668 $\times 10^{-22}$ & 3.749 $\times 10^{-22}$ & 0 \\ 
		750-800 & 0.758 $\times 10^{-13}$ & 0.0048 & 33.578 $\times 10^{-22}$ & 22.145 $\times 10^{-22}$ & 5.091 $\times 10^{-22}$ & 0 \\ 
		800-850 & 1.625 $\times 10^{-13}$ & 0.0048 & 16.992 $\times 10^{-22}$ & 16.631 $\times 10^{-22}$ & 3.498 $\times 10^{-22}$ & 0 \\ 
		850-900 & 3.537 $\times 10^{-13}$ & 0.0057 & 20.249 $\times 10^{-22}$ & 8.562 $\times 10^{-22}$ & 4.554 $\times 10^{-22}$ & 0 \\ 
		900-950 & 3.000 $\times 10^{-13}$ & 0.0050 & 9.680 $\times 10^{-22}$ & 12.817 $\times 10^{-22}$ & 1.315 $\times 10^{-22}$ & 0 \\ 
		977.02 & 4.400 $\times 10^{-13}$ & 0.0039 & 2.240 $\times 10^{-22}$ & 18.730 $\times 10^{-22}$ & 0 & 0 \\ 
		950-1000 & 1.475 $\times 10^{-13}$ & 0.0044 & 50.988 $\times 10^{-22}$ & 21.108 $\times 10^{-22}$ & 0 & 0 \\ 
		1025.72 & 3.500 $\times 10^{-13}$ & 0.0052 & 0 & 1.630 $\times 10^{-22}$ & 0 & 0 \\ 
		1031.91 & 2.100 $\times 10^{-13}$ & 0.0053 & 0 & 1.050 $\times 10^{-22}$ & 0 & 0 \\ 
		1000-1050 & 2.467 $\times 10^{-13}$ & 0.0044 & 0 & 1.346 $\times 10^{-22}$ & 0 & 0 \\ 
		\hline
	\end{tabular}
	\caption{Flux irradiance and absorption cross sections for each species throughout the EUV spectrum.}
	\label{tb:EUVcrosssections}
\end{table}

\section{Simulation setup for GITM and TIE-GCM} \label{app:CCMC}
The numerical results for GITM~\cite{Ridley2006} and TIE-GCM~\cite{Roble1982,Roble1988,Richmond1992,Qian2014} were obtained through the CCMC simulation service, which allows users to run space weather models on request.
Whereas minimal user intervention is required, a few choices are possible and these are displayed in Table~\ref{tb:CCMCsetup}. Any additional parameter required for the simulation was left at its default value.

\begin{table}[htbp]
	\centering
	\begin{tabular}{|l|l|l|}
		\hline
		Model version & GITM 21.11 &TIE-GCM 2.5 \\ \hline
		Period of study & 20-24/03/2014 & 20-24/03/2014 \\ \hline
		Resolution & 2.5$^\circ$ latitude, 5$^\circ$ longitude (default) & 1.25$^\circ$ latitude and longitude \\ \hline
		EUV model & EUVAC & Default \\ \hline
		$F_{10.7}$& 150 & 150 \\ \hline
		Ionospheric electrodynamic model & Weimer \cite{Weimer2005} & Heelis \cite{Heelis1982} \\ \hline
		Auroral precipitation model & FTA \cite{Wu2021} & Default \\ \hline
		Magnetic field/activity & $\bm{B} = (0,0,-5)$ nT,  $v_x = 400$km/s (default) & $K_p=1$  \\ \hline
	\end{tabular}
	\caption{Simulation setup for GITM and TIE-GCM in the CCMC run-on request service.}
	\label{tb:CCMCsetup}
\end{table}

\bibliographystyle{elsarticle-num} 
\bibliography{ref_SW}

\end{document}